\documentclass[12pt]{JHEP3}
\usepackage{amsmath,epsfig}
\usepackage{amssymb,amsfonts}
\usepackage{latexsym}
\usepackage[latin1]{inputenc}
\usepackage{xcolor}

\usepackage{graphicx}
\usepackage{longtable}

\relax
\renewcommand{\theequation}{\arabic{section}.\arabic{equation}}
\def\be{\begin{equation}}
\def\ee{\end{equation}}
\def\bea{\begin{eqnarray}}
\def\eea{\end{eqnarray}}

\newcommand\fverb{\setbox\pippobox=\hbox\bgroup\verb}
\newcommand\fverbdo{\egroup\medskip\noindent%
                        \fbox{\unhbox\pippobox}\ }
\newcommand\fverbit{\egroup\item[\fbox{\unhbox\pippobox}]}

\newcommand{\bear}{\begin{eqnarray}}

\newcommand{\eear}{\end{eqnarray}}

\newcommand{\bsea}{\begin{subeqnarray}}
\newcommand{\esea}{\end{subeqnarray}}
\newbox\pippobox

\def\6{\partial}

\def\a{\alpha}
\def\g{\gamma}

\def\pa{\partial}

\def\s{\sigma}

\def\sp{\;\;\;,\;\;\;}

\def\sq
\def\a{\alpha}

\def\tr{{\rm Tr}}

\def\hri#1#2{\href{http://arxiv.org/abs/#1}{[ArXiv:#1]#2}}

\title{Scaling of the Holographic AC conductivity for non-Fermi liquids at criticality}
\author{\Large   Elias Kiritsis$^{a,b}$, Francisco Pe\~na-Benitez$^{a}$\\
~\\
~\\
$^a$\href{http://hep.physics.uoc.gr}{Crete Center for Theoretical Physics},
Department of Physics, University of Crete, 71003 Heraklion, Greece.
~\\
$^b$Univ Paris Diderot, Sorbonne Paris Cit\'e, \href{http://www.apc.univ-paris7.fr/APC_CS/}{APC},
UMR 7164 CNRS, F-75205 Paris, France.
~\\
E-mail: \href{http://hep.physics.uoc.gr/~kiritsis/}{http://hep.physics.uoc.gr/~kiritsis/}, pena@physics.uoc.gr
}

\preprint{CCTP-2015-09\\
CCQCN - 2015 - 57}

\abstract{The frequency dependence of the AC conductivity is studied in a holographic model of a non-fermi liquid that is amenable to both analytical and numerical computation. In the regime that dissipation dominates the DC conductivity, the AC conductivity is described well in the IR by a Drude peak despite the absence of quasiparticles. In the regime where pair-production-like processes dominate the conductivity there is no Drude peak.
A scaling tail is found for the AC conductivity that is independent of the charge density and momentum dissipation.
Evidence is given that this scaling tail $\sigma_{AC}\sim \omega^m$ appears generically in quantum critical holographic systems and the associated scaling exponent $m$ is calculated in terms of the Lifshitz and conduction critical exponents.
}

\begin{document}
%\maketitle

\section{Introduction}\label{intro}

Insulators and superconductors are central concepts in strongly correlated electron materials giving rise to high T$_c$ superconductivity, \cite{zaanen}. The region in which the insulator-metal transition happens, \cite{mit}, is characterized by novel features of the electronic conductivity, \cite{gap}. Moreover, the scaling of the AC conductivity, \cite{ACScaling} has long been one of the benchmark features of such materials.

Holography has provided a new paradigm and a new arena for theoretical models that address physics at strong coupling. The setup is semiclassical (enforced by a large-N limit involved)\footnote{Useful reviews for condensed matter physics applications can be found in \cite{sachdev}, \cite{rev} and section 2 of \cite{kkp}.}.
It is a natural framework to describe quantum critical systems at zero and finite density. It is very convenient to describe conductivity, a major observable in condensed matter.
A study of holographic ground states has indicated that they are very diverse in their properties, \cite{Hartnoll:2008kx}, \cite{Horowitz:2009ij,Gubser:2009cg}, \cite{cgkkm,gk1,gk2}, \cite{helical}.

Most holographic systems analyzed at finite density are translationally invariant.
Exceptions also exist, using D-brane defects and magnetic vortices, \cite{def}, but such systems have been much more difficult to analyze so far. The standard symmetry argument indicates that the real part of the AC conductivity will have a $\delta(\omega)$ contribution as in a translationally invariant system a constant electric field generates an infinite current. The $\delta$ function is related, because of unitarity and causality using the dispersion relations to a ${1\over \omega}$ pole in the imaginary part of the conductivity.\footnote{This issue was analyzed in holography in  \cite{rev,rev2}.}
This $\delta$-function is distinct from the one appearing in superfluid/superconducting phases.

 The interaction with momentum dissipation agents has been discussed in rather general terms in  \cite{ Hartnoll:2012rj,impurities}. When the interaction with momentum dissipating centers is IR irrelevant, a perturbative IR calculation can determine the scaling of the IR DC conductivity. When the dissipation is IR relevant, it can change the nature of the saddle point, turning the system into an insulator, \cite{hd}-\cite{DGc}.

The formula obtained for the holographic DC conductivity  is a sum of two contributions,  \cite{massive},\cite{DGc},\cite{G},\cite{dgk}:
\be
\sigma_{DC}=\sigma_{DC}^{pc}+\sigma_{DC}^{diss}
\label{eq:1}\ee
The first term, $\sigma_{DC}^{pc}$,  has been interpreted, \cite{KO},  as a quantum-critical pair-creation contribution as it persists at zero charge density. For the RN black hole it is a constant proportional to the inverse of the bulk gauge coupling constant that counts the relative density of charge-carrying degrees of freedom to the neutral ones in the strongly-coupled plasma. More recently, it was verified in \cite{Donos:2014cya} that the first term in (\ref{eq:1}) is the electric conductivity in the absence of a heat current. The interpretation as a pair creation term is then natural since charged pairs are created with zero total momentum and therefore not contributing to a net matter flow.

The second contribution in (\ref{eq:1}) is due to the effects of dissipating momentum. When translation-breaking operators are irrelevant, the system is expected to be metallic and this term is expected to give the leading contribution to the DC conductivity. Then, a description of momentum relaxation in terms of the memory matrix formalism is appropriate and shows that the conductivity takes a Drude-like form, though no quasi-particle description is assumed \cite{Hartnoll:2012rj}. Moreover, in \cite{dgk} it was shown that in several cases that were studied, the drag related part of the conductivity had the form ${Q^2\over \Gamma_1+\Gamma_2+\cdots}$ where $Q$ is the charge density and $\Gamma_i$ are diffusion rates of the various diffusion mechanisms at play. It is strongly  suspected that this result is general, \cite{dgk}.
 However, there are other features that are less clear and the recent work \cite{dgy} has cast doubt on this simple characterization of the two terms in the conductivity formula. In this work we will keep using the previous characterizations (pair-production and Drude or dissipative part) in lack of better ones.

This general form of the DC conductivity was seen already in pure metric backgrounds in \cite{KO} and was generalized to dilatonic backgrounds in \cite{cgkkm}. In both cases, as the gauge field action is the DBI action, (\ref{eq:1}) is replaced by
\be
\sigma_{DC}=\sqrt{(\sigma_{DC}^{css})^2+(\sigma_{DC}^{diss})^2}
\label{a2}\ee
giving results compatible with (\ref{eq:1}) in the regimes where pair creation or drag diffusion dominates the conductivity.
In general, we expect a nonlinear formula that reflects the bulk action of the gauge field. In the probe DBI cases the momentum dissipation is due to the fact that charge degrees of freedom are subleading compared to uncharged ones. This means that there is a momentum conserving $\delta$-function but its coefficient is hierarchically suppressed.

In \cite{cgkkm} it was observed, based on (\ref{a2}), that for running scalars other than the dilaton and in 2+1 boundary dimensions, the drag DC resistivity, when it dominates,  is proportional to the electronic entropy. This seems to be a general property of strange metals where both the measured electronic entropy and resistivity   are linear in temperature. This  was extended in \cite{DSZ} to more general cases using the massive graviton theory, and most importantly provided a kinetic explanation for the correlation suggesting a more general validity.

The general properties of holographic conductivity were further corroborated recently by a careful study of the holographic current-current correlators and the associated properties of their poles in the complex plane, \cite{dgx,dgy}.
On the other hand the search for scaling regimes in the AC conductivity has been less studied. In \cite{cgkkm} it was suggested that generalized scaling geometries can have large-$\omega$ tails that are scaling and the scaling exponent was calculated in EMD hyperscaling violating solutions. This was further analyzed in \cite{G1,G11}.

In \cite{lattice1} it was claimed, based on the numerical solution of the conductivity equations in the presence of a charged lattice that a scaling of the AC conductivity was found with the phenomenologically relevant power $\s_{AC}\sim \omega^{-{2\over 3}}$. This suggestion did not survive however more accurate numerical study, \cite{donos}.

The above properties of holographic DC conductivity prompted the construction of holographic models with strange metal behavior, \cite{pol}, \cite{cgkkm},\cite{kkp}, \cite{pal}.
In particular, in \cite{kkp} a DBI action was assumed to be describing the dynamics of charge which otherwise moves in a AdS-Schwarzschild black hole (in light-cone coordinates). There was a light-cone electric field $F_{y+}=E$ whose inverse played the role of a doping-like  parameter that controls the phase diagram.
The formalism of \cite{KO} was used to compute the resistivity as well as the magneto-resistance in this system.

The $T+T^2$ behavior of the resistivity
in \cite{Cooper2009} and the $T+T^2$ behavior of the inverse Hall angle, observed in \cite{mackenzie} at {\em very low temperatures} $T<30K$, where a single scattering rate is present, were successfully described.
The model was  in accord with the distinct origin of the criticality at very low temperatures advertised in \cite{Hussey2009},
while the higher temperature, $T>100K$, scaling has different behaviors between
the linear temperature resistivity and the quadratic temperature inverse Hall angle, signaling two
scattering rates \cite{Tyler1997}.
In addition to the resistivity and inverse Hall angle, very good agreement was  also found with experimental
results of the Hall Coefficient, magnetoresistance and K\"ohler rule on various high-$T_c$ cuprates,
\cite{Cooper2009}-\cite{Daou2009}.
The model provided also a  change of paradigm from the notion of a quantum critical point, as it is quantum critical at $T\to 0$ on the entire overdoped region.

In this paper  we compute the AC conductivity of the holographic strange metal in \cite{kkp}. In this model the DC conductivity depends on two parameters: a scaling variable $t$ proportional to the temperature that also depends on doping and a scaling variable controlling the  total charge density $J$. They are defined in (\ref{scaling}) in terms of the parameters of the holographic theory.
Note that both scaling variables depend on the ``doping" related variable, $E$.
In \cite{kkp} it was argued that $x-x_o\simeq E^{-{1\over 2}}$ has the same dependence as ``doping" when the model is compared to experimental data.
In particular the $E\to 0$ limit corresponds to large doping while the $E\to \infty$ limit corresponds to optimal doping.
Therefore, at fixed temperature $T$ and charge density $J_+$, both scaling variables $t$ and $J$ increase with $x$.

There are two main regimes on the $(t,J)$ plane, shown in the middle plot of figure \ref{fig:regions}. The one in the upper left corner we call the quantum critical (QC) regime (also Pair-Production regime or Charge Conjugation Symmetric Regime (CCSR)). It is characterized by the fact that the DC conductivity in this regime is dominated by the pair production/charge conjugation symmetric contribution.
The regime in the lower-right corner is the Drude regime (DR). It is characterized by the fact that the DC conductivity in this regime is dominated by the dissipation (drag) contribution, \cite{KO}.

We have defined a parameter $q$ in (\ref{q}) to distinguish between the two regimes. $q\gtrsim 1$ denotes the DR while $q\lesssim 1$ denotes the QC/PP regime.
As the drag contribution to the conductivity is proportional to charge density, it follows that at zero charge density ($J=0$) we are always in the QC/PP regime.

We can vary the  total charge density $J$ to enter any of the two regimes. Inside each regime there are two regions with distinct behavior of the DC conductivity

\begin{itemize}

\item In the Drude regime ($q\gg 1$), when $t\ll 1$ the resistivity is linear in $t$ (and consequently in the temperature). We call this regime the {\em linear regime}. When $t\gg 1$, the resistivity is quadratic in $t$. We call this regime the {\em quadratic regime}.

\item In the QC/PP  regime,  ($q\ll 1$), when $t\ll 1$  the resistivity behaves as $\rho\sim t^{-{3\over 2}}$. We call this regime, the {\em regime I}.
    When $t\gg 1$ the resistivity behaves as $\rho\sim t^{-{1\over 2}}$. We call this regime, the {\em regime II}.

\end{itemize}

In the $t\to 0$ limit the theory has an effective Lifshitz exponent $z=2$ while as $t\to\infty$ it crosses over to an effective relativistic Lifshitz exponent, $z=1$, \cite{kkp}.
What we find in our analysis is as follows:

\begin{enumerate}

\item The characteristic temperature scale that controls the AC conductivity is an effective temperature $T_{eff}$ (with an associated scaling effective temperature $t_{eff}$ defined analogously).
    This is distinct from the system temperature $T$ and is due to the existence of a hierarchy of interactions. Similar effects have been observed in electronic systems and in holography, \cite{eff1}-\cite{eff4}.

The relation between $t_{eff}$ and $t$ is plotted in figure \ref{fig:temperature}. In the small $t$ regime $t_{eff}\sim {1\over \sqrt{t}}$ while at large enough $t$, $t_{eff}\simeq t$.
As was observed in \cite{eff2} it is always $t_{eff}\geq t$.

\item We define a {\em generalized relaxation time} $\tau$ by the IR expansion of the AC conductivity,
     \be
     \sigma(\omega) \approx \sigma_{DC}\left(1+ i ~\tau~\omega + \mathcal{O}(\omega^2)\right)
\ee
In the presence of the Drude peak, this is the conventional definition of an associated relaxation time. When there is no Drude peak present, $\tau$ is still well-defined, although in that case the interpretation as a relaxation time is lost.

    We give an analytical formula for $\tau$ in (\ref{eq:tau_final}).
    It takes a simple form for large and small values of the scaling temperature variable $t$.  In the regime I (see fig. \ref{fig:regions})  we obtain
\be
\tau\sim \sqrt{t}
\ee
while in the regime II (with $t\gg 1$) it is set by the inverse of the temperature
\be
\tau\simeq \frac{1}{ t}
\ee
These behaviors are shown graphically in figure \ref{fig:tau} for the various regimes.

\item In the Drude regime ($q\gg 1$) where the dominant mechanism for the conductivity is momentum dissipation, there is a clear Drude peak as seen for example in figures \ref{fig:sigmas}.

    In the PP/QC  regime it is also clear from the same figures that no Drude peak is to be seen in the IR of the AC conductivity.

\item At zero charge density (PP/QC regime) there is a scaling tail for the AC conductivity that behaves as
    \be
    |\sigma|\sim \left({\omega\over t_{eff}}\right)^{-{1\over 3}}\sp Arg(\sigma)\simeq {\pi\over 6}
\ee
For finite charge density this tail survives not only in the Charge Conjugation Symmetric regime  but also in part of the Drude regime, as seen in the various plots of figure \ref{fig:sigma1} as well as the ones of figure \ref{fig:abssigma1}.
The qualitative reason for this is that in the presence of the Drude peak, its tail falls off as ${1\over \omega}$ and this is faster than $\omega^{-{1\over 3}}$. Therefore the scaling tail will eventually win over the Drude peak for $\omega\geq \omega_0$ and the only condition that it is visible is that the UV structure of the theory kicks-in at $\omega_{UV}\gg \omega_0$.

\item This scaling tail of the AC conductivity generalizes to more general scaling holographic geometries, as previously described in \cite{cgkkm}. The equation that determines the conductivity is given in (\ref{3}) and the equivalent Schr\"odinger problem has a potential of the form $V_{eff}=V_1+\rho^2 V_2$ where $\rho$ is the IR charge density (that is proportional to the UV charge density).

  In particular, for a metric with Lifshitz exponent $z$, hyperscaling violation exponent $\theta$ and conduction exponent $\zeta$ with $d$ spatial boundary dimensions,  we find that in general
   \be
    |\sigma|\sim {\omega}^{m}\sp Arg(\sigma)\simeq -{\pi~m\over 2}
\ee
with
\be
 m = \left|{z+\zeta-2\over z}\right|-1\,,
\label{mm}\ee
There are several constraints in the parameters of this formula that are detailed in section \ref{acs}.
Moreover this formula is valid when the associated charge density does not support the IR geometry.

\item There are some special cases of (\ref{mm}) that deserve mentioning.
For an AdS$_2$ IR geometry, the exponent can be obtained by the $z\to\infty$ limit in (\ref{mm}) giving $m=0$.

For hyperscaling violating semilocal geometries we must take $\theta\to \infty$, $z\to \infty$ with ${\theta\over z}=-\eta$ fixed and obtain
\be
m=\big|{d-2\over 2}\eta+1\Big|-1={d-2\over d}~\eta
\ee

Finally, for the gauge field conformal case\footnote{This corresponds to the bulk coupling constant function $Z(\phi)$ of the gauge field asymptoting to a non-zero constant in the IR.} we obtain $m=0$ when $d=2$.

\item We find that for two spatial dimensions, negative values for the exponent $m$ are correlated with the sign of the Lifshitz exponent $z$ and the strength of the gauge field interaction in the bulk. Parametrizing the IR asymptotics of gauge coupling function $Z$ as
    \be
    Z\sim r^{\kappa}\sp r\to \infty
    \label{110}\ee
    in conformal coordinates, we obtain that $z\kappa>0$ for negative values of $m$ to be possible.

\item In the special case where the associated gauge field seeds the IR scaling geometry, $\kappa$ in equation (\ref{110}) is fixed as a function of $z,\theta$ and the exponent $m$ takes the value
    \be
    m=\Big|{3z-2+d-\theta\over z}\Big|-1
   \label{eq_m_int1} \ee
and is always positive. The special case where the IR geometry is AdS$_2$ can be obtained from the $z\to\infty$ limit of (\ref{eq_m_int1}) giving $m=2$. For  hyperscaling violating semilocal geometries, we must take, $\theta\to \infty$, $z\to \infty$ with ${\theta\over z}=-\eta$ fixed.
In this case we obtain
\be
m=|3+\eta|-1=2+\eta
\ee

\item An important issue is whether the scaling of the AC conductivity described above for the general scaling geometries is controlled by the dynamics of the charge density, or it is decided by the neutral system.
What we find is as follows: the effective Schr\"odinger potentials that controls the calculation of the conductivity has two parts. One that is independent of charge density and one that is proportional to the square of the charge density. In the generic case it is the first that controls the UV scaling of the AC conductivity described above. Only if the charge density is supporting the IR geometry, then the second part is of the same order as the first and it is the sum that controls the scaling of the AC conductivity. This special case is also the only one we found where the exponent $m$ in    (\ref{m}) is always positive. In the other cases it can also be negative, but unitarity implies always that $m\geq -1$.

\end{enumerate}

These findings suggest that this is a generic source of scaling tails in the AC conductivity in holographic systems. Moreover, in generic systems this scaling is expected to be  independent of the mechanism of momentum dissipation.

The structure of this paper is as follows:

In the next section we review the holographic model for a strange metal that we will study in this paper.

In section  \ref{Kubo} we derive the linearized equations leading to the calculation of the AC conductivity and establish the presence of an effective temperature for the charge dynamics.

In section \ref{sec:dc_cond} we analyzed the equations for the AC conductivity and derive an analytic formula for the generalized relaxation time.

In section \ref{sec:AC} we derive analytic formula for the scaling regime and numerically compute the AC conductivity for different values of the model parameters.

 In section \ref{acs} we do a general analysis of AC conductivity scaling in generic quantum critical saddle points.

Finally section \ref{out} contains our conclusions and outlook.

There are several appendices that provide technical details. Appendix \ref{ap:efflag} contains the effective DBI action and the equations of motion.
Appendix \ref{ap:wsm} contains the computation of the open string metric while appendix \ref{ap:change_coord} contains a change of coordinates in the world-sheet induced black hole metric. Appendix \ref{ap:fluctuations} derives in detail the equations for the fluctuations. In appendix \ref{ape} we present the computational details of the asymptotics of the AC  DBI conductivity.
Finally in appendix \ref{f0} we analyze theories with two charge densities and the associated scaling of the AC conductivity.

\section{A holographic Model For a Strange metal}

In this section we will review the holographic model introduced in \cite{kkp}. In the same reference a pedestrian introduction to the holographic idea and its potential condensed matter applications was given.
The charged matter is described by a  Dirac-Born-Infeld (DBI)  action\footnote{Details on the DBI action and its origin can be found in \cite{book}. The type of charge degrees of freedom it describes has been analyzed in \cite{cgkkm}.},  in the probe limit on a fixed black-hole  background. This background is a solution  of the Einstein action

\begin{equation}
I_E = \frac{1}{16\pi G_5}\int d^5 x\sqrt{-g}\left( R + \frac{12}{\ell^2}\right)\,,
\label{einstein}\end{equation}
where $g$, $R$ and $\ell$ are the determinant of the metric, Ricci scalar and the length scale of the model which controls the cosmological constant, respectively.
This solution is the standard AdS-Schwarzschild solution, \cite{witten}.

 The total action is given by the sum $I_E+S$, where $S$ is the
DBI action in (\ref{dbi})
 The bulk gravitational theory lives in 4+1 dimensions, while the dual boundary theory lives in 3+1 dimensions at the boundary of the bulk space-time.
However we will consider a non-relativistic limit (equivalently a null reduction in gravity by taking the light-cone $x^+$ as time). In this limit the remaining symmetry is a $z=2$ Schr\"odinger symmetry, one spatial dimension is lost and the boundary dual quantum field  theory lives in 2+1 dimensions.
For details we refer the reader to the original reference, \cite{kkp}.

 A crucial ingredient of the approach in \cite{kkp} was to write the AdS-Schwarzschild solution in light-cone coordinates
\begin{equation}
ds^2 = g_{++}(dx^+)^2 + g_{--}(dx^-)^2 + 2g_{+-}dx^+ dx^- +\sum_{i=2}^{i=3} g_{ii}(dx^i)^2 + g_{uu}(du)^2,
\end{equation}
with the coordinates ordered  as $x^M=(x_+,x_-,y,z,u)$. The components read
\begin{eqnarray}
g_{++}(u) = \frac{1}{4 b^2}g_{ii}(u) (1-h(u)) \qquad &,&\qquad  g_{--}(u) =  b^2 g_{ii}(u) (1-h(u))\nonumber \\
g_{uu}(u) = \frac{1}{4 u^3 g_{ii}(u) h(u)} \qquad &,&\qquad  g_{ii}(u) = \frac{1}{\ell^2 u} ,
\end{eqnarray}
where the blackening factor is
\be
\label{eq:blackening}
h =1 - (u/u_0)^2\;.
 \ee
In these coordinates the boundary is located at $u=0$ and the horizon at $u=u_0$. Notice that the radial coordinate $u$ is dimensionless  and to ensure the non relativistic scaling $z=2$ we assign $[b]=-1$ (in mass units).
Although the Ads-Schwarschild solution has the full conformal symmetry, in this coordinate system only the z=2 Schr\"odinger symmetry is manifest.

Another peculiarity of the system is that the coordinate $x^+$ will be considered as the field theory time. Such background interpolates between a $z=1$ CFT at high $T$ and a $z=2$ Lifshitz-like nonrelativistic system at $T\sim 0$. The black-hole temperature is given by $b^2\pi^2\ell^4T^2u_0=1$.

The background is a $5-$dimensional space time in light-cone coordinates $x^\pm = x\pm t$. The spatial coordinate $x$ plays a special role. From  a field theory viewpoint this coordinate frame can be seen as an infinite boost in the $x$ direction. Therefore  the $x$ dependence disappears and the non-trivial dynamics happens in the two transverse spatial dimensions $y,z$, \cite{kkp}

 The charged matter degrees of freedom are described by a DBI action
\begin{equation}
S = -\mathcal{N}\int d^5x \sqrt{-\det{(g+F)}},
\label{dbi}\end{equation}
where $g$ represents the background metric and $F$ the gauge field dual to the U(1) conserved current. It is possible to find a solution of the DBI equations of motion of the form

\begin{equation}
\label{eq:A0}
A_{(0)} = (E y + h_+(u) )dx^+ + (b^2E y + h_-(u) )dx^- +(b^2E x^- + h_y(u) )dy ,
\end{equation}
In the solution above,  $E$ represents an electric field in the $y$ direction. In  appendix A of reference \cite{kkp} is computed and shown  the full form of the background fields for this solution. Using the work of Karch-O'Bannon,  \cite{Karch:2007pd}  the nonlinear DC conductivity can be found, \cite{kkp}

\begin{eqnarray}
\label{eq:sigmadc1}
\sigma_{DC} &=&\frac{1}{\ell^2 u_0^2 }\sqrt{u_\star(E,u_0) \left(b^2 \mathcal N^2 + J_+^2 l^6 u_0^2 u_\star(E,u_0))\right)}
\end{eqnarray}
where $J_+$ is the charge density and $u_\star(E,u_0)$ is the critical point in the radial direction where both the numerator and denominator under the square root change sign, \cite{Karch:2007pd}. It will coincide, as we will show later-on, with the world-volume horizon.
It  is given as
\begin{equation}
\label{eq:ustar}
u_\star(E,u_0) = \frac{1}{2^{1/2}b E \ell^2}\left\{ \left[(2bE\ell^2u_0)^2+1\right]^{1/2} - 1\right\}^{1/2}\, ,
\end{equation}
where $u_\star(0,u_0)=u_0$ is satisfied.

To describe the key steps of the procedure to obtain  (\ref{eq:sigmadc1}) and (\ref{eq:ustar}), we remind the reader first that in the holographic set up, the boundary value of $A_{(0)}$ corresponds to a background gauge field switched on in the dual field theory.
We also note that after inserting (\ref{eq:A0}) in (\ref{dbi}), the action will depend only on the radial derivative of $h_\pm$ and $h_y$. Therefore, the system will have three $u$-independent quantities that can be identified with the one-point functions of the conserved quantities of the dual field theory, \cite{Karch:2007pd}
\begin{equation}
J_\pm = -\frac{\delta S}{\delta h_\pm'}\qquad ,\qquad J_y = -\frac{\delta S}{\delta h_y'}\, ,
\end{equation}
$J_+$ is interpreted as the charge density, $J_-$ related to the particle number and $J_y$ with the electric current in the $y$-direction.

The quantity inside the square root of the DBI action (\ref{dbi})  is not positive definite. If we demand reality of the action at all the points of the bulk space-time.
 several conditions must be satisfied. The first condition is the emergence of a surface besides the horizon satisfying the following equation\footnote{This hypersurface was called in the literature ''singular shell" \cite{Mas:2009wf} and afterwards was proven to be the location of the induced horizon of the world volume metric \cite{Kim:2011qh}.}
 \begin{equation}
 u-u_\star(E,u_0)=0\,,
 \end{equation}
 the second condition relates the particle number with the charge density as follows
 \begin{equation}
 J_- = \rho_- J_+ \qquad , \qquad \rho_- = \frac{2 u_0^2 - u_\star^2}{2 b^2 u_\star^2}\,,
\end{equation}
and the last one is Ohm's law
\begin{equation}
J_y = \sigma_{DC} E\,,
\end{equation}
with $\sigma_{DC}$ defined as in equation (\ref{eq:sigmadc1}).

The  conductivity receives two independent contributions, therefore we will split it as follows

\begin{equation}
\label{eq:sigmaDC}
\sigma_{DC}^2 =\sigma_0^2(\sigma_{DR}^2+\sigma_{QC}^2) \,,
\end{equation}
with the following redefinitions
\begin{eqnarray}
\sigma_0 &=&  2^{3/4}\mathcal N b \sqrt{b E} \\
\sigma_{DR}^2 &=& \frac{J_+^2 \ell^4 u_\star^2}{2^{3/2}\mathcal N^2 b^3 E} \\
\sigma_{DR}^2 &=& \frac{ u_\star}{2^{3/2} b  E\ell^2u_0^2}\,.
\end{eqnarray}

The term $\sigma_{DR}$ is proportional to the  charge density. It has been interpreted, \cite{Karch:2007pd}, as the term coming from the drag force acting on heavy charge carriers, in a  Drude-like paradigm (although there are no identifiable quasi-particles here). This interpretation was extended to more general dilaton-dependent DBI actions, \cite{cgkkm}.

The term $\sigma_{QC}$ is present even in the absence of charge density and it is interpreted as a contribution to the conductivity due to  pair creation of charges, \cite{Karch:2007pd}. This interpretation is not without pitfalls, but if fits the pair-production paradigm in more than one ways, \cite{dg,dgk}.
 We will call  from now on  the regime in which $\sigma_{DR}\gg \sigma_{QC}$  the  Drude regime (DR). We will call the regime in which the pair-production conductivity dominates
,  $\sigma_{DR}\ll \sigma_{QC}$,  the Quantum Critical regime (QC). The reason for those names will be better motivated in section \ref{sec:AC}.

We introduce scaling variables $t,J$ to describe the temperature and charge density respectively
\begin{equation}
t = \frac{\pi \ell b T}{2\sqrt{bE}} \quad , \quad J^2 = \frac{J_+^2}{(2\mathcal Nb)^2\sqrt{2}(bE)^3} \, ,
\label{scaling}\end{equation}
The physical quantities depend on these variables. The conductivities, %$\sigma_0$,
 $\sigma_{DR}$ and $\sigma_{QC}$ read in term of $t,J$ as
\begin{eqnarray}
%\sigma_0^2 &=&  \frac{2b^2\mathcal N^2}{\ell^2 u_\star} \frac{1}{t}\sqrt{A(t)-2t^2} \\
\sigma_{DR}^2 &=& \frac{J^2}{t^2A(t)}\sp A(t) =  t^2 + \sqrt{1 + t^4}  \\
\sigma_{QC}^2 &=& \frac{t^3}{\sqrt{A(t)}}\;.
\end{eqnarray}

\begin{figure}[tbp]
\begin{center}
\includegraphics[width=\textwidth]{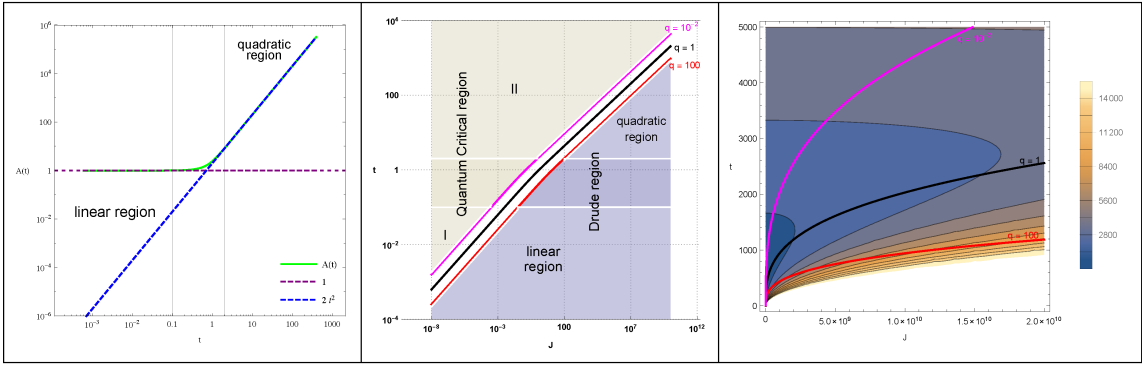}
\caption{
{\bf Left:} Comparison of the function $A(t)$ with its low and high $t$ behavior.
{\bf Center:} Location of the four regimes in the space of parameters $(J,t)$ log-log scale. The black line $q=1$, separates the DR respect to the QC regime. The magenta line represents the region with $q=10^{-2}$  and the red one $q=100$.
{\bf Right:} Contour Plot of the DC conductivity as a function of the scaling charge density and the temperature.}
\label{fig:regions}
\end{center}
\end{figure}

The conductivity formula (\ref{eq:sigmaDC}) has a complicated dependence on the parameter $t$. It can be however analyzed in  different regions of the parameter space, depending on whether  we are in a drag dominated regime (DR) or a pair-creation dominated regime (QC). To characterize the later we will define the ratio among Drude and QC conductivities,
\begin{equation}
q(t,J) = \frac{\sigma_{DR}^2}{\sigma_{QC}^2}\,.
\label{q}\end{equation}
These two regimes are illustrated in the second log-log plot of figure \ref{fig:regions}.
 The  DR and QC regimes can be split into two extra regions, namely $t\gg 1$ and $t\ll 1$ respectively.

 In the first plot of the figure \ref{fig:regions} we illustrate in a log-log plot the function $A(t)\sim 1$ for $t<1$ and $A(t)\sim 2t^2$ for $t>1$. The asymptotic behaviors imply that the resistivity $\rho = 1/\sigma$ will have a temperature dependence of the form
\begin{equation}
\rho = \left\{
\begin{array}{c}
\left.
\begin{array}{lcr}
(\sigma_0 J)^{-1} t \quad & q\gg 1, t\ll 1 \qquad \text{linear}~~~~~\\
\sqrt{2}(\sigma_0J)^{-1}  t^2 \qquad & q\gg 1, t\gg 1 \qquad \text{quadratic}
\end{array}
\right| \text{ Drude regime (DR)}\\
\left.
\begin{array}{lcr}
\sigma_0^{-1}t^{-3/2} \quad & q\ll 1, t\ll 1 \qquad \text{regime ~I}\\
\sqrt{2}\sigma_0^{-1} t^{-1/2} \quad & q\ll 1, t\gg 1\qquad \text{regime~II}
\end{array}
\right|  \text{Quantum critical (QC)}
\end{array}\right. \quad.
\label{regime}\end{equation}

We observe in the DR (large enough $J$) that conductivity is linear at low temperatures and becomes quadratic at higher temperatures.
In general the DC conductivity for a background with dynamical exponent $z$ is $\sigma_{DC}\sim T^{2\over z}$, \cite{pol}.
The background in question here has a $z=2$ effective Schr\"odinger symmetry in the IR which is restored to a $z=1$ conformal symmetry in the UV.

The theory has an extra parameter, the Electric field, $E$. This acts as an external parameter, that is similar to doping, pressure, electric and magnetic fields in strange metals. Varying $E$, from (\ref{scaling}) we observe that this changes $t$ and $J$ and therefore affects the crossover of the conductivity from linear to quadratic. Indeed, in \cite{kkp} it was shown that varying $\infty<E<0$ produces a phase diagram for the drag-related conductivity that is qualitatively similar to that measured recently in cuprates, \cite{Cooper}. Here $E\to\infty$ corresponds to optimal doping while $E\to 0$ corresponds to the overdoped limit. The cross-over scale between linear and quadratic regimes asymptotes to zero as $E\to 0$ while it diverges when $E\to \infty$.

In \cite{kkp} the general magnetoresistivity was also computed. It was shown that it reproduces the $T+T^2$ behavior of the inverse Hall angle observed in \cite{mackenzie}
at {\em very low temperatures} $T<30K$, where a single scattering rate is present.

In addition to the resistivity and inverse Hall angle, very good agreement was also found with experimental
results of the Hall Coefficient, magnetoresistance and K\"ohler rule on various high $T_c$ cuprates
\cite{Cooper,mackenzie,Hussey2009,Tyler1997,Ong1991,Takagi1992,Kendziora1992,hwang,harris,Hussey1996,tyler,Naqib2003,Nakajima2004,E1,E2,AndoBoebinger,Daou2009}.
This  model provides a change of paradigm from the notion of a quantum critical point,
as it is quantum critical as $T\to 0$ on the entire overdoped region.
In this sense this work breaks apart from other holographic approaches \cite{Cubrovic:2009ye, Liu:2009dm,Faulkner:2009wj},
where the measured transport is due to loop fermion effects. As such, it is applicable to a more general class of materials
{\it e.g.,} $d$ and $f$-electron systems, where the low temperature resistivity varies as $T + T^2$ \cite{stewart} and
exhibit a quantum critical line \cite{Cooper,zaum}.

\section{Stationary transport from Kubo formulas and the effective temperature\label{Kubo}}

We will use linear response theory to study the electric properties of the system. To do so  we need to switch-on fluctuations for the gauge field
on top of the background discussed before $A= A_{(0)}+a$ with $a$ infinitesimal. These fluctuations will evolve with a dynamics determined by the effective Lagrangian \cite{Kim:2011qh} (see appendix \ref{ap:efflag} for a derivation)
\begin{equation}
L_{eff}=-\mathcal{N}\sqrt{-\det{s}}\left( \frac{1}{4g_5^2}f_{MN}f^{MN} + \frac{1}{8\sqrt{-\det{s}}}\epsilon^{MNPQR}f_{MN}f_{PQ}Q_R \right),
\end{equation}
where  $f=\mathrm d a$, the effective (open string) metric is  defined as $s_{MN} = g_{MN} -  (F_{(0)}g^{-1}F_{(0)})_{MN}$, the effective coupling is $g_5^2=\sqrt{-\det{s}}/\sqrt{-\det{ (g + F_{(0)})}}$ and the vector present in the Chern Simons is
\begin{equation}
Q_R = -\frac{\sqrt{-\det{ (g + F_{(0)})}}}{8}\epsilon_{MNPQR}~\theta^{MN}\theta^{PQ},
\end{equation}
\be
\theta^{MN} = {1\over 2}\left[[( g + F_{(0)})^{-1}]^{MN}-(M\leftrightarrow N)\right]\;.
\ee

Notice that we are raising  indices with the open string metric $s_{MN}$. In particular the line element is
\begin{eqnarray}
\nonumber ds^2 &=& s_{++}(dx^+)^2  + 2s_{+-}dx^+dx^- + 2s_{+u}dx^+du + 2s_{+i}dx^+dx^i  +  s_{--}(dx^-)^2 +  \\
\label{eq:WSmetric}&& 2s_{-i}dx^-dx^i + s_{ij}dx^idx^j + s_{uu}du^2 +2s_{ui}dudx^i \, ,
\end{eqnarray}
where the exact form of the components of the world-volume metric are presented in appendix \ref{ap:wsm} in terms of the background metric and gauge field. Finally, the equations of motion for the fluctuation are
\begin{equation}
\label{eq:MaxFluct}
\partial_M\left( \frac{\sqrt{-\det s}}{g_5^2}f^{MN} \right) -\frac{1}{2}\epsilon^{NMRPQ}\partial_M Q_R f_{PQ} = 0\, .
\end{equation}

\subsection{The effective temperature}
\label{sec:change_coord}

  We will now show that the open string metric has a horizon in a modified location with respect to the background horizon located at $u_0$. In particular we will show that $u_\star$ represents the location of the horizon for the effective metric $s$.

As a consequence of this fact, the charged matter is at a temperature different than the bulk system.
This can happen in other holographic systems, \cite{ct}, \cite{gkmn}, \cite{Karch-h} and is a peculiarity of the large-$N$ limit implicit in holographic theories, that generates a hierarchy of interactions.

\begin{figure}[t!]
\begin{center}
\includegraphics[height=0.32\textwidth]{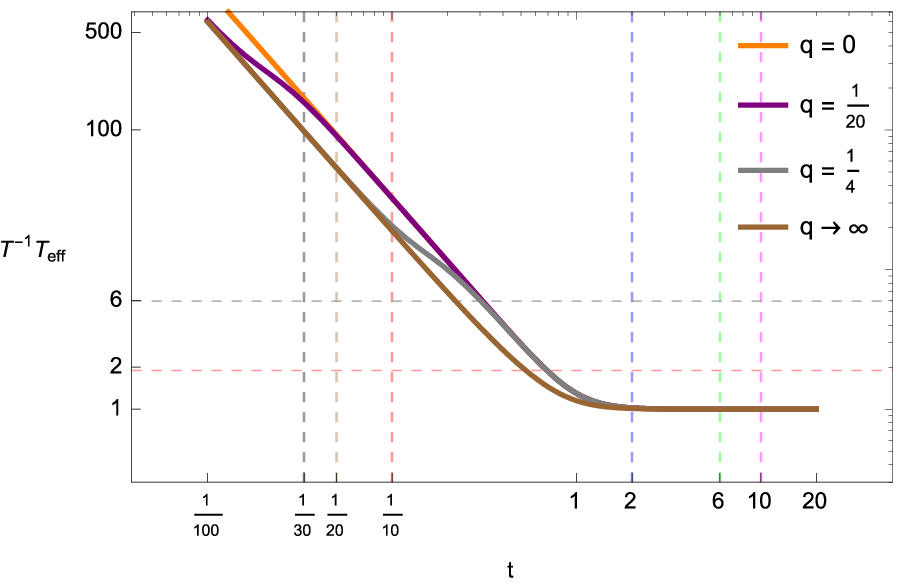}
\includegraphics[height=0.32\textwidth]{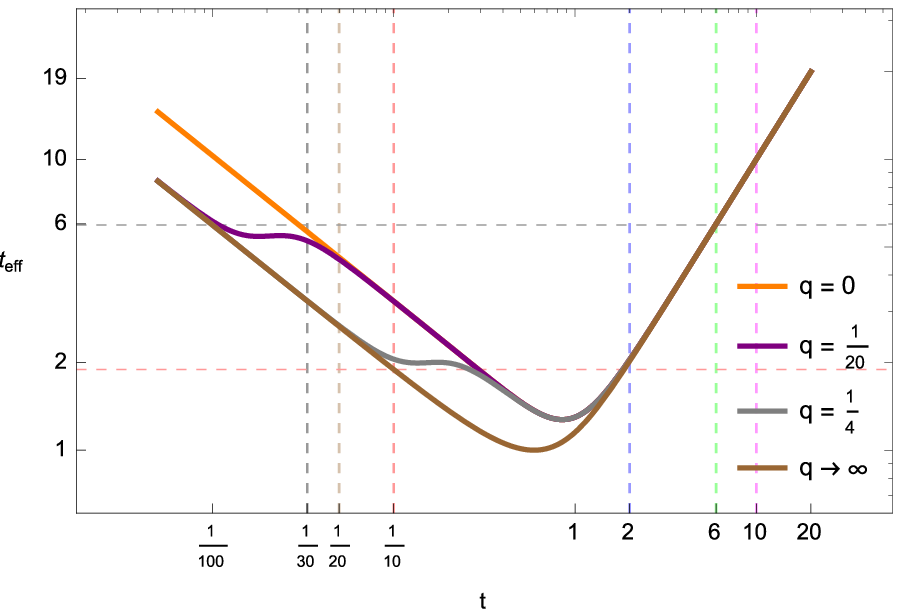}
\caption{ Dependence of the world-volume temperature $T_{eff}$ with the parameters $J$ and $t$, the effective temperature is always bigger than the background temperature $T$.}
\label{fig:temperature}
\end{center}
\end{figure}

 To show the previous statement  we will diagonalize the metric (\ref{eq:WSmetric}). In doing so we will be able to extract the location of the horizon and also we will be able to compute the temperature of the induced black-hole. We change coordinates as follows
\begin{eqnarray}
\nonumber \mathrm d x^+ &=& \mathrm d \tau + f^+_-(u) \mathrm dX^- + f^+_u(u) \mathrm d u   \, ,\\
\nonumber \mathrm dx^- &=& \mathrm d X^- + f^-_\tau(u) \mathrm d\tau + f^-_u(u) \mathrm d u   \, ,\\
\nonumber \mathrm dy &=& \mathrm dY + f^y_\tau(u) \mathrm d\tau + f^y_u(u)\mathrm d u   \, , \\
\label{eq:coordinatesTrans} \mathrm dz &=& \mathrm dZ + f^z_\tau(u) \mathrm d\tau  \, ,
\label{coor}\end{eqnarray}
and also rescale the radial coordinate as $u\to u_\star u$. The full expression for the functions $f$'s can be found in  appendix \ref{ap:change_coord}.
The diagonal metric becomes
\begin{equation}
\mathrm d s^2 = \tilde s_{uu} \mathrm du^2 + \tilde s_{--} (\mathrm d X^-)^2 + \tilde s_{\tau\tau}\mathrm d \tau  +\tilde  s_{yy}\mathrm dY^2 +\tilde  s_{zz}\mathrm dZ^2 \, .
\end{equation}
where again the component functions can be found in appendix \ref{ap:change_coord}.  From the new form of the metric we can find the location of the open string horizon and the effective temperature associated to this black hole. We find that  $\tilde s_{\tau\tau}(1) = \tilde s_{uu} ^{-1}(1) = 0$, which indicates that the new horizon  is located at the position $u=1$ or equivalently  at $ u_\star$. Expanding around this point we obtain the near-horizon expressions
\begin{eqnarray}
\label{eq:H1approx}
 \tilde s_{uu} &\approx & \frac{\ell^2 u_0^2 \left(2 u_0^2 - u_\star^2 \right)}{4 (1-u) \left(2 (\tilde J_+)^2 u_0^3 u_\star + 6 u_0^4 - 5 u_0^2 u_\star^2 + u_\star^4\right)} \, ,\\
 \label{eq:H3approx}
\tilde s_{\tau\tau} &\approx & -\frac{2 (1-u) u_0 \left(2 u_0^2 - u_\star^2\right)}{ b^2 \ell^2 u_\star^3 \left((\tilde J_+)^2 u_\star + u_0\right)} \, , \\
\tilde s_{--} &\approx & \frac{4 b^2 u_\star}{\ell^2 u_0^2} \, , \\
\tilde s_{ij} &\approx & \frac{1}{\ell^2 u_\star}\delta_{ij}  \, ,
\end{eqnarray}
where we have introduced the redefinition $ \tilde J_+ = \frac{u_0^{3/2}\ell^3}{b\mathcal N}J _+ $. Following \cite{eff4} we compute the effective temperature $T_{eff}$ from these formulae
\begin{equation}
T_{eff} = T\sqrt{\frac{6u_0^4 + 2(\tilde J^+)^2 u_0^3 u_\star - 5u_0^2u_\star^2 + u_\star^4}{2u_\star^3(u_0 + (\tilde J^+)^2 u_\star)}}.
\end{equation}
The previous equation written  in term of the scaling variables (\ref{scaling}) reads
\begin{equation}
t_{eff} = t \sqrt{ \frac{ J^2 A(t)^2 \sqrt{ A(t) - 2 t^2} + t^5 \left( t^2 A(t) + 2 t^4 + 3 \right) }{2 \sqrt{2} t^3 \left( t^5 \sqrt{A(t)} + J^2 \right) } }\, .
\end{equation}
We illustrate in  Fig. \ref{fig:temperature} the $t$ dependence of $t_{eff}$.

Note that the function $t_{eff}(J,t)$ has a minimum around $t\sim1$.
It is possible to find its exact location in the limits of $J\to 0$ and $J\to\infty$
\begin{eqnarray}
t_{eff}(t_{min}) &=& 1 \quad , \quad t_{min}(J=\infty) = 2^{-3/4} \approx 0.59 \\
\nonumber t_{eff}(t_{min}) &\approx& 1.274 \quad , \quad t_{min}(J=0) = 2^{-3/4}\frac{\sqrt[4]{\sqrt{133} \cos (\phi )-1}}{\sqrt{2} \sqrt[4]{3}} \approx 0.85
\end{eqnarray}
where $\phi =\frac{1}{3} \left(\pi -\tan ^{-1}\left(\frac{6 \sqrt{13443}}{1367}\right)\right)$\footnote{There is also a second minimum that satisfies $t_{min}^{(2)}\leq t_{min}$ and  $t_{min}^{(2)}\to 0$ when $q\to 0$.}.

 It is worth mentioning that $t_{eff}\geq t$ always, as first observed for holographic systems in \cite{gkmn}.  Therefore the presence of the electric field heats up the system above the temperature of the heat bath given by the bulk black hole.

If we analyze $t_{eff}$ in the four distinct regimes of the parameters $J$ and $t$ (defined in (\ref{regime})) we obtain:

\begin{equation}
t_{eff}  \approx \left\{
\begin{array}{lr}
2^{-3/4} t^{-1/2}   & \text{linear regime}    \\
t   & \text{quadratic regime}   \\
2^{-3/4}\sqrt{3} t^{-1/2} & {\rm regime}~\text{I}   \\
t  & {\rm regime}~\text{II}
 \end{array}
\right. \quad.
\end{equation}

\section{The fluctuation equations}

In the previous section we derived the equations of motion  for fluctuations on the black hole background with Schr\"odinger symmetry in equation  (\ref{eq:MaxFluct}).  We will work in the new coordinates introduced in (\ref{coor}). As the background electric field is pointing in the $y$-direction we will switch-on a fluctuation in the $z$-direction which will be transverse. We will solve this equation and from the source and vev of the solution we will determine the two point function of the dual current and from it the AC conductivity using the Kubo formula.

 To compute the frequency dependence of the conductivity it is only necessary to study the time dependence in the fields. Without loss of generality we will switch on  a fluctuation in the $z$ direction $a_z=a_z(u,x^-,\tau)$\footnote{In appendix \ref{ap:fluctuations} we write the full system of equation of motion}. Since in the dual field theory the time coordinate is given by $x^+$, we need to take into account this fact. To do so, we write   the near boundary expansion of the change of coordinates in Eqs. (\ref{eq:coordinatesTrans}) as
\begin{eqnarray}
\nonumber  x_b^+(\tau,X^-) = x^+ &=&  \tau - \frac{1}{\rho_-} X^-   \quad , \quad x^- = X^- + \rho_- \tau    \, ,\\
\label{eq:coordinatesTrans_exp} y &=& Y   \quad , \quad z = Z  \, ,
\end{eqnarray}
We therefore introduce the following plane wave ansatz  to evolve the fields with the time $x^+$
\begin{equation}
 f(u,X^-,\tau) \to f(u)e^{-i\omega x_b^+(\tau,X^-)}\, .
 \end{equation}
 The only non-trivial Maxwell equation is
\begin{equation}
\label{eq:az_q0}
\left( g_5^{-2}\sqrt{-\tilde s} \tilde s^{uu}\tilde s^{zz} a_z'(u )\right)'  - g_5^{-2}\sqrt{-\tilde s}  \tilde s^{zz}\omega^2 \left( \frac{\tilde s^{--}}{\rho_-^2}  + \tilde s^{\tau\tau} \right)a_z(u) = 0  \,,
\end{equation}
and  the Chern-Simons term vanishes for this sector. In order to use the holographic prescription to compute the retarded Green function, we need to impose as a boundary conditions $a_z(0)=a_0$ and to select the infalling mode at the horizon.

Performing a near-horizon expansion of  Eq. (\ref{eq:az_q0}) we obtain
\begin{equation}
\label{eq:fluct_hor}
\frac{\omega^2/(4\pi T_{eff})^2}{(u-1)^2}a_z(u) +\frac{1}{ u- 1}a_z'(u)+ a_z''(u) = 0\;.
\end{equation}
The near-horizon solutions are $a_z(u)\sim (1-u)^{\pm i w}$, with the dimensionless frequency $w$ defined as
\begin{equation}
w = \frac{\omega}{4\pi T_{eff}}\;.
\end{equation}
Choosing the negative sign  ensures the selection of the infalling boundary condition. Notice that the scaling frequency is naturally defined with $T_{eff}$.

\subsection{The DC conductivity and the relaxation time}
\label{sec:dc_cond}
To solve the  differential equation for arbitrary frequency it is necessary to use numerical tools. The low-frequency conductivity  however can be obtained  using  perturbation theory.  To do so we will expand the field $a_z$ up to second order in $w$ and the infalling boundary condition will be set by expanding the appropriate solution of  eq. (\ref{eq:fluct_hor}) as follows
\begin{equation}\label{eq:field_exp}
a_z(u) = (1-u)^{- i w}\left(\mathcal A^{(0)}_z(u) + w \mathcal A^{(1)}_z(u) + w^2 \mathcal A^{(2)}_z(u) \right)\, .
\end{equation}
Using the previous ansatz and the Kubo formula we can write the conductivity as
\begin{eqnarray}\label{eq:Kubo}
\sigma_{zz}(\omega) &=& - \frac{ 2b\mathcal N \pi T }{ \ell u_\star }\frac{ i }{ \omega } \frac{ a'_z(0) }{ a_z(0) } \\
\nonumber &=& -i \frac{b \mathcal N \sqrt{u_0} }{2\ell u_\star} \left( \frac{4  \pi   T  }{  \omega } \frac{ \mathcal A_z^{'(0)}(0) }{ a_0 }  +  \frac{   T  }{   T_{eff} }  \left( \frac{ \mathcal A_z^{'(1)}(0)}{ a_0 }  + 1 \right)  +  \frac{  T  \omega  }{4 \pi  T_{eff}^2 }\frac{\mathcal A_z^{'(2)}(0)}{ a_0} \right)\,.
\end{eqnarray}

In order for the system to have  a finite DC conductivity $\mathcal A_z^{'(0)}(0)$ must vanish.
 We do know  already that the present model has a finite conductivity, from its calculation in \cite{kkp}, therefore assuming $\mathcal A_z^{'(0)}(0)=0$ and setting $\omega=0$ we obtain the DC conductivity in terms of the solution in (\ref{eq:Kubo})\footnote{In any case we will solve the system and we will confirm that $\mathcal A_z^{'(0)}(0)=0$.}
\begin{eqnarray}
\label{eq:Kubo_dc} \sigma_{zz}^{DC} &=& -i \frac{b \mathcal N \sqrt{u_0} }{2\ell u_\star}   \frac{   T  }{   T_{eff} }  \left( \frac{ \mathcal A_z^{'(1)}(0)}{ a_0 }  + 1 \right) \, ,
\end{eqnarray}
There is another transport coefficient with units of time that can be read off eq. (\ref{eq:Kubo}).
We write the AC conductivity as
\begin{eqnarray}
\sigma(\omega) &\approx& \sigma_{DC}\left(1+ i \tau\omega + \mathcal{O}(\omega^2)\right) \, ,\\
\label{eq:Kubo_relax} \tau &=& -\frac{1 }{4\pi}\frac{i}{\sigma_{zz}^{DC}}   \frac{  T   }{  T_{eff}^2 }\frac{\mathcal A_z^{'(2)}(0)}{ a_0} \, .
\end{eqnarray}
If the conductivity is gives as a sum over poles in the upper half plane, then $\tau$ is determined by the pole with the smallest imaginary part. 
In a regime which is of the Drude type, $\tau$ can be interpreted as the relaxation  time. We will call it the {\em generalized relaxation time}.
 In other regimes bit cannot be interpreted strictly speaking as a relaxation time.

To finally compute the conductivity and the generalized relaxation time it is necessary to calculate the boundary value of the derivative of the first and second order fields.

 Substituting the expansion (\ref{eq:field_exp}) into  eq. (\ref{eq:az_q0}) the system can be rewritten as follows
\begin{equation}
\label{eq:perturb1}\partial_u\left(\alpha_0(u)\alpha(u)\partial_u\mathcal A^{(i)}_z(u)\right) = \mathcal S_z^{(i-1)}(u)\, ,
\end{equation}
where the index $i$ takes values $i=0,1,2$.  $\mathcal S_z^{(i)}$ is a source term  depending on the solutions at lower order in the perturbative expansion, and with the first term vanishing $\mathcal S_z^{(-1)}=0$. The $\alpha$ functions are defined as
\be
\alpha_0(u) = \frac{2\mathcal N}{\ell^3 u_0^2}(-1+u^2) \sp
\alpha(u) =  \frac{1}{\alpha_0(u)}g_5^{-2}\sqrt{-\tilde s}\tilde s^{uu}\tilde s^{zz}\,.
\ee
The function $\alpha(u)$  was normalized so that $ \alpha(u)\left|_{J_+\to 0 \, ,\, E\to 0}\right.=1$.
 The  source terms   and the explicit calculation to solve the differential equation can be found in appendix \ref{ap:fluctuations}. The general solution for eq. (\ref{eq:perturb1}) can be written as
\begin{equation}
\mathcal A_z^{(i)}(u) = C_1^{(i)} + \int \mathrm du \frac{1}{\alpha_0(u)\alpha(u)}\left(C_2^{(i)} + \int \mathrm du \, \mathcal S_z^{(i-1)}(u)\right)\, ,
\end{equation}
the integration constants $C_2^{(i)}$ are fixed demanding regularity for the solution at the horizon and $C_1^{(i)}$ is directly identified as the non-normalizable mode.

Finding the solution for \ref{eq:perturb1} and substituting  into Eqs. (\ref{eq:Kubo_dc}) and (\ref{eq:Kubo_relax}) we obtain the expected value for the static conductivity Eq. (\ref{eq:sigmaDC}) and the following expression for the generalized relaxation time
\begin{eqnarray}
\nonumber T_{eff}~\tau &=& \frac{\log{2}}{2\pi} - \sigma_{DC}\frac{ \ell   u_{0}^{3/2}}{4 \pi  b \mathcal N  u_\star}\frac{T_{eff}}{T}\int_{0}^{1}\mathrm dx \, \frac{\alpha'(x)}{\alpha(x)^2} \log{\left(\frac{1-x}{1+x}\right) } +\\
\label{eq:tau_final}&&- \frac{1}{\sigma_{DC}}\frac{b \mathcal N  u_\star }{4 \pi  \ell  u_{0}^{3/2} }\frac{T_{eff}}{T}\int_{0}^{1}\mathrm dx \, p'(x) \log{\left(1-x^{2}\right) } \, ,
\end{eqnarray}
where
$$
p(u) = -2g_5^{-2}\sqrt{-\tilde s}\frac{(-1+u^2)u_0}{b^2\ell \mathcal N u} \tilde s^{zz}\left( \frac{\tilde s^{--}}{\rho_-^2}  + \tilde s^{\tau\tau} \right).
$$
The formula (\ref{eq:tau_final}) gives an analytic expression for the generalized relaxation  time.
In the limit where  both the charge density and the electric field asymptote to zero, the integrals vanish because $p(u)=\alpha(u)=1$.  In this limit the generalized relaxation time takes the constant (QC) value
\begin{equation}
\tau =  \frac{\log{2}}{2\pi T}\, .
\end{equation}
When the charge density $J=0$ there are also two  interesting limits we can study.
The first one is $t\ll 1$ where
\begin{eqnarray}
\alpha(u) &\approx& \frac{1}{2 t^2}\sqrt{\frac{u^4+u^3+2 u^2+u+1}{u+1}} \\
p(u) &\approx& \sqrt{2} t \sqrt{\frac{u+1}{u^4+u^3+2 u^2+u+1}} \, ,
\end{eqnarray}
 and substituting these functions in (\ref{eq:tau_final}) we can write the generalized relaxation time as
\begin{equation}
\label{eq:tauJ0:tsmall}
T\tau = \frac{t^{3/2}}{2^{1/4} \pi T}\int _0^1\frac{1}{ \sqrt{(1+u)\left(u^2+1\right) \left(u^2+u+1\right)}} \approx 0.1565~t^{3/2} \qquad , \qquad t\ll 1 \, .
\end{equation}

The other regime we can study is $t\gg 1$ where the functions $\alpha$ and $p$ read
\begin{eqnarray}
\alpha(u) &\approx& 1+ \frac{u (u (u+2)+2)+2}{8 t^4 (u+1)} \\
p(u) &\approx& 1-\frac{u (u+1)^2+1}{8 t^4 (u+1)} \,.
\end{eqnarray}
In this case the generalized relaxation time is given by
\begin{equation}
\label{eq:tauJ0:tbig}
T\tau = \frac{\log 2}{2 \pi } +\frac{\log (32)-2}{32 \pi }t^{-4}+{\cal  O}(t^{-8}) \qquad ,\qquad t\gg 1 \, ,
\end{equation}
Finally we summarize in the Table \ref{teb_tau} the two asymptotic values obtained from the general Eq. (\ref{eq:tau_final}) and obtained above.
\begin{table}
\begin{center}
\begin{tabular}{|c|c|}
\hline
$\tau$ & $J=0$ \\
\hline
$\sim T^{-1}$ & $t\gg1$\\
$\sim \left({T\over E}\right)^{1/2}$ & $t\ll1$\\
\hline
\end{tabular}
\end{center}
\label{teb_tau}
\caption{High and low-temperature behavior of the generalized relaxation time}
\end{table}

\section{The AC conductivity}
\label{sec:AC}
To understand the full frequency dependence of the optical conductivity it is necessary to resort to the numerical solution of the equation.
In order to have  analytical control we can study the high-frequency behavior using WKB techniques after writing the equation as a Schr\"odinger system. We will start this section by analytically studying the high-frequency regime of the system and then we will study the model for arbitrary frequency using numerical methods.

\subsection{High Frequency behavior}
\label{sec:efectPotential}

In order to have an idea of the behavior of the conductivities at high frequency we can rewrite the equation of motion as the Schr\"odinger equation, from which we can read the effective potential that can provide insight on the solutions.
To do so we rewrite  Eq. (\ref{eq:az_q0}) as
\begin{equation}
-\frac{H_0 }{H_1}\left(H_1 a_z'  \right)' -  w^2 a_z=0\,,
\end{equation}
with
\begin{align}
H_0(u) = \frac{1}{(4\pi T_{eff})^2}\frac{4(1-u^2)^2}{b^2\ell^4u_0u}\frac{\alpha(u)}{p(u)} \qquad , \qquad H_1(u) = \alpha_0(u)\alpha(u)\, ,
\end{align}
If we redefine the radial coordinate and the field as
\begin{equation}
\label{eq:redef}
dr = H_0(u)^{-1/2}du \qquad , \qquad a_z(u) =  \phi(u)\psi(u)
\end{equation}
with
\begin{equation}
\phi(u) = \sqrt[4]{\frac{H_0(u)}{H_1(u)^2}}
\end{equation}
the equation can be rewritten as
\begin{align}\label{eq_Schro1}
-\psi''(r) + \left( V(r) - w^2\right)\psi(r) = 0
\end{align}
and the Schr\"odinger potential reads
\begin{align}
\label{eq:potential}
V(u) = -\frac{H_0(u)}{\phi (u)}\left(\frac{ H_1'(u) \phi '(u)}{H_1(u) } -  \phi ''(u)\right)
\end{align}
where the derivatives are taken with respect to the coordinate $u$.
\begin{figure}[h!]
\begin{center}
\includegraphics[width=1\textwidth]{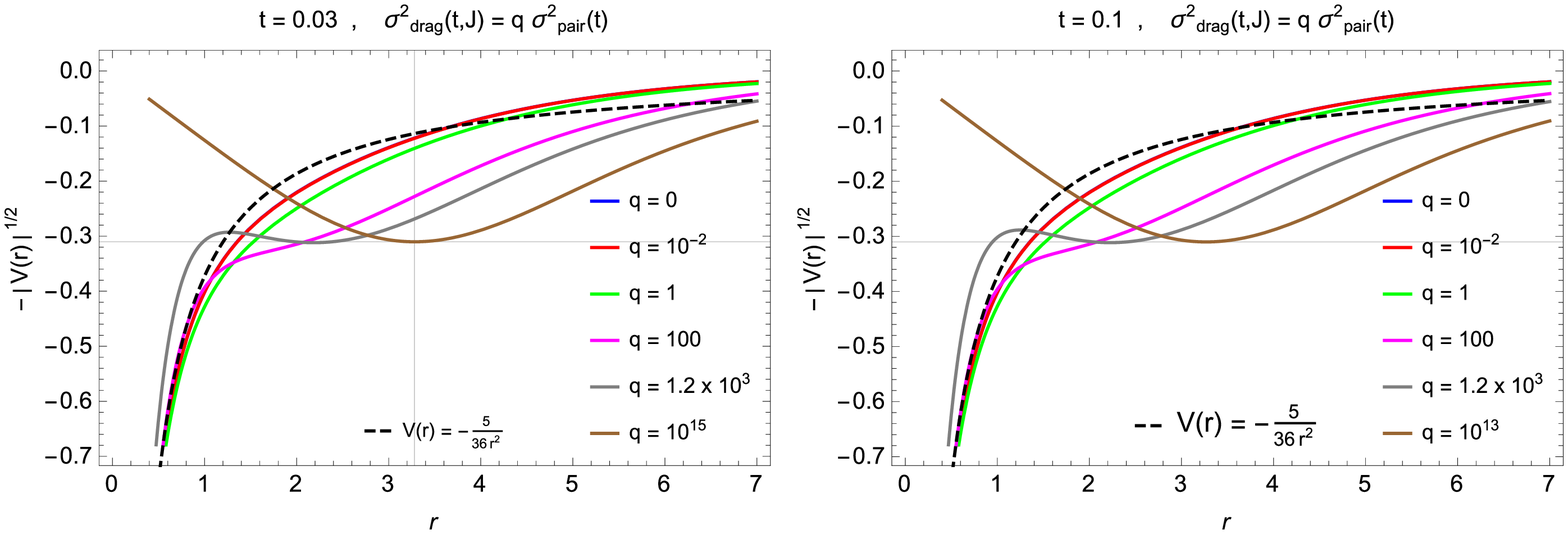}
\includegraphics[width=1\textwidth]{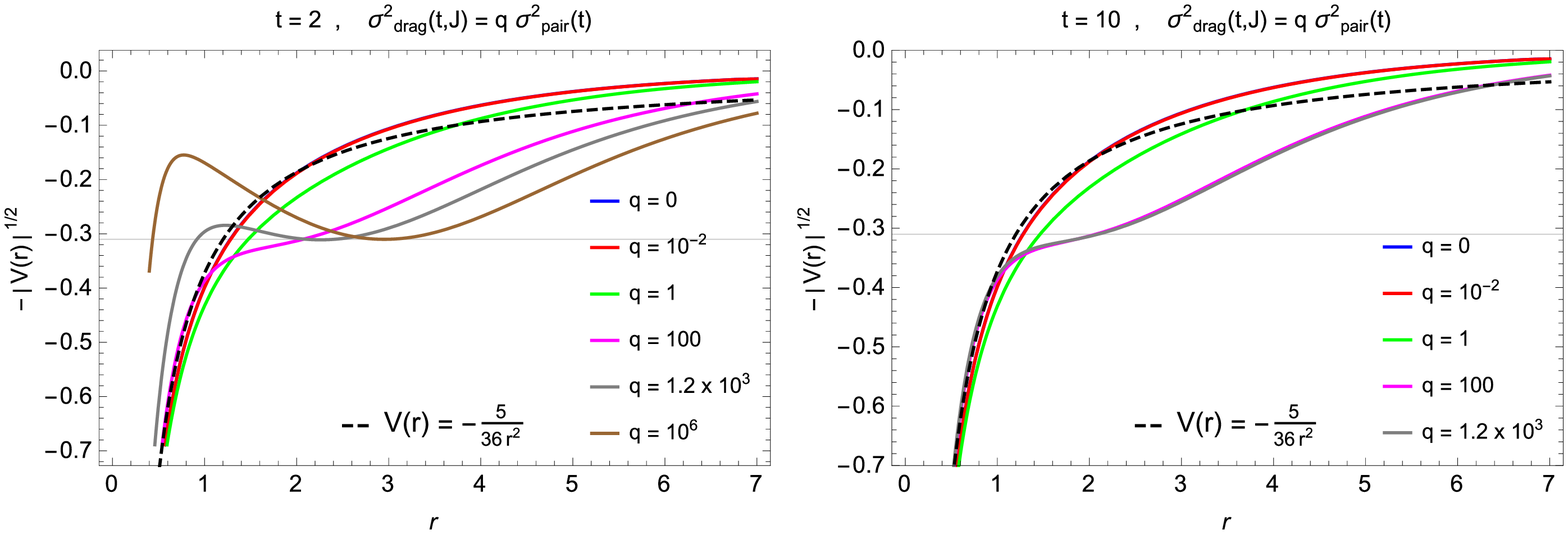}
\caption{Schr\"odinger potential for the different regimes of interest.}
\label{fig:potential}
\end{center}
\end{figure}

 It is not possible to do an analytic integration of  Eq. (\ref{eq:redef}), however  we can obtain the near horizon and near boundary behavior as
\begin{align}
\label{eq:ruH}
 r \sim -\log(1-u) \qquad , \qquad u\lesssim 1 \, , \\
\label{eq:ruB}
 r \sim  \frac{4b\ell^2\pi T_{eff}}{3u_0}(u_\star u)^{3/2} \qquad , \qquad u\ll 1\,.
\end{align}
In the new coordinates the boundary and horizon are at $0$ and $\infty$ respectively. We can write the asymptotic potential
\begin{eqnarray}
\label{eq:potentialB}
  V &\sim& -c(t,J)e^{-r} \quad\, , \quad r \gg  1 \, , \\
 V &\sim& -\frac{5}{36 r^2} \qquad\, , \quad r \ll 1\,,
\end{eqnarray}
where the constant $c(t,J)$ has a complicated dependence with $t$ and $J$. Nevertheless the formula simplifies in the QC  and DR
\begin{eqnarray}
c(t,J) &\simeq& \frac{1}{4}\left(1 + q(t,J) \right) \qquad,\qquad  q(t,J) \ll 1\\
c(t,J) &\simeq & \frac{1}{2}\left(1 - \frac{1}{2q(t,J)} \right)\qquad,\qquad  q(t,J) \gg 1\, .
\end{eqnarray}

 In figure \ref{fig:potential} we show the Schr\"odinger potential for all the regimes of interest.

In the QC regime,  $q\ll 1$, the qualitative behavior is similar in the two sectors I to II that correspond to $t\ll 1$ and $t \gg 1$ respectively (see figure \ref{fig:regions}). Indeed, close to the boundary the potential diverges  as $r^{-2}$ and then  approaches the horizon monotonically. The behavior changes in the DR regime  $q\gg 1$ because an intermediate well is formed for high enough values of $q$.
It is  worth mentioning that for $q\to\infty$ the $r^{-2}$ region in the potential is squeezed towards the boundary. Minimizing Eq. (\ref{eq:potential}) we can compute the location and depth of the well in the limit $q\to\infty$
\begin{eqnarray}
V(u) &\approx & \frac{1}{4} u \left(u^2-1\right) \qquad , \qquad q\to\infty \\
u_{min} &=& \frac{1}{\sqrt{3}}\,, \\
V_{min} &=& -\frac{1}{2\times 3^{3/2}} \approx -(0.310)^2\,,
\end{eqnarray}
with $r\approx 2 \left(\tan ^{-1}\left(\sqrt{u}\right)+\tanh ^{-1}\left(\sqrt{u}\right)\right)$.
From  figure \ref{fig:potential} we observe that the location of the minimum satisfies
\be
r_{min}(q)<r_{min}(\infty)
 \ee
 and that the depth of the well does not depend on the value of $q$.

\begin{figure}[t!]
\begin{center}
\includegraphics[width=.5\textwidth]{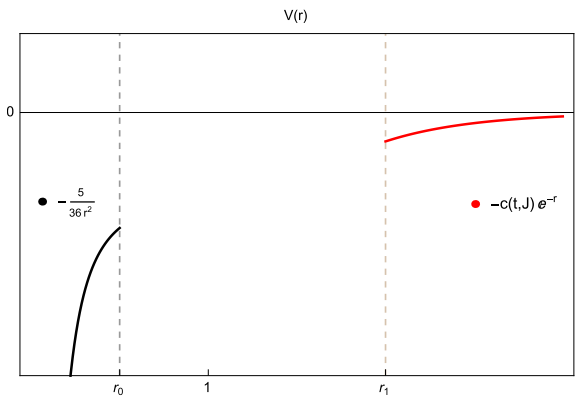}
\caption{In this plot the asymptotic forms of the potential are indicated. The near boundary region is to the left while the near horizon region is to the right. The system has analytic solutions in both asymptotic regions.}
\label{fig:potentialapprox}
\end{center}
\end{figure}

Having the asymptotic expansion of the  potentials we proceed to solve the Schr\"odinger Eq. (\ref{eq_Schro1}) analytically in these two regions (see figure \ref{fig:potentialapprox}) and we will then match them assuming that the frequency is much bigger than  $\sqrt{|V_{min}|}$.

The asymptotic solutions read
\begin{eqnarray}
\label{eq:sol3}
\psi(r) &=& c(t,J)^{-i w}J_{-2iw}\left(2c(t,J)^{1/2}e^{-r/2}\right)\Gamma(1-2iw) \quad \,,\quad r> r_1 \\
\label{eq:sol1}
\psi(r) &=& a_1 r^{1/2} J_{1/3}(rw) + a_2 r^{1/2} Y_{1/3}(rw) \quad \,, \quad r< r_0
\end{eqnarray}
where we have fixed the integration constants in the near horizon solution selecting the infalling boundary condition and normalizing the solution as $\psi(r\to\infty) = e^{-irw}$. The constants $a_1$ and $a_2$ will be fixed by the matching conditions of the wave function. Before doing so, we will use Eq. (\ref{eq:sol3}) and the Kubo formula to write a formula for the optical conductivity in terms of the integration constants $a_1$ and $a_2$
\begin{equation}
\left(b^2\ell \mathcal NT\right)^{-1}\sigma(\omega) = i\left(\frac{ 2\pi}{3}\right)^{7/3} \frac{\left(\sqrt{3} a_1+ a_2\right)}{ \Gamma \left(1/3\right)\Gamma \left(7/3\right)a_2}\left(\frac{\omega}{T}\right)^{-1/3}
\end{equation}

The previous asymptotic solutions are valid for arbitrary frequencies. For frequencies that are much bigger than the lower value of the potential inside the region $r\in(r_0,r_1)$ (see fig. \ref{fig:potentialapprox}) we can go further. In this case, the intermediate solution will be a combination of plane waves
\begin{equation}
\psi(r) = \tilde a_1 e^{wr} + \tilde a_2 e^{-wr}\, .
\end{equation}
Matching this solution with the wave function (\ref{eq:sol3}) we fix the constants
\begin{eqnarray}
\tilde a_1 &=& 1 -i\frac{e^{-r_1}c(t,J)}{2w}  - \frac{e^{-2 r_1} c(t,J)^2}{4 w (2 w+i)} \\
\tilde a_2 &=& -\frac{e^{2 i r_1 w}c(t,J)}{2w(2 w+ i )}\left(e^{-r_1}  -\frac{i e^{-2 i r_1 } c(t,J)}{2 (w+i) }\right)\,.
\end{eqnarray}
Following the same matching process we can fix the constants $a_1$ and $a_2$ and we finally obtain the analytic expression for the wave function in the near-boundary region.
\be
\label{eq:a1}
a_1 = \frac{\pi e^{-i x} }{12 \sqrt{r_0}} \left(  Y_{\frac{1}{3}}(x) \left(\tilde a_1 e^{2 i x} (1-6 i x)+6 i \tilde a_2 x+\tilde a_2\right)+6   x Y_{-\frac{2}{3}}(x) \left(\tilde a_2+\tilde a_1 e^{2 i x}\right)\right)
\ee
\be
\label{eq:a2}
a_2 = -\frac{\pi e^{-i x} }{12 \sqrt{r_0}} \left( J_{\frac{1}{3}}(x) \left(\tilde a_1 e^{2 i x} (1-6 i x)+6 i \tilde a_2 x+\tilde a_2\right) +   6x J_{-\frac{2}{3}}(x) \left(\tilde a_2+\tilde a_1e^{2i x}\right) \right)\,,
\ee
with $x=r_0~w$. Our analysis is valid for $w^2\gg |V_{min}|$, although $x$ can be either $x\gg 1$ or $x\ll1$ because $r_0$ can become small  if $q$ is high enough.

We expand Eqs. (\ref{eq:a1}) and (\ref{eq:a2}) for $x\ll 1$ first
\begin{equation}
\left.\begin{array}{lcl}
a_1 &=& \frac{-5 \sqrt[3]{2} x^{2/3}  \Gamma \left(-\frac{1}{3}\right)+4 (6 i x-1)  \Gamma \left(\frac{1}{3}\right)}{24\ 2^{2/3} \sqrt{r_0} \sqrt[3]{x}}e^{i x} \\
a_2 &=& -\frac{5 \pi  \sqrt[3]{x} e^{i x}}{12 \sqrt[3]{2} \sqrt{r_0} \Gamma \left(\frac{4}{3}\right)}
\end{array} \right\}\qquad ,\qquad r_0w\ll 1 \, , w^2\gg |V_{min}| \, ,
\end{equation}
and obtain the AC conductivity of the form
\begin{equation}
\label{eq:sigmaDC_interm}
\sigma(\omega) \approx \frac{8 \mathcal N  \Gamma \left(1/3\right) }{15\ell^{5/3} \Gamma \left(7/3\right)}\left(3^{-2/3 } \sqrt[3]{\frac{2}{\pi }}\frac{r_0}{T_{eff}}+\frac{2 \sqrt[3]{2}  i   }{3   }\left(\frac{\pi }{3}\right)^{2/3}\omega^{-1}\right)\left(\frac{bT_{eff}}{r_0u_0}\right)^{2/3}\,.
\end{equation}
This behavior describes an intermediate regime for very small $r_0$.  In the opposite  case,  $x\gg1$ the constants asymptote to
\begin{equation}
\left.\begin{array}{lcl}
a_1 &=& \frac{(-1)^{5/12} }{\sqrt{r_0}}  \sqrt{\frac{\pi x}{2}}\\
a_2 &=& \frac{(-1)^{11/12} }{\sqrt{r_0}}\sqrt{\frac{\pi x}{2}}
\end{array} \right\}\qquad ,\qquad r_0w\gg 1\, , w^2\gg |V_{min}| \, ,
\end{equation}
which produce the following expression for the UV optical conductivity
\begin{equation}
\label{eq:sigmaUV}
\left(b^2\ell \mathcal NT\right)^{-1}\sigma(\omega) \approx \left(\frac{ 2\pi}{3}\right)^{7/3} {\left(\sqrt{3}+i\right)
\over  \Gamma \left(1/3\right) \Gamma \left(7/3
\right)}\left(\frac{\omega}{T}\right)^{-1/3}\,.
\end{equation}
This last formula is the correct  estimate for the conductivity for large enough $\omega$.

\subsection{The numerical computation of the AC conductivity}
\label{sec:numcomp}
To compute the full frequency dependence of the conductivity it is necessary to find numerical solutions. In order to implement properly the infalling boundary condition at the horizon we must find first an approximate solution near the horizon $a_z(u)\approx a^s_z(u)$.

We redefine the gauge field as $a_z^{s}(u)=(1-u)^{- i w}\mathcal{A}_z(u)$, and using the Eq. (\ref{eq:az_q0}) we find an analytic and regular solution for $\mathcal A_z(u)$ in the region of $u\sim 1$ as an expansion in $(u-1)$ up to fourth order.

 We then integrate numerically Eq. (\ref{eq:az_q0}) from the point $u=1-\epsilon$ to the boundary located at the cut-off $\epsilon(J,t)$, using as a boundary conditions $a_z(1-\epsilon)=a^s_z(1-\epsilon)$ and $a_z'(1-\epsilon)=a^{s'}_z(1-\epsilon)$. After doing the integration of the differential equation and using the holographic dictionary we extract the conductivities in the regimes of interest. In order to plot dimensionless quantities we normalized the conductivity as described below.

The form of a Drude peak in the low frequency AC ($\omega\ll\omega_{UV}$) conductivity is given by
\begin{equation}
\label{eq:drude_function}
\sigma(\omega) = \frac{\sigma_{DC}}{1-i\tau\omega},
\end{equation}
where $\sigma_{DC}$ is the DC value of the conductivity and $\tau$ is the relaxation time. If the validity of (\ref{eq:drude_function}) extends to a region in which $\tau\omega_0 \gg 1$ with $\omega_0\ll\omega_{UV}$, the conductivity will obey the following power law in this regime
\begin{equation}
\label{eq:Drude_expand}
\sigma(\omega) \approx \frac{\sigma_{DC}}{\omega^2} + \frac{i\tau^{-1}\sigma_{DC}}{\omega}\approx \tau^{-1}\sigma_{DC}\,\omega^{-1} e^{i\pi/2}\, .
\end{equation}
In what follows we will state that the conductivity has a Drude peak if it  can be fitted with the formula (\ref{eq:drude_function}) up to some frequency satisfying $\tau\omega_0 \gtrsim 1$ ($\omega_0\ll\omega_{UV}$).

On the other hand, from the UV behavior of the system under consideration, the conductivity  (\ref{eq:sigmaUV}) satisfies a power law with exponent $-1/3$ and phase $30^\circ$.We reproduce the asymptotic formula here,
\begin{equation}
\label{eq:sigmaabspredict}
\tilde\sigma \approx 2\left(\frac{ 2\pi}{3}\right)^{7/3}  \Gamma \left(1/3\right)^{-1} \Gamma \left(7/3
\right)^{-1} \left(\frac{\omega}{T}\right)^{-1/3}e^{i\pi/6} \approx 2\times0.67261\left(\frac{\omega}{T}\right)^{-1/3}e^{i\pi/6} \,,
\end{equation}
with $
\tilde\sigma(\omega) = \left(b^2\ell \mathcal NT\right)^{-1}\sigma(\omega)$.

\begin{figure}[t]
\begin{center}
\includegraphics[width=1\textwidth]{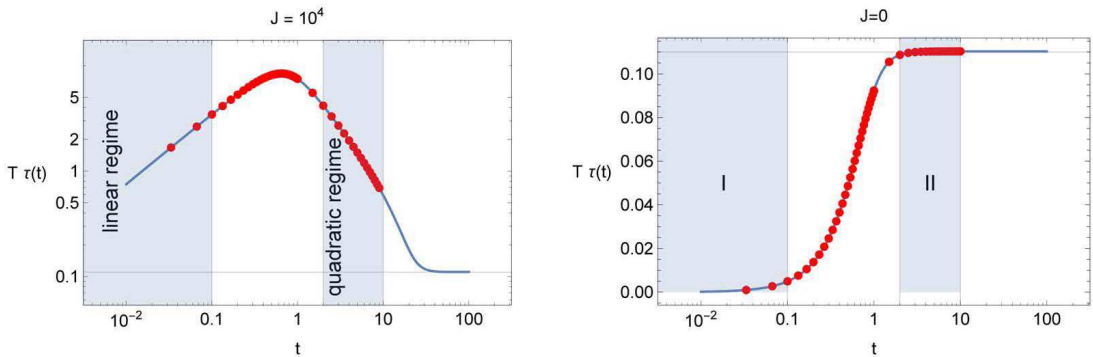}
\caption{The generalized relaxation time $\tau$ as a function of the scaling temperature variable $t$.The left plot shows the behavior of the dimensionless quantity T$\tau$ (T is the temperature) deep in the Drude regime (DR). The right plot shows the behavior in the  Quantum Critical regime (QCR) at zero density. Dots show numerical data and the continuous line is the analytic formula obtained in the text using perturbation theory .}
\label{fig:tau}
\end{center}
\end{figure}
To infer the existence of a Drude peak in the IR, we estimate the value of $\omega_0\gtrsim 1/\tau$ and we will compare it with the UV cut-off $r_0$\footnote{From figure \ref{fig:potential} we estimate the location $r_0$ where the potential deviates from its UV behavior $V\sim -5/(36r^2)$.} (as shown in figure  \ref{fig:potentialapprox}). This analysis is done  for  three different values of $t$ in the DR and in the QC regime. To do so, it is necessary to compute the generalized  relaxation time. In figure \ref{fig:tau} we plot $\tau(t)$  for the the values $J=10^4$ (DR) and $J=0$ (QC). We use the analytic formula (\ref{eq:tau_final}) and the numerical data. From the plots we observe a perfect agreement among the analytical data and the numerical one. We take this result as a non trivial check of our code.

We now  define $\omega_{UV}\sim1/r_0$  and we plot it in table \ref{tab1}  together with $\omega_0$  for $t=1/30,1/10,2$  in the QC and in the DR regimes\footnote{Note that $q(1/30,10^4)\sim 10^{15}$, $q(1/10,10^4)\sim 10^{13}$ and $q(2,10^4)\sim 10^{3}$}. In the QC regime we observe that $\omega_0\gg\omega_{UV}$, in consequence we should not expect to see a Drude peak in such regime. On the contrary in the DR we observe that $\omega_0\ll\omega_{UV}$, and a Drude peak forms in the present system as we shall see.
\begin{table}[h!t]
\begin{center}
\begin{tabular}{|l|c|c|}
\hline
 & $J= 0$ (QC)  & $J=10^4$ (DR) \\
 \hline
$t=1/30$ & $w_0\gtrsim1000,\quad  w_{UV} \sim 2 $ & $w_0\gtrsim0.6,\quad  w^{-1}_{UV} \sim 0 $\\
$t=1/10$ & $w_0\gtrsim200,\quad\,\,  w_{UV} \sim 2 $ &$w_0\gtrsim0.3,\quad  w^{-1}_{UV} \sim 0 $\\
$t=2$ & $w_0\gtrsim9,\qquad  w_{UV} \sim 2 $ &$w_0\gtrsim0.2,\quad  w^{-1}_{UV} \sim 0 $\\
\hline
\end{tabular}
\end{center}
\caption{$w_0$ for different values of $t$ in the QC and DR regimes}
\label{tab1}
\end{table}

\begin{figure}[t!]
\begin{center}
\includegraphics[width=1\textwidth]{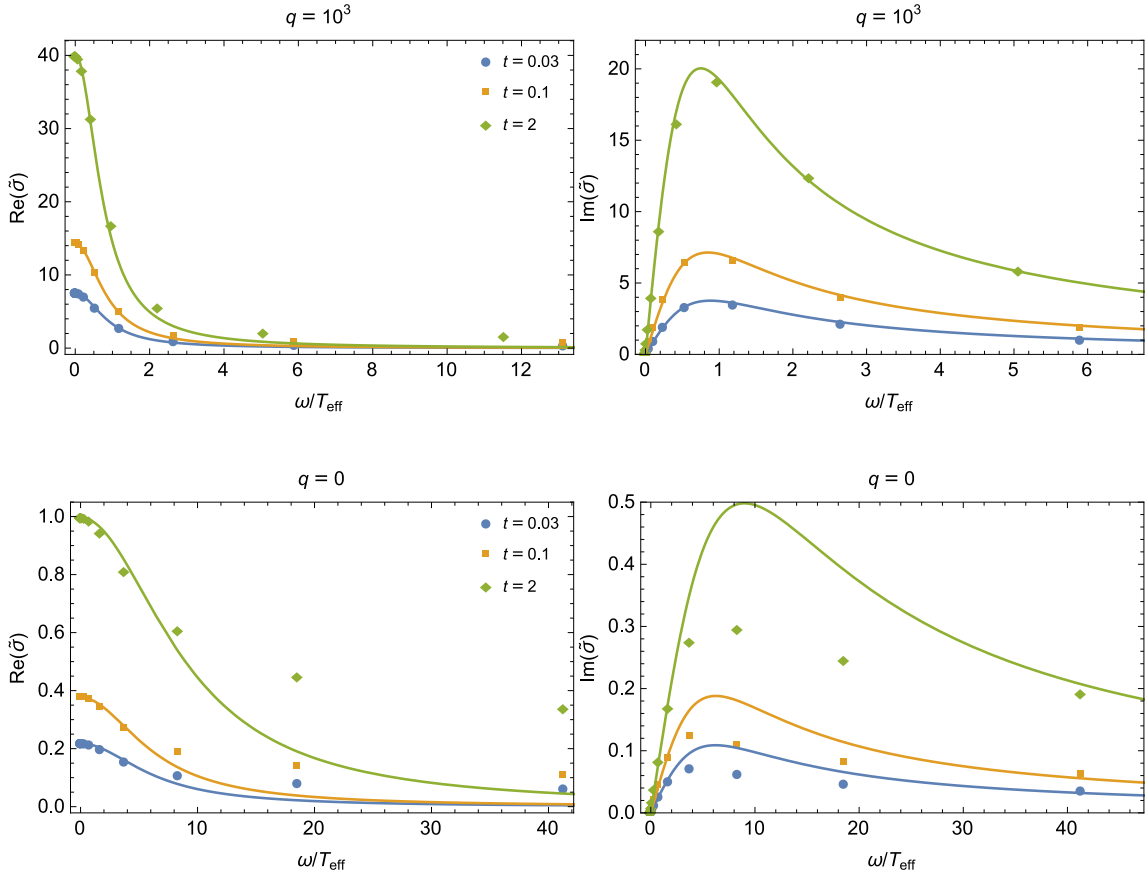}
\caption{Conductivity for three different temperatures in the DR (up) and in the QC (down)  regimes. The left figures show the real part of the conductivity while the right figures the imaginary part of the conductivity. Dots represent the numerical data, continues lines correspond to a fit to the Drude peak formula. }
\label{fig:sigmas}
\end{center}
\end{figure}

With the previous analysis we proceed to study the conductivities. In figure \ref{fig:sigmas}, we plot the real (left) and imaginary (right) part of the conductivity for $q=10^3$ (DR) and for $q=0$ (QC). We also fitted the Drude function (\ref{eq:drude_function}) on top of the numerical data. As expected from the previous analysis, in the QC regime we do not see the formation of a Drude peak. The point in which the conductivity deviates from the the Drude fitting is much less than the $\omega_0$ values estimated in the table \ref{tab1}. We observe such deviations because the UV physics start dominating  at those scales. On the other hand for $q=10^3$ (DR) the numerical data almost agrees with the Drude fitting in all the plotted region. That confirms the fact that the UV scale happens to satisfy $\omega_{UV}\gg\omega_0$, and the Drude paradigm seems to be valid in this dragging regime even though there are no quasi-particles.

\begin{figure}[t!]
\begin{center}
\includegraphics[width=.9\textwidth]{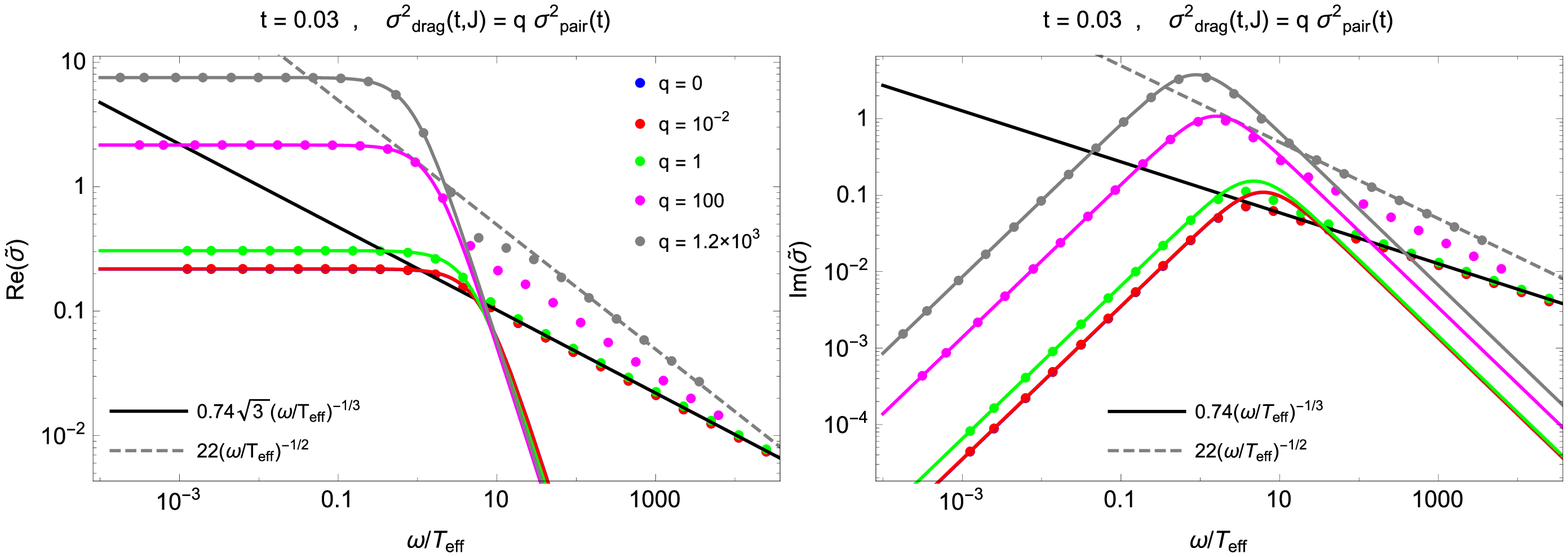}
\includegraphics[width=.9\textwidth]{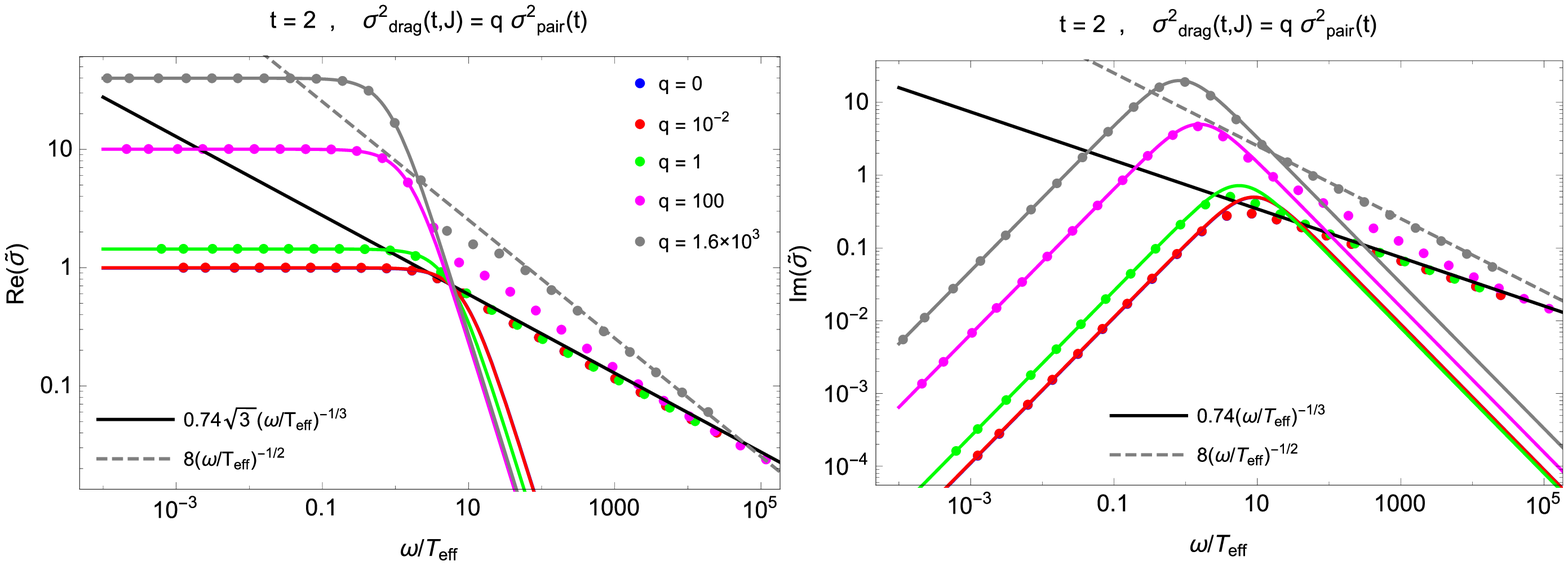}
\caption{Real (left) and imaginary (right) part of the conductivity. In the top figures the conductivity is plotted for different values of the parameter $q$ ranging from the QC regime
 ($q\ll 1$) to the DR regime ($q\gg 1$) for a fixed value of the scaling temperature variable $t$  with $t\ll 1$.
In the bottom plots $q$ is again varied in the full range at a fixed value of $t$, with $t>1$.
The colored continuous lines show the Drude fitting using the analytic computation of the generalized relaxation time. Straight black and dashed gray lines
show the UV and the  intermediate power law behavior of the AC conductivity.}
\label{fig:sigma1}
\end{center}
\end{figure}

In the following we will analyze the UV properties of the conductivities using our numerical results. In figure \ref{fig:sigma1} we show in log-log plots the real (left)  and imaginary (right) parts of the conductivity. The parameters in these plots are properly adjusted to display the behavior in each of four regimes of (\ref{regime}).
\begin{itemize}

\item We set first $t=1/30$ (upper plots of figure \ref{fig:sigma1}) which corresponds to either the regime (I) or the linear regime. The values of $q$ used for (I) are $q=0$ and $q=10^{-2}$ and for the linear regime are $q=100$ and $q=1.2\times 10^3$. The conductivity was also computed for the transition value $q=1$.

\item In the lower plots  of figure \ref{fig:sigma1} we show the conductivities for $t=2$ and $q=0$,  $q=10^{-2}$ (II regime), $q=1$ (transition)  and $q=100$, $q=1.2\times 10^3$ (quadratic regime).

\end{itemize}

From figure \ref{fig:sigma1} we observe a similar qualitative behavior in  the regimes (I) and (II), the same happens among the linear and quadratic regime. The location of the UV physics in all the regimes is at $\omega_{UV}\sim 10T_{eff}$.
We observe in the QC the absence of the Drude peak and we also verify that the conductivity obeys the power law obtained in subsection \ref{sec:efectPotential},  Eq. (\ref{eq:sigmaUV}).

In the Drude regime we confirm the appearance of the Drude peak in the IR. However, unlike the QC,  we observe a transition from the peak to a power law  of the form $\omega^{-1/2}$, indicating the existence of an intermediate regime which we were not able to extract analytically from the analysis of the Schr\"odinger potential.
We note that we observe the appearance of the intermediate regime when a well in the Schr\"odinger potential is formed (see figure \ref{fig:potential}), so it seems to be related to it. The exponent $-{1\over 3}$ in the DR should show up at much higher frequency, but we were not able to numerically verify this because at high frequencies our numerical code was no longer  reliable.

\begin{figure}[t!]
\begin{center}
\includegraphics[width=.9\textwidth]{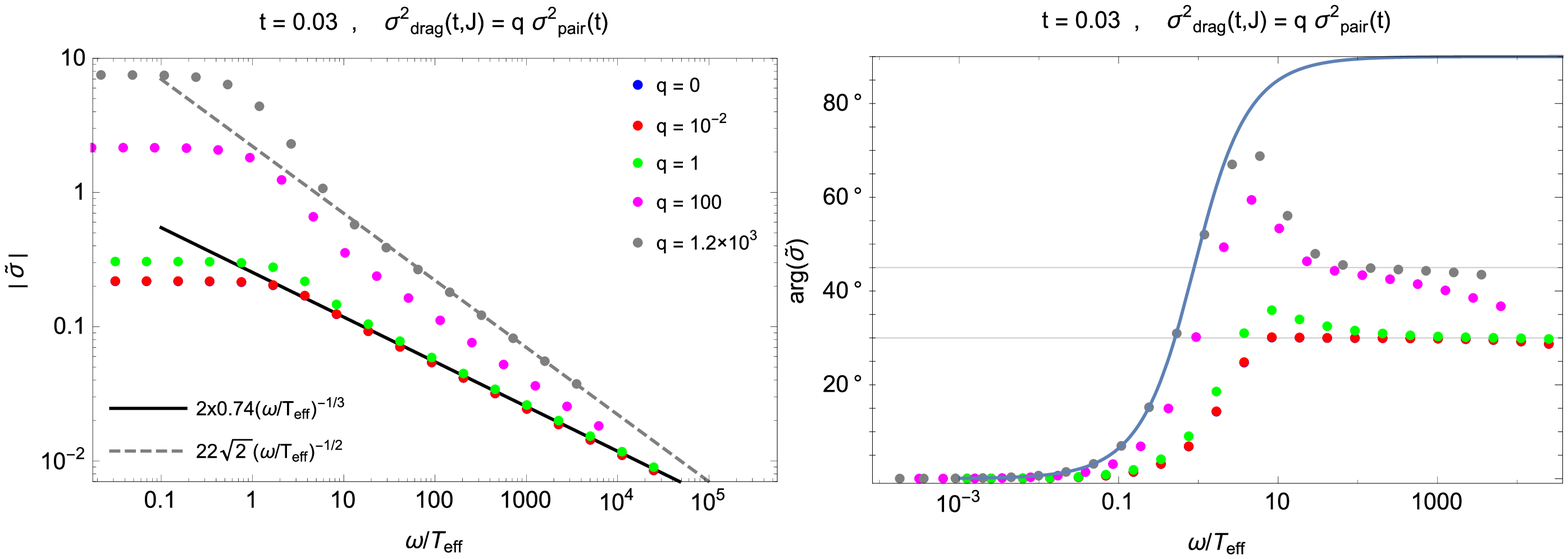}
\includegraphics[width=.9\textwidth]{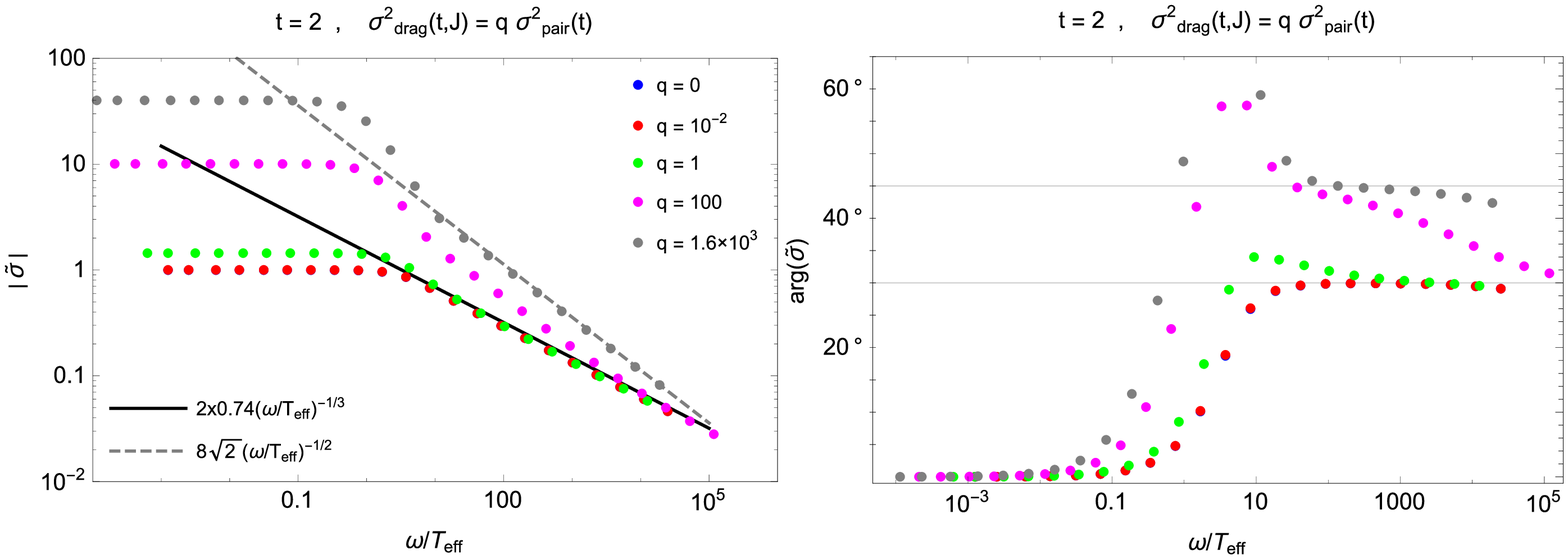}
\caption{Absolute value (left) and argument (right) part of the conductivity. In the top figures the conductivity is plotted for different values of the parameter $q$ ranging from the QC regime
 ($q\ll 1$) to the DR regime ($q\gg 1$) for a fixed value of the scaling temperature variable $t$  with $t\ll 1$.
In the bottom plots $q$ is again varied in the full range at a fixed value of $t$, with $t>1$.
The blue continuous line shows the Drude fit using the analytic computation of the generalized relaxation time. Straight black and dashed gray lines
show the UV and the  intermediate  power law behavior of the AC conductivity.}
\label{fig:abssigma1}
\end{center}
\end{figure}

In Fig. \ref{fig:abssigma1} we plot  the absolute value and the phase of the conductivity. The plots helps verify independently the presence of the power law for the intermediate frequency regime of a similar nature as was seen in the cuprates in \cite {ACScaling}.

In the QC  regime,  $q\ll 1$,  we observe a power-law decay  with an exponent of $-1/3$ and a constant of proportionality independent of $t$ and $J$ consistent with Eq. (\ref{eq:sigmaabspredict}).
The phase approaches asymptotically the value of $30^\circ$. This is again reminiscent of the power law seen in the cuprates,  \cite {ACScaling}.

In the DR we observe that the bigger the value of $q$, the closer to $90^\circ$ the phase gets. We interpret this behavior as the consequence of the competition among the drag and the critical pair-production physics. As the drag contribution is dominating  with the increasing $q$, the Drude description of the AC conductivity is becoming  better.

 For $q\sim 10^3$ (large), the conductivity behaves as $\omega^{-{1\over 2}}$ for $\omega >100T_{eff}$ (linear regime and quadratic regime). Note that the UV behavior of the conductivity for such a large value for $q$ in the linear regime is
\be
\sigma\sim 22(1+i){\omega\over T_{eff}}
\ee
and in the quadratic regime it is
\be
\sigma\sim 8(1+i){\omega\over T_{eff}}
 \ee(see Fig. \ref{fig:sigma1}). In view of this fact  the conductivity is expected to approach a constant phase of $45^\circ$ and this is what  we actually observe in figure \ref{fig:abssigma1}.
 This scaling happens in an  intermediate regime and we expect that for much larger values of the frequency it will asymptote to the $\omega^{-{1\over 3}}$ behavior that we found by the matching conditions method.
This was verified numerically for $q=100$ and we observed the tendency of the data to approach the value $30^\circ$ after almost a constant phase of $45^\circ$ for some  range of frequencies.

\begin{figure}[t!]
\begin{center}
\includegraphics[width=.5\textwidth]{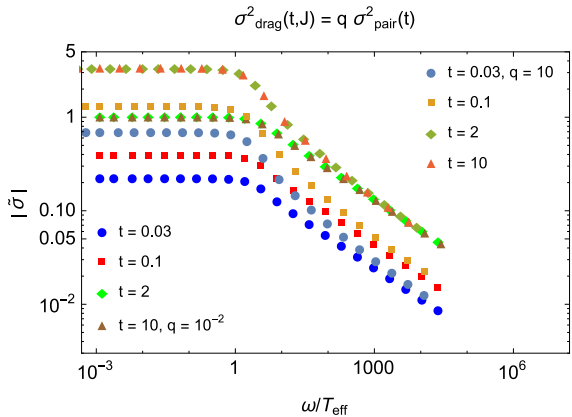}
\caption{Absolute value of the conductivity in the QC ($q=10^{-2}$) and in the DR ($q=10$) regimes for the scaling temperatures $t = 0.03$, $t = 0.1$, $t = 2$ and $t = 10$. We observe the same power-law,  independent of the temperature and regime, in the variable $\omega/T_{eff}$ for high enough values of the later ratio.}
\label{fig:comparesigma}
\end{center}
\end{figure}

 In figure \ref{fig:comparesigma} we show a check the result of the asymptotic scaling of the AC conductivity indicated in the analytic formula  in Eq. (\ref{eq:sigmaabspredict}) for different values of $t$ and $q$.  We have obtained a consistent result between analytic and numerics, a fact that provides a credibility check for our numerical calculations.

\begin{figure}[t]
\begin{center}
\includegraphics[width=.9\textwidth]{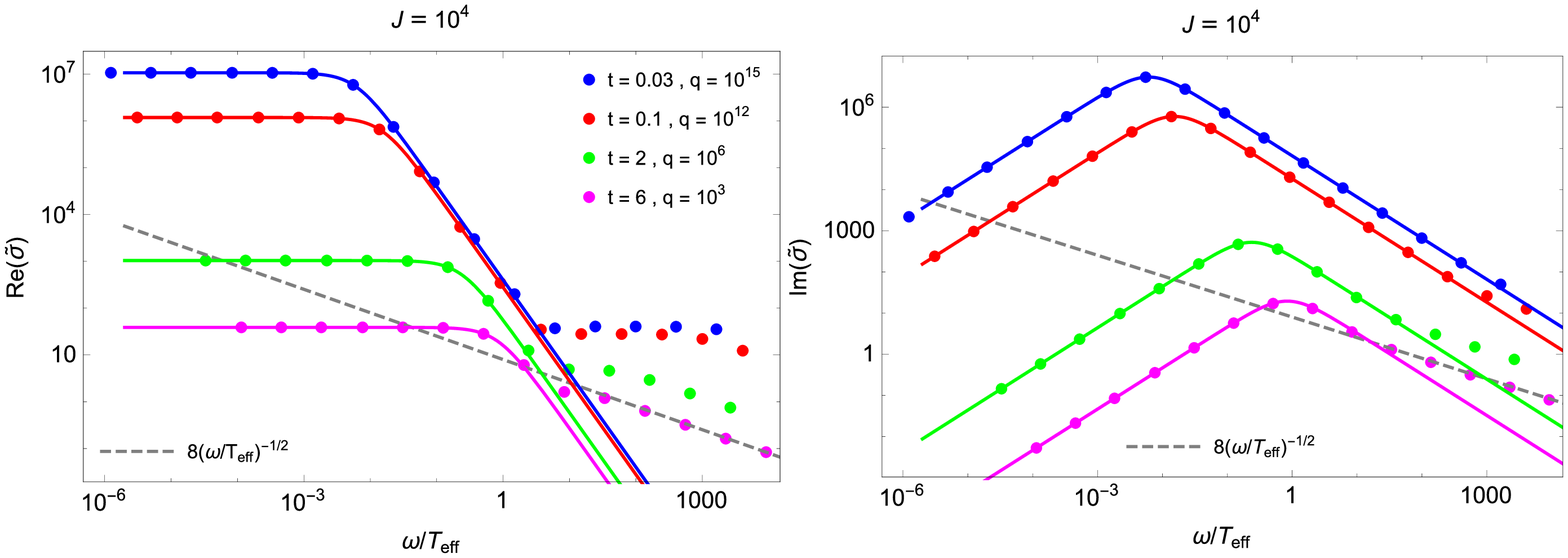}
\includegraphics[width=.9\textwidth]{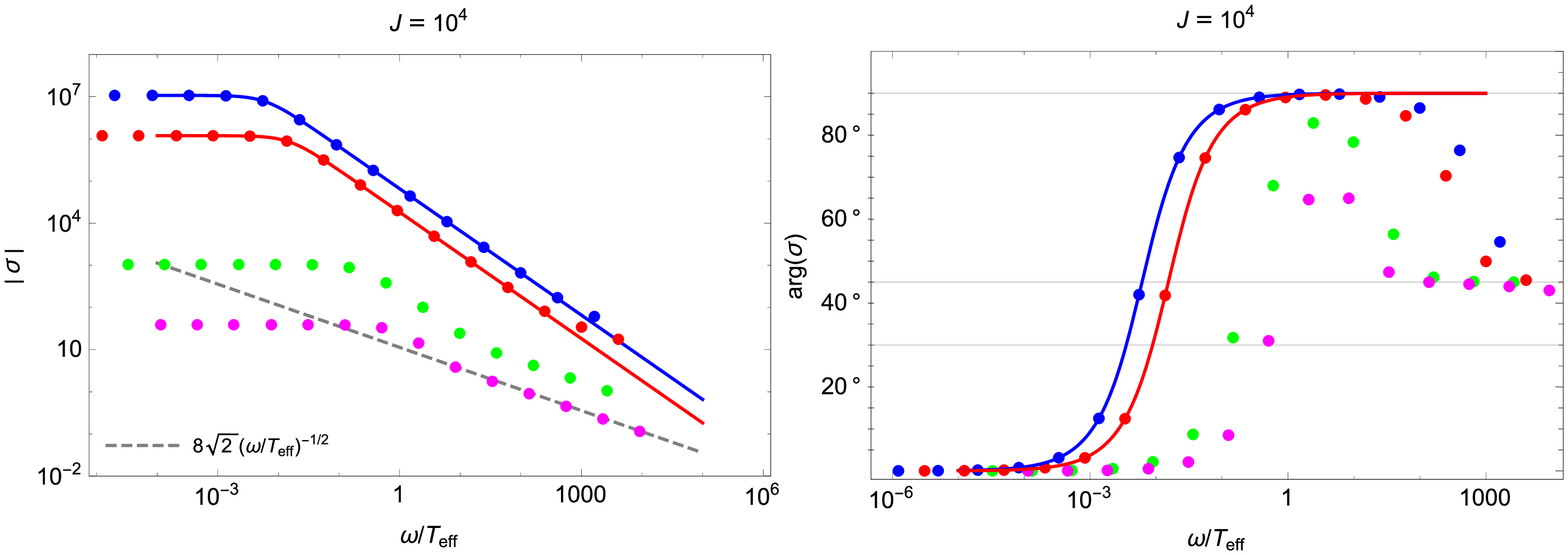}
\caption{Conductivity for  $J=10^4$. Continues lines show the Drude fitting. Gray dashed line is a fitting with exponent $-1/2$ }
\label{fig:bigJ}
\end{center}
\end{figure}

Our analytic computation of the high-frequency behavior of the conductivities shows an intermediate regime in which the real part of the conductivity behaves as a constant when $r_0w\ll 1$ but $w\gg 1$ (Eq. (\ref{eq:sigmaDC_interm})). To reproduce such a regime it is necessary to go deep in the DR  (see  Fig. \ref{fig:potential}), so we fixed $J=10^4$ and computed for different values of $t$ both in the linear and quadratic regime. We show in Fig. \ref{fig:bigJ} the conductivity for $t=0.03,0.1,2,6$.
The upper plots are the real and imaginary part of the conductivity and the plots below are its absolute value and its argument.

We observe that in the linear regime ($t<1$) a plateau in the real part of the conductivity is formed at  $\omega>100T_{eff}$ and the imaginary part has a $\omega^{-1}$ behavior as we showed in section \ref{sec:efectPotential}\footnote{The continuous line in figure \ref{fig:bigJ} represents the Drude fit. We do know however from Eq. (\ref{eq:Drude_expand}) that at high frequencies the imaginary part of the Drude form asymptotes to $\omega^{-1}$.}. For $\omega<100T_{eff}$ the conductivity is very well-fitted by the Drude formula. Note for example that the phase reaches $90^\circ$ as we would expect for a Drude behavior, and then presumably decreases to $45^\circ$ and stays there for some range of the frequencies because of the $\omega^{- 1/ 2}$ behavior.

The quadratic regime ($t>1$) just shows the intermediate regime with conductivity scaling as $\omega^{- 1/ 2}$ and phase $45^\circ$.

In summary, we associate the presence of the Drude peak to a regime in which the drag related  dissipation associated with strongly coupled physics  dominates over the quantum critical pair-production mechanism for conductivity. In the QC regime the scaling (UV) physics ($\sigma\sim\omega^{-1/3}$) appears at relatively low frequencies compare with the associated
UV scales of the DR ($\omega>10^4T_{eff}$). Consequently the scaling tail of the conductivity is controlled by the quantum critical physics.

\section{AC conductivity scaling and quantum critical saddle-points\label{acs}}

We will now consider the AC conductivity in holographic systems with a critical IR geometry characterized by a hyperscaling violating exponent $\theta$ and Lifshitz exponent $z$,
\cite{cgkkm,gk1,sh}, in the absence of momentum dissipation.

The extremal hyperscaling-violating background is given by
\be
ds^2=-D ~dt^2+B~dr^2+C~ (d\vec x)^2=r^{2\theta\over d}\left[-{dt^2\over r^{2z}}+B_0{dr^2\over r^2}+{(d\vec x)^2\over r^2}\right]\sp A_t \sim Q r^{\zeta-z}
\label{eq_hypermet}\ee
Here  $d$ is for the number of space dimensions of the dual boundary field theory, $z$ is the Lifshitz exponent, $\theta$ is the hyperscaling-violation exponent, $\zeta$ is the conduction exponent, \cite{gk2,G1,GHMO} that controls the gauge field  and $Q$ is proportional to the UV charge density. $B_0$ is a positive constant that can be absorbed into a redefinition of the generalization of the AdS scale (set to 1 here) and the boundary space-time coordinates.

It is a solution to the EMD action,  \cite{cgkkm},
\be
 S_{A}=\int d^{d+2}x\sqrt{g}\left[R-{1\over 2}(\partial\phi)^2+V(\phi)-{ Z(\phi)\over 4}F^2\right]\, ,
\label{2}\ee
where the background electric field generated by the boundary charge density  is
\be
A_t'={Q\sqrt{DB}\over ZC^{d\over 2}}\,,
\label{4}\ee
In generic scaling solutions, the bulk gauge coupling constant and scalar potential
\be
Z(\phi)\sim r^\kappa\qquad ,\qquad V(\phi) \sim r^{-\rho}
 \label{Z}\ee
are $r$ dependent as they depend on the running scalar. In the case where in the IR  $Z\sim e^{\gamma\phi}$ and $V\sim e^{-\delta\phi}$, the dilaton $\phi$ behaves in the IR  as $\phi\sim  \alpha \log r$ therefore  $\kappa=\alpha\gamma$, $\rho=\alpha\delta$.
Then, the previous equation determines the conduction exponent, \cite{gk2,G1,karch}  to be
\begin{equation}
\zeta=d-\theta - \kappa + \frac{2\theta}{d}\,.
\end{equation}
Note that this relation is only valid if the U(1) symmetry is unbroken.

The exponents $\theta$ and $\zeta$ track the presence of the violation of naive scaling in the classical solution. When $\theta\not=0$ the metric violates hyperscaling.
When $\zeta\not=d$, the gauge field profile violates naive scaling.

When $\theta=\zeta=0$ there is a genuine Lifshitz scaling symmetry in the theory
and the various observables have canonical scaling. For example the energy scales with mass-dimension $z$, spatial momenta with mass-dimension one, and the charge density $\rho$ with mass dimension $d$ while the current density has dimension $z+d-1$. Moreover, all observables of the dual quantum field theory have naive scaling and the generalization  of the AdS scale $\ell$\footnote{We define $\ell$ in general by substituting $r\to r/\ell$ in the metric (\protect\ref{eq_hypermet}).} never appears in QFT formulae.

This is to be contrasted with the case where at least one of the exponents $\theta,\zeta$ is non-zero. In that case the following are true:
\begin{enumerate}

\item  There is always a scalar field that runs in the solution, $\phi=f\left({r\over \ell}\right)$ and it is typically responsible for the nontrivial values of the exponents $\theta,\zeta$.

\item Physical observables are still scaling but they violate hyperscaling and/or naive scaling\footnote{Note the here we take a more general definition for hyperscaling violation compared to the condensed matter literature:
    any physical observable that scales with an anomalous exponent, violates hyperscaling according to our definition.} the scale $\ell$ appears in physical observables, to correct for the unusual dimensions. For example the entropy is
\begin{equation}
S\sim \ell^{-\theta}~T^{d_\theta\over z} \qquad , \qquad d_\theta = d -\theta \,,
\end{equation}
with $d_\theta$ appearing as the effective spatial dimensionality.
In this way $S$ has physical dimension $d$.

\end{enumerate}

In some string theory examples, the hyperscaling violation scale $\ell$ can be interpreted as a Kaluza-Klein scale, \cite{gk1}, namely the radius of an internal dimension that has been compactified. In this case the higher-dimensional theory is a theory that preserves hyperscaling.

In the generic case, the hypescaling-violation scale $\ell$ may be distinct from other UV scales that drive the flow of the theory. In the simplest possible gravitational example with a conventional relativistic (ie AdS) fixed point and a hyperscaling violating solution in the IR, we must have a potential in (\ref{2}) with two exponentials and a gauge coupling function $Z$ with one exponential. In that case the dynamical UV scale $\Lambda_{UV}$ can be mapped uniquely to the IR hyperscaling violating scale $\ell$. In all hyperscaling-violating solutions we will therefore
denote  $\ell$ by $\Lambda_{IR}$.

In the near-extremal case there is also a blackness factor $f(r)$ in the metric
\be
ds^2=r^{2\theta\over d}\left[-{f(r)dt^2\over r^{2z}}+B_0{dr^2\over f(r)~r^2}+{(d\vec x)^2\over r^2}\right]\sp f(r) = 1 - (r/r_h)^{d-\theta+z}
\ee
as well as an appropriate modification of the gauge field.

From the Einstein equation we conclude that the only way for the gauge field to support\footnote{This means that the contribution of the gauge field is of the same order as the gravitational contribution to the bulk equations of motion.} the gravitational  IR background is to have $d_\theta+\zeta=0$. If $d_\theta+\zeta<0$ the charge will not backreact to leading order and the IR solution will not depend on the presence of this charge.
 Finally, the case  $d_\theta+\zeta>0$ is not allowed as it is incompatible with the  equations of motion (the gauge field contribution is larger than gravity). We therefore have the general inequality
\be
d_\theta+\zeta\equiv d+\zeta-\theta ~~\leq~~ 0
\label{ine}\ee
Note that this inequality implies for EMD solutions that in the presence of a non-trivial charge density only the AdS$_2$ solution is truly scale invariant. All other solutions have hyperscaling violation.

The fluctuation equation for a time-dependent gauge field $\delta A_i\equiv a_i(r)e^{i\omega t}$ is (after solving for the associated metric perturbation,
 $\delta g_{ti}$, \cite{cgkkm})
\be
Z^{-1}C^{2-d\over 2}\sqrt{\frac{B}{D}}\pa_r\left(ZC^{d-2\over 2}\sqrt{D\over B}a_i'\right)+\left[\frac{B}{ D} \omega^2-{Q^2B\over ZC^{d}}\right]a_i=0
\label{3}\ee
The last term is proportional to the charge density, as the charge  $Q$ enters in the invariant combination, $Q\over C(r)^{d\over 2}$, which is the charge density at any given $r$. It  is responsible for the presence of a non-trivial Drude weight in the absence of momentum dissipation, \cite{DG2,KR}.

In the near extremal scaling background (\ref{eq_hypermet}) the fluctuations obey the equation\footnote{It was shown in \cite{gk1} that near extremality, the equation can rewritten in-terms of the scaling variables $r/r_h$ and $\omega/T$  showing that the conductivity, up to an overall scale is a function of $\omega/T$.}
\be
a_i''+\left({3-z-\zeta  \over r}+\frac{f'}{f}\right)a_i' + {B_0\over r^2}\left({\omega^2\over f^2}~r^{2z} - {Q^2\over f} r^{d_\theta+\zeta} \right)~a_i=0\,,
\label{5}\ee
We will absorb the positive constant $B_0$ into a redefinition of $\omega$ and $Q$, therefore  from now on we set $B_0=1$.
We will also set the temperature to zero (that sets $f=1$ in (\ref{5})).

This equation is of the form
\be
\left(e^{A(r)}a'\right)'+e^{A(r)+2B(r)}(\omega^2-G(r))~a=0
\label{6}\ee
where a prime stands for a derivative with respect to $r$.
We define a new variable $u$ and a new function $\psi$ so that
\be
{du\over dr}=e^B\sp a=e^{-{1\over 2}(A+B)}~\psi
\label{7}\ee
The equation (\ref{6}) now becomes a Schr\"odinger-like equation
\be
-\ddot\psi+V\psi=\omega^2\psi
\label{10}\ee
with
\be
 V={1\over 2}(\ddot A+\ddot B)+{1\over 4}(\dot A+\dot B)^2+G\,,
\label{11}\ee
where a dot is a derivative with respect to $u$. For all gapless solutions described here the IR is either at $r\to\infty$ or $r\to 0$ and the Schr\"odinger coordinate $u\sim r^z$ it will behave always as $u\to\infty$.

 We now turn to our concrete problem in (\ref{5}), where the Schr\"odinger potential can be calculated asymptotically to be
\begin{eqnarray}
\label{eq_pot}V &=&{\nu^2-{1\over 4}\over u^2} + \frac{Q^2}{z^2u^2}\left(z u \right)^{\frac{d_\theta + \zeta }{z}}\\
\nonumber \nu^2 &=& \frac{\left(2 \theta +d^2-d (\kappa +\theta +2)\right) (2 \theta +d (d_\theta+2 z-2-\kappa ))}{4d^2 z^2}+\frac{1}{4} .\\
\end{eqnarray}
The leading form of the potential in the IR depends on the value of the charge density $Q$ and the exponents. More precisely:

\begin{enumerate}
\item[\bf I] $d_\theta + \zeta < 0$ or $Q=0$.
In both of these cases the leading form of the potential will be
\begin{equation}
V ={\nu^2-{1\over 4}\over u^2}\,,
\end{equation}
and is scale invariant. The small $\omega$ behavior of the  AC conductivity is, \cite{kachru,cgkkm,gk1}
\be
\sigma(\omega)\sim \omega^m\sp m=2|\nu|-1
\label{emd2}\ee
\item[\bf II] $d_\theta + \zeta = 0$. When this condition is satisfied the second term in the potential in (\ref{eq_pot}) has the same $u^{-2}$ behavior as the first term and in this case
    \be
\sigma(\omega)\sim \omega^m\sp m=2|\bar\nu|-1\sp \bar\nu^2=\nu^2+{Q^2\over z^2}
    \ee
    Note that in this case, the gauge field is contributing non-trivially to the IR equations of motion. Also the term $Q^2$ appearing in the exponent is not the UV charge density (that is a free parameter) but the IR charge density which in this case is fixed completely and is a function of $z,\theta$, \cite{cgkkm}.

\item[\bf III] The remaining case $d_\theta + \zeta >0$  is incompatible with (\ref{ine}) imposed by  the equations of motion  as we argued  previously.

\end{enumerate}

In view of this classification we may now specialize (\ref{emd2}) to two different cases:
\begin{itemize}

\item The charged IR solution of \cite{cgkkm,gk1}.
In this case the IR background is conformal to Lifshitz with hyperscaling violation. The  gauge field has an action $e^{\gamma\phi}F^2$, but $\gamma$ is determined in this case by the exponents $z$ and $\theta$,   \cite{cgkkm}.
The dilaton is given by
\be
e^{\phi}\sim r^{\frac{2\theta}{d\delta}}\sp \alpha=\frac{2\theta}{d\delta}
\ee
 there is an asymptotic dilaton  potential $e^{-\delta\phi}$ with
\be
\delta^2=\frac{2 \theta ^2}{dd_\theta (d(z-1)-\theta)},
\ee
and the charge of the solution is given by
\begin{equation}
Q^2 =  \frac{2  (z-1)  }{d_\theta-1 +z }.
\end{equation}
%{\bf\Large What is $V_0$?}

In this case the Gubser criterion reads
\begin{equation}
\label{eq_GubsBound}
\frac{ z - d_\theta}{d(z-1)-\theta}>0,\frac{ d_\theta-1+z }{d(z-1)-\theta}>0,\frac{ ( z-1) }{d(z-1)-\theta}>0,
\end{equation}
and the thermodynamic stability condition
\begin{equation}
\label{eq_ThermBound}
\frac{ z}{(d-1)(z-1)-\theta}>0.
\end{equation}
when satisfied it is correlated to the existence of a gapless spectrum at zero and finite temperature, \cite{cgkkm,gk1}.

For these solutions of EMD
\be
\kappa=\alpha\gamma =2 d-\frac{2 \theta  (d-1)}{d}\sp m=\Big|  3 - \frac{ 2}{ z} + \frac{ d_\theta}{ z} \Big|-1\,.
\label{emd1}\ee
In this case, both terms in the potential (\ref{eq_pot}) scale as ${1\over u^2}$, the IR charge density $Q$ is fixed from the equations of motion and is a function of the exponents $z,\theta$ and we are in case {\bf II} above.

A few special cases of (\ref{emd1}) deserve to be mentioned.  The AdS$_2$ case is obtained when $z\to \infty$. In that case $m=2$.
The hyperscaling violating, semilocal geometries are obtained from the limit $z\to\infty$, $\theta\to-\infty$, with  ${\theta\over z}=-\eta$ held fixed. For these geometries,
\be
m=|3+\eta|-1=2+\eta~>~0
\ee

\begin{figure}[t]
\begin{center}
\includegraphics[width=.49\textwidth]{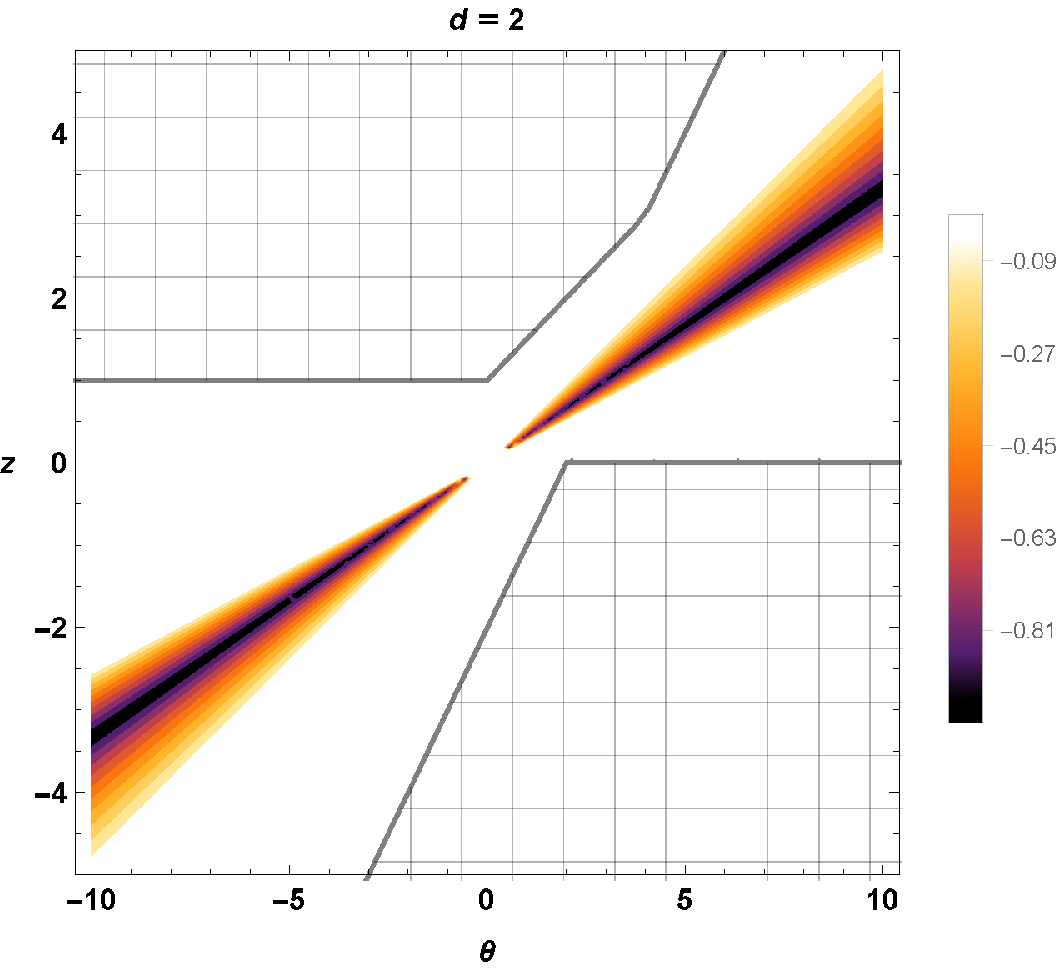}
\includegraphics[width=.49\textwidth]{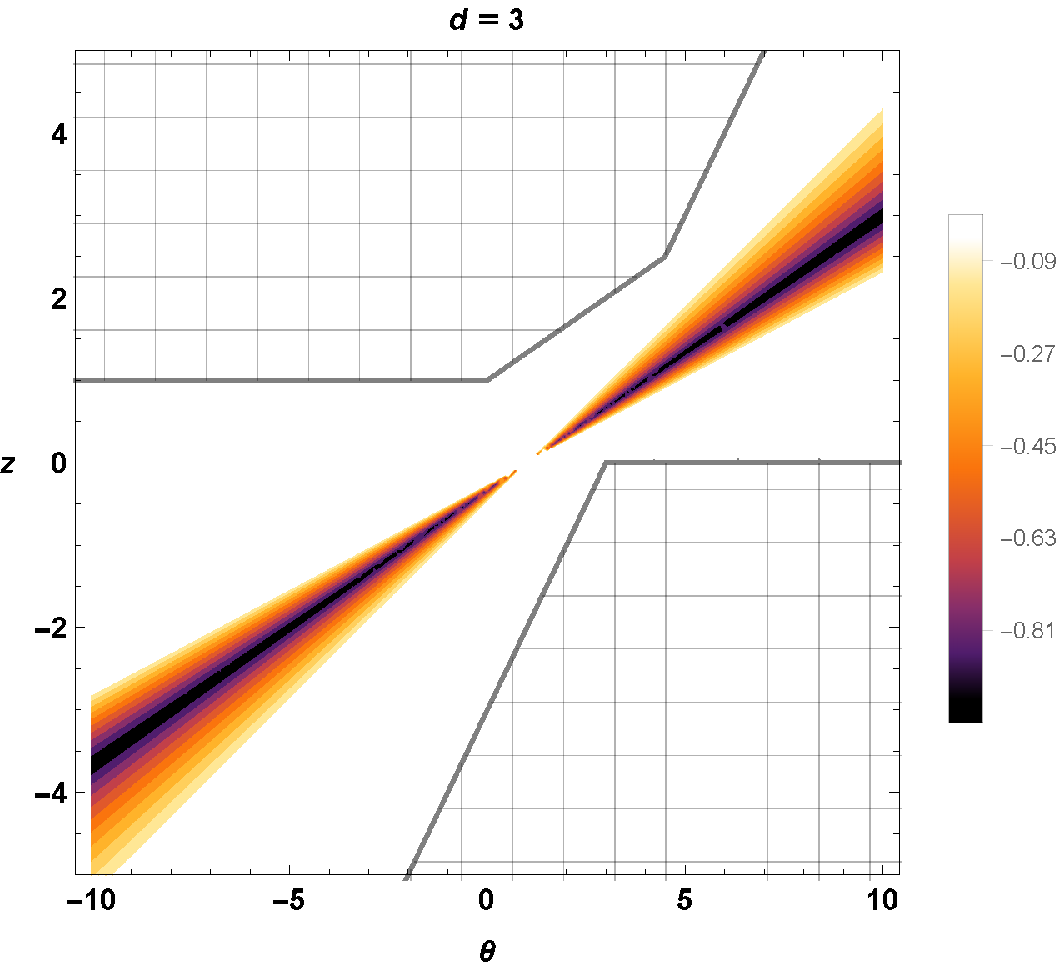}
\caption{Contour plots to illustrate the region in the parameter space where the exponent $m$ takes negative values for the single charged model.   Left: Conductivity in the charged case for $d=2$. Right: Conductivity in the charged regime for $d=3$. The allowed values for the parameters are bounded by the gray mesh. The negative values for $m$ are outside the permitted region.}
\label{fig:contoursigma2}
\end{center}
\end{figure}
In  Fig. \ref{fig:contoursigma2} we plot the exponent $m$ as a function of $z,\theta$. We observe that the region where $m$ takes negative values in the  $\theta-z$ plane, is always outside the region bounded by the Gubser criterion and the thermodynamic stability, for both $d=2,3$. This was already observed in \cite{cgkkm} which also gave a plot of the allowed values of $m$. It is therefore not possible to have a negative exponent $m$ in this case.

\item The only  other possibility that remains to be studied is when the second term in (\ref{eq_pot}) is subleading to the first term.
     This happens when the relevant gauge field is subleading in the IR equations of motions and does not backreact to the IR  geometry to leading order.
To do this we will need at least two gauge fields in the holographic theory, one to seed a general scaling geometry and the other to be subleading (and therefore treated as a probe in the IR geometry)\footnote{Clearly this generalizes to several gauge fields, with similar quantitative conclusions. However there could be other possibilities for seeding the leading solution that we do not address in this paper.}. We introduce therefore the  gauge fields $A_\mu,V_\mu$. $A_\mu$ will be nontrivial and will support the hyperscaling violating solution as in the previous cases. $V_{\mu}$ will have a smaller coupling in the IR and will be therefore subleading in the equations of motion.

The gauge field action is now
\be
\delta S_{V}=-{1\over 4}\int d^{d+2}x\sqrt{g} \left[Z_A(\phi) F_A^2+ Z_V(\phi)F_{V}^2\right]
\ee
and the scaling of the  conductivities has been extensively analyzed in appendix \ref{f0}. We report the results here and we distinguish two cases.

\begin{enumerate}
\item
One of the gauge fields (say A) supports the IR geometry while the other (say V) is subleading in the IR. This was analyzed in \ref{f1}.

In this case the conductivity $\sigma_A$ is as the previous example, (\ref{emd1})
The conductivity for $V$, $\sigma_V$ scales differently. We define
\be
 Z_V\sim r^{\bar\kappa},
\ee
in particular we parametrize $\kappa$ in the following way
$$
\bar\kappa = \kappa + \delta\kappa.
$$
The charge exponent for $V$ is
\be
\bar\zeta=d_\theta -\bar\kappa+{2\theta\over d}\,,
\label{555}\ee
 The exponent  $m$ in the conductivity $\sigma_V$ now reads
\be
 m = \left|\frac{\theta  (d-2)}{dz}-\frac{1}{z} \left(d+z-2-\bar\kappa \right)\right|-1=\Big|{z+\bar\zeta-2\over z}\Big|-1\,,
 \label{eq_m2c}
\ee
$\theta$ and $z$ must obey the Gubser and thermodynamic stability bounds (\ref{eq_GubsBound}), (\ref{eq_ThermBound}) as well as $d_\theta+\bar\zeta<0$.

Note that the exponent $m$, written in terms of the two hyperscaling violation exponents $\theta,\zeta$, does not depend on $\theta$. Therefore it is sensitive only to hyperscaling violation originating in the charge sector.

A few special cases of (\ref{eq_m2c}) deserve to be mentioned.  The AdS$_2$ case is obtained when $z\to \infty$ and $m=0$ in this case.
The hyperscaling violating, semilocal geometries are obtained from the limit $z\to\infty$, $\theta\to-\infty$, with  ${\theta\over z}=-\eta$ held fixed. For these geometries,
\be
m={d-2\over d}~\eta\sp \eta>0\sp d\geq 2
\ee
and $m$ is always positive.

In figure \ref{fig_2charges11} we show the values of the exponent $m$ for $d=2$ in a charged background ($Q_A,Q_V\neq 0$) where  the backreaction of $Q_V$ can be neglected. In general $m$ depends on $z,\theta,\bar \kappa$.

The plot represents $m$ for  $|\bar\kappa-\kappa|=3$. It is always possible to obtain a region with negative values of $m$, either with $\bar\kappa-\kappa >0$ and $z>0$ or $\bar\kappa-\kappa<0$ and $z<0$ (see appendix \ref{f0}). Note that from Eq. (\ref{eq_m2c})  $m=0$ for $\bar\kappa=0$ which is the case when the bulk gauge coupling constant flows to a constant value in the IR.

We conclude that in order to have $m<0$ in $d=2$ two conditions must be satisfied:
\begin{enumerate}

\item The charge density in question must not backreact on the metric.
\item  $\bar\kappa>0$, namely the bulk gauge coupling must be IR free for $z>0$ and UV-free for $z<0$.

\end{enumerate}
The inverse correlation to the sign of $z$ is also correlated to   the relative flow between energy and momentum, controlled by ${g_{tt}\over g_{xx}}~\sim~ r^{2(1-z)}$. Taking into account the constraints this ratio for $z>1$ vanishes in the IR while for $z<1$ it diverges in the IR.

\begin{figure}[t]
\begin{center}
\includegraphics[height=.6\textwidth]{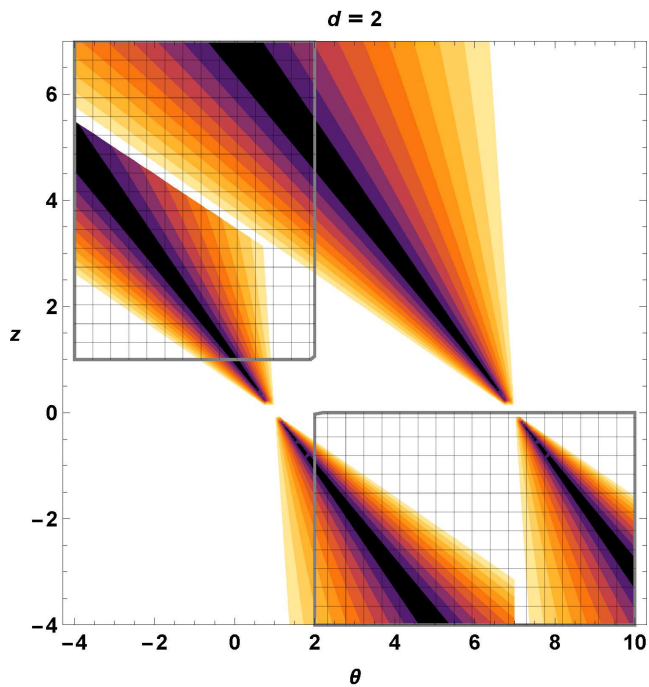}
\caption{Contour plots to illustrate the region in the parameter space where the exponents $m$ takes negative values for the $V$ gauge field. The plot was made for $d=2$. The negative values for $m$ are inside the permitted region. We have set $|\delta\kappa|=3$}
\label{fig_2charges11}
\end{center}
\end{figure}

\item A different case involves the two gauge fields being equally important in the IR, but that their coupling constants, $Z_A$ and $Z_V$, become different as one flows to the UV. In that cases $Z_A-Z_V$ is much smaller than $Z_A+Z_V$ in the IR.
    This cases was analyzed in \ref{f2}.
    What was found is that $\sigma_{A}$ and $\sigma_V$ contain at low frequency, a scaling term of the type shown in (\ref{emd1}) and another as in (\ref{eq_m2c}). There is also a cross-conductivity $\sigma_{AV}$ that is partly negative and contains the difference of the two terms above.

\end{enumerate}

\end{itemize}

\section{Outlook\label{out}}

We have analyzed the AC conductivity in a holographic model of a strange metal proposed in \cite{kkp}. We found some interesting scaling limits of the conductivity that seem to generalize to general critical holographic systems. We have also found that this scaling regime is controlled by the uncharged system and seems to be generically independent on the mechanisms of momentum dissipation.

The detailed results are as follows:

\begin{enumerate}

\item We define a parameter $q$ that distinguishes the Drude regime ($q\gg 1$) where the DC conductivity is dominated by dissipation and the Charge Conjugation symmetric regime ($q\ll 1$) where dissipation is negligible.

\item The characteristic temperature scale that controls the AC conductivity is an effective temperature $T_{eff}$ (with an associated scaling effective temperature $t_{eff}$ defined analogously).
    This is  a characteristic effect for the system in question and more generally for systems where charge is described by a D-brane system. It is not expected in more general cases.

    The effective temperature found  is distinct from the system temperature $T$ and is due to the existence of a hierarchy of interactions. Similar effects have been observed in electronic systems and in holography, \cite{eff1}-\cite{eff4}.

\item We have defined a {\em generalized relaxation time} $\tau$ by the IR expansion of the AC conductivity,
     \be
     \sigma(\omega) \approx \sigma_{DC}\left(1+ i ~\tau~\omega + \mathcal{O}(\omega^2)\right)
\ee
In the presence of the Drude peak, this is the conventional definition of an associated relaxation time. When there is no Drude peak present, $\tau$ is still well-defined, although in that case the interpretation as a relaxation time is not clear.

    We gave an analytical formula for $\tau$ in (\ref{eq:tau_final}).
    It takes a simple form for large and small values of the scaling temperature variable $t$.  In the regime I (see fig. \ref{fig:regions})  we obtain
\be
\tau\sim \sqrt{T}
\ee
while in the regime II (with $t$ typically large) it is set by the inverse of the temperature
\be
\tau\sim \frac{1}{ T}
\ee

\item In the Drude regime ($q\gg 1$) where the dominant mechanism for the conductivity is dissipative, there is a clear Drude peak as seen for example in figures \ref{fig:sigmas}.

    In the Charge Conjugation Symmetric regime it is also clear from the same figures that no Drude peak is to be seen in the IR of the AC conductivity.

\item At zero charge density (Charge Conjugation Symmetric regime) there is a scaling tail for the AC conductivity that behaves as
    \be
    |\sigma|\sim \left({\omega\over t_{eff}}\right)^{-{1\over 3}}\sp Arg(\sigma)\simeq {\pi\over 6}
\ee
For finite charge density this tail survives not only in the Charge Conjugation Symmetric regime but also in part of the Drude regime, as seen in the various plots of figure \ref{fig:sigma1} as well as the ones of figure \ref{fig:abssigma1}.
The qualitative reason for this is that in the presence of the Drude peak, its tails falls off as ${1\over \omega}$ and this is faster than $\omega^{-{1\over 3}}$. Therefore, the scaling tail will eventually win over the Drude peak for $\omega\geq \omega_0$ and the only condition that it is visible is that the UV structure of the theory kicks-in at $\omega_{UV}\gg \omega_0$.

\item This scaling tail of the AC conductivity generalizes to more general scaling holographic geometries, as previously described in \cite{cgkkm}. The equation that determines the conductivity is given in (\ref{3}) and the equivalent Schr\"odinger problem has a potential of the form $V_{eff}=V_1+\rho^2 V_2$ where $\rho$ is the IR charge density and is proportional to the UV charge density.

    In particular for a metric with Lifshitz exponent $z$, hyperscaling violation exponent $\theta$ and conduction exponent $\zeta$ with $d$ spatial boundary dimensions,  we find that in general
   \be
    |\sigma|\sim {\omega}^{m}\sp Arg(\sigma)\simeq -{\pi~m\over 2}
\ee
with
\be
 m = \left|{z+\zeta-2\over z}\right|-1\,,
\label{m}\ee
There are several constraints in the parameters of this formula that are detailed in section \ref{acs}.

\item There are some special cases of (\ref{m}) that deserve mentioning.
For an AdS$_2$ IR geometry the exponent can be obtained by an $z\to\infty$ limit in (\ref{m}) giving $m=0$.

For hyperscaling violating semilocal geometries we must take $\theta\to \infty$, $z\to \infty$ with ${\theta\over z}=-\eta$ fixed and obtain
\be
m=\big|{d-2\over 2}\eta+1\Big|-1={d-2\over d}~\eta
\ee

Finally, for the gauge field conformal case  we obtain $m=0$ when $d=2$.

\item We find that for two spatial dimensions, negative values for the exponent $m$ are correlated with the sign of the Lifshitz exponent $z$ and the strength of the gauge field interaction in the bulk. Parametrizing the IR asymptotics of gauge coupling function $Z$ as
    \be
    Z\sim r^{\kappa}\sp r\to \infty
    \ee
    in conformal coordinates, we obtain that $z\kappa>0$ for negative values of $m$ to be possible.

\item In the special case where the associated gauge field seeds the IR scaling geometry, $\kappa$ is fixed as a function of $z,\theta$ and the exponent $m$ takes the value
    \be
    m=\Big|{3z-2+d-\theta\over z}\Big|-1
   \label{eq_m_int_2} \ee
and is always positive. In the case where the IR geometry is AdS$_2$ can be obtained for $z\to\infty$ limit of (\ref{eq_m_int_2}) giving $m=2$. For
hyperscaling violating semilocal geometries, we must take, $\theta\to \infty$, $z\to \infty$ with ${\theta\over z}=-\eta$ fixed.
In this case we obtain
\be
m=|3+\eta|-1=2+\eta
\ee

\item An important issue is whether the scaling of the AC conductivity described above for the general scaling geometries is controlled by the dynamics of the charge density, or it is decided by the neutral system.

In the example we have analyzed  the effective Schr\"odinger potentials that control the calculation of the conductivity have two parts. One that is independent of charge density and one that is proportional to the square of the charge density. In the generic case it is the first that controls the UV scaling of the AC conductivity described above. Only if the charge density is supporting the IR geometry, then the second part is of the same order as the first and it is the sum that controls the scaling of the AC conductivity. This special case is also the only one we found where the exponent $m$ in    (\ref{m}) is always positive. In other cases it can also be negative, but unitarity implies always that $m\geq -1$.

\end{enumerate}

These findings suggest that there is a generic source of scaling tails in the AC conductivity in holographic systems. Moreover, in generic systems this scaling is expected to be  independent of the mechanism of momentum dissipation. On the other hand, whether it is visible it depends on the details of the momentum dissipation mechanism.

An important part of this story is the dependence of the T=0 result on scales. For this we must consider a hierarchy of cases as a function of the properties of the system. It also matters whether the charge density is interacting strongly with the gravitational system or is a probe whose backreaction on the system can be neglected. We will discuss all these cases in turn

\begin{itemize}

\item In the truly scale invariant fixed points ($\zeta=\theta=0$) with a general Lifshitz scaling exponent, $z$, the AC conductivity will scale with $\omega$ with the canonical dimension,
    \be
    \sigma\sim \omega^{d-2}
    \ee
    for any $z$ and no intrinsic scale enters in this relation.
    Note that in the EMD case, apart from the AdS$_{d+2}$ solution (z=1), there are no other such scale-invariant solutions. To obtain general scale invariant Lifshitz solutions, one should use the more general EMD action, \cite{gk2} where the U(1) symmetry is generically broken.

\item In the presence of hyperscaling violation $\theta\not=0$, and/or $\zeta\not=0$,  the IR theory contains a nontrivial scale $\Lambda_{IR}$ that appears in many relations and is responsible for the change of scaling exponents and the violation of hyperscaling, \cite{gk1}. In some cases which can become regular by embedding the dual bulk theory to a higher-dimensional relativistic scale invariant theory such a scale is the radius of the extra dimensions. This scale appears in scaling relations in order to ``correct" the naive scaling dimension. For example
    \be
    \sigma(\omega)\sim \Lambda_{IR}^{d-2-m}~\omega^{m}
\label{out2}\ee

\item In the cases where the charge density acts as a probe in the IR geometry (and its effects are therefore subleading), it introduces a new IR scale that is set by the associated charge density. However we have seen that this new scale does not enter to leading order in the scaling relation for the associated conductivity. We are therefore again in the situation described in the previous cases.

    \item It is expected that hyperscaling violating geometries should have a resolution of the their naked singularities via an $AdS_2$ or a stringy geometry in the IR, and they will be only intermediate scaling regimes, \cite{resolve}. In such a case $\Lambda_{IR}$ is the associated scale to this mid-IR regime.

\end{itemize}

To try and extend qualitatively our general scaling analysis of section \ref{acs} we must reason as follows:
There are two further modifications to the setup above so that this result is embedded in a complete RG flow.
The first step is to match these IR asymptotics to an AdS UV geometry. The second it to turn on finite temperature.
We will now discuss such modifications in turn.

The non-trivial RG flow at $T=0$ introduces in the simplest of cases a transition  scale $\Lambda_{UV}$. In the hyperscaling violating geometries this is related to the characteristic scale of the IR geometry. In scaling geometries it is related to the UV operator that drives the flow. Quite generically this can be the charge density.
Our general AC scaling found above is valid in the region $\omega\ll \Lambda_{UV}$.

In the presence of a finite temperature, $T$, it was shown in \cite{gk1} that the AC conductivity is a function of $\omega/ T$.
There are two ratios that control the behavior of the AC conductivity: $T/\Lambda_{UV}$ and $ \omega/ T$. In the hyperscaling violating case, $\Lambda_{IR}$ also enters. However, in the limit $\Lambda_{UV}\to\infty$, $\Lambda_{IR}$ enters trivially as in (\ref{out2}). We have the following regimes

\begin{itemize}

\item ${T\over \Lambda_{UV}}\gg 1$. The theory is in the UV scaling region and the AC conductivity is characterized by the UV fixed point alone.

\item  ${T\over \Lambda_{UV}}\ll 1$. In this case, there are three regimes for the AC conductivity as a function of frequency.
\begin{itemize}

\item $\omega \ll T$ This is the IR regime controlled by the properties of the black-hole horizon. In this regime the mechanisms of momentum dissipation and hydrodynamics are
controlling the behavior, \cite{HKMS}. If the drag dissipation is present, it determines the Drude peak (that maybe pronounced or less prominent depending on the strength)
and this is the main characteristic of this region.

\item $T\ll \omega \ll \Lambda_{UV}$. In this regime the scaling behavior found earlier, at $T=0$ is prominent.  As the power evolution of the AC conductivity is always $\sim \omega^m$ with $m> -1$, it will always dominate over the Drude $1/\omega$ decay for large enough frequency. The only possibility where this will {\em not} be visible is if the cross over scale is of the order of or larger than $\Lambda_{UV}$.

\item $\omega \gg \Lambda_{UV}$. In this regime the AC conductivity is determined by the UV fixed point.

\end{itemize}
\end{itemize}

The type of AC conductivity scaling described here is very reminiscent of the one seen in the cuprates in \cite{ACScaling}, in the intermediate region $T\ll \omega\lll \Lambda_{UV}$. Indeed the scaling exponent found there is $m=-{2\over 3}$ which suggests in view of our discussion the presence of hyperscaling violation.

There are several problems that remain open in view of our results

\begin{enumerate}

\item We should back up our scaling analysis and the description of different regimes above with a complete calculation of the AC conductivity from numerical complete RG flows that connect AdS UV fixed to either IR or intermediate geometries with general scaling exponents ($z,\theta,\zeta$).

    \item The analysis of scaling described here was based on geometries that preserve the U(1) symmetry. This analysis needs to be extended to the general class of U(1)-breaking extremal scaling geometries found in \cite{gk2}. Part of this analysis was already done in \cite{G1} but it should be extended to more general cases. We expect that our results will extend smoothly to these more general solutions. The aim is a universal formula for the AC exponent $m$ that controls the scaling of the AC conductivity.

        \item   Although the scaling tails described above are expected to be independent of the mechanism of momentum dissipation, it is interesting to study their interaction with momentum dissipation and different types of low-frequency behavior. In particular it would be interesting to study this in geometries with helical symmetry as the general class presented in \cite{dgk}.

\item The scaling exponent of the AC conductivity is expected to depend on the charge interactions captured by the bulk (self)-interactions of the U(1) gauge bosons. Understanding how is an interesting question.

    \item The eventual comparison with cases realized experimentally as in \cite{ACScaling} is of direct interest.

\end{enumerate}

We are planning to address some of these questions in the near future.

\section{Acknowledgements}

We would like to thank F. Aprile,  B. Doucot, B. Gouteraux, T. Ishii, D. van der Marel, J. Ren,  C. Rosen, X. Zotos,  and especially J. Zaanen for fruitful discussions. We would also like to thank B. Gouteraux for a critical reading of the manuscript.

This work was supported in part by European Union's Seventh Framework Programme
under grant agreements (FP7-REGPOT-2012-2013-1) no 316165, the EU program ''Thales" MIS 375734
and was also cofinanced by the European Union (European Social Fund, ESF) and Greek national funds through
the Operational Program ''Education and Lifelong Learning" of the National Strategic
Reference Framework (NSRF) under ''Funding of proposals that have received
a positive evaluation in the 3rd and 4th Call of ERC Grant Schemes".

 \newpage
\appendix

 \renewcommand{\theequation}{\thesection.\arabic{equation}}
\addcontentsline{toc}{section}{Appendices}
\section*{APPENDIX}

%\section{Background Solution}
%\label{ap:background}

%{\bf\Large This is missing. Can you add.}

\section{Effective Lagrangian  and Equation of Motion for Gauge fluctuations}
\label{ap:efflag}
In this appendix we present the technical details of the derivation of the equation that determines holographically the two-point
correlator for the current and the associated AC conductivity.

We substitute for the gauge field the background value as well as a linearized perturbation $A= A_{(0)}+\epsilon a$. $\epsilon\ll 1$ is a small parameter that we
use to derive the linearized perturbation equations.
\begin{multline}
\mathcal{L} = -\mathcal{N}\sqrt{-\det{\left(\gamma+\epsilon f\right)}}\\
\simeq -\mathcal{N}\sqrt{-\det{\gamma}}\left(1 + \frac{\epsilon}{2}\tr\left[ \gamma^{-1}f\right] + \epsilon^2\left(\frac{1}{8}\tr\left[\gamma^{-1}f\right]^2 - \frac{1}{4}\tr\left[(\gamma^{-1}f)^2\right] \right) \right)
\label{eq_LagrangeExpand}
\end{multline}
where we have defined
\be
\gamma=g+F_{(0)}\sp f=\mathrm d a\;.
\ee
We can split the matrix
 $\gamma^{-1}$ into its symmetric and antisymmetric parts $\gamma^{-1} = s^{-1} + \theta$ respectively.
The symmetric part of $\g^{-1}$ is as usual interpreted as the inverse open-string metric.
This induced world-volume metric has a horizon at the surface $u=u_\star$ instead of $u=u_0$, and it is given by
\begin{equation}
%\label{eq:WSmetric}
s = g -  F_{(0)}g^{-1}F_{(0)}.
\end{equation}

The quadratic part $L_{eff}$ in $\epsilon$ of (\ref{eq_LagrangeExpand}) will determine the dynamics
of the fluctuation $a$. It can be rewritten in the form of a Maxwell theory with
 a Chern-Simons contribution as follows
\begin{equation}
L_{eff}=-\mathcal{N}\sqrt{-\det{s}}\left( \frac{1}{4g_5^2}f_{MN}f^{MN} + \frac{1}{8\sqrt{-\det{s}}}\epsilon^{MNPQR}f_{MN}f_{PQ}Q_R \right),
\end{equation}
where the effective coupling is
\be
g_5^2=\sqrt{-\det{s}}/\sqrt{-\det{\gamma}}\sp \epsilon^{txyzu}=-\epsilon_{txyzu}=1
\ee
\begin{equation}
\label{eq:Q}
Q_R = -\frac{\sqrt{-\det{\gamma}}}{8}\epsilon_{MNPQR}\theta^{MN}\theta^{PQ}.
\end{equation}

Notice that now we are raising  indices with the open string metric $s$. In particular the induced line element is
\begin{eqnarray}
\nonumber ds^2 &=& s_{++}(dx^+)^2  + 2s_{+-}dx^+dx^- + 2s_{+u}dx^+du + 2s_{+i}dx^+dx^i  +  s_{--}(dx^-)^2 +  \\
&& 2s_{-i}dx^-dx^i + s_{ij}dx^idx^j + s_{uu}du^2 +2s_{ui}dudx^i \, ,
\end{eqnarray}
The components of the world-volume  metric above are given in appendix \ref{ap:wsm} in terms of the background metric. Finally, the equation of motion for the perturbation is
\begin{equation}
%\label{eq:MaxFluct}
\partial_M\left( \frac{\sqrt{-\det s}}{g_5^2}f^{MN} \right) -\frac{1}{2}\epsilon^{NMRPQ}\partial_M Q_R f_{PQ} = 0\, .
\end{equation}
% where the corresponding $Q_R$ is given by
% \begin{equation}
% Q = -E_y g_{ii}h_-'dz
% \end{equation}

\section{The Open String Metric}
\label{ap:wsm}

In this appendix we give the explicit expression of the open string metric in terms of the background metric and the other background fields.
The charge density and the background electromagnetic field modify the induced geometry on the probe D-branes. Using eq. (\ref{eq:WSmetric}) we can write the non-vanishing components of the world-volume metric as
\begin{align}
\nonumber s_{++} = g_{++} + \frac{|\vec E|^2}{g_{ii}} + \frac{h_+'^2}{g_{uu}} \quad , \quad s_{+-} = g_{+-} + \frac{h_-'h_+'}{g_{uu}} \quad , \quad s_{+i} = \frac{h_+'h_i'}{g_{uu}} \quad , \quad  \\
\nonumber  s_{+u} = -\frac{E_y h_y' + E_z h_z'}{g_{ii}} \quad , \quad  s_{--} = g_{--} + \frac{h_-'^2}{g_{uu}} \quad , \quad s_{-i} = \frac{h_-'h_i'}{g_{uu}} \qquad\quad \\
\nonumber s_{ij} = g_{ij} + \frac{h_i'h_j'}{g_{uu}}  + \frac{E_iE_jg_{--}}{G_{+-}} \quad , \quad s_{iu} =  - \frac{E_i}{G_{+-}}\left(g_{+-}h_-' - g_{--}h_+' \right)\qquad\\
 s_{uu} = g_{uu} + \frac{(h_y')^2+(h_z')^2}{g_{ii}} + \frac{ g_{--} (h_+') ^2 + g_{++}(h_-')^2 -  2 g_{+-} h_-'h_+'}{G_{+-}}\, ,\qquad
\end{align}
with $i=y,z$ and where we use through all the appendices the notation $\vec E = (E_y,E_z)$, $\vec h = (h_y,h_z)$, $\vec E\cdot \vec h = E_yh_y + E_zh_z$ and $\vec E\times \vec h = E_yh_z - E_zh_y$. Please do not confuse $\vec h$ with the blackening factor $h(u)$ (see Eq. (\ref{eq:blackening})).

The components of  the antisymmetric matrix are
\begin{align}
\nonumber \theta^{+-} = -\frac{g_{ii}h_-' }{\det \gamma} \vec E\cdot \vec h' \quad\, ,\quad  \theta^{+u} = \frac{ g_{ii}^2}{\det \gamma}(g_{--}h_+' - g_{+-}h_{-}' )\,,\\
\nonumber \theta^{+y} = E_y \left(\frac{g_{ii} \left(g_{--} g_{uu}+h_-'^2\right)}{\det\gamma}+\frac{g_{--} h_z'^2}{\det\gamma}\right) -h_y\frac{ E_z' h_z' g_{--}}{\det\gamma}  \,,\\
\nonumber \theta^{+z} = E_z \left(\frac{g_{ii} \left(g_{--} g_{uu}+h_-'^2\right)}{\det\gamma}+\frac{g_{--} h_y'^2}{\det\gamma}\right) -h_z\frac{ E_y' h_y' g_{--}}{\det\gamma}\,,\\
\nonumber  \theta^{-y} = E_y \left(-\frac{g_{ii} \left(g_{+-} g_{uu}+h_-' h_+'\right)}{\det\gamma}-\frac{g_{+-} h_z'^2}{\det\gamma}\right)+\frac{E_z g_{+-} h_y' h_z'}{\det\gamma}\,,\\
\nonumber    \theta^{-z} = E_z \left(-\frac{g_{ii} \left(g_{+-} g_{uu}+h_-' h_+'\right)}{\det\gamma}-\frac{g_{+-} h_y'^2}{\det\gamma}\right)+\frac{E_y g_{+-} h_z' h_y'}{\det\gamma}\,,\\
\nonumber\theta^{-u} = \frac{|\vec E|^2 g_{ii} h_-'}{\det\gamma}+g_{ii}^2 \left(\frac{g_{++} h_-'}{\det\gamma}-\frac{g_{+-} h_+'}{\det\gamma}\right) \quad , \\
\nonumber\theta^{yu} = -\frac{E_y E_z g_{--} h_z'}{\det\gamma}+\frac{E_z^2 g_{--} h_y'}{\det\gamma}+\frac{G_{+-} g_{ii} h_y'}{\det\gamma} \,,\\
\nonumber\theta^{zu} = -\frac{E_y E_z g_{--} h_z'}{\det\gamma}+\frac{E_y^2 g_{--} h_y'}{\det\gamma}+\frac{G_{+-} g_{ii} h_z'}{\det\gamma} \,,\\
\theta^{yz} = -\frac{\vec E\times\vec h' \left(g_{+-} h_-'-g_{--} h_+'\right)}{\det\gamma}\,.
\end{align}
Finally having the components for $s$ and $\theta$ we can compute the vector $Q$ (\ref{eq:Q}), which reads
\begin{equation}
Q = -\vec E\times \vec h' (g_{+-}\mathrm dx^+ + g_{--}\mathrm dx^-) +g_{ii}h_-'\vec E\times\mathrm d\vec x .
\end{equation}

\section{Change of Coordinates}
\label{ap:change_coord}

In this appendix we show the $f's$ functions appearing in the change of coordinates  used %in the subsection \ref{sec:change_coord}
 to diagonalize the open string metric (\ref{eq:WSmetric}). For completeness we write again the transformation

\begin{eqnarray}
\nonumber \mathrm d x^+ &=& \mathrm d \tau + f^+_-(u) \mathrm dX^- + f^+_u(u) \mathrm d u   \, ,\\
\nonumber \mathrm dx_- &=& \mathrm d X^- + f^-_\tau(u) \mathrm d\tau + f^-_u(u) \mathrm d u   \, ,\\
\nonumber \mathrm dy &=& \mathrm dY + f^y_\tau(u) \mathrm d\tau + f^y_u(u)\mathrm d u   \, , \\
\label{eq:coordinatesTrans1} \mathrm dz &=& \mathrm dZ + f^z_\tau(u) \mathrm d\tau + f^z_u(u)\mathrm d u \, ,
\end{eqnarray}
with the radial coordinate rescaled as $u\to u_\star u$.  The functions $f's$  read

\begin{eqnarray}
f^+_u &=& b^2(h-1)\frac{ u_\star  \vec E\cdot \vec h'}{h \left(b^2 \vec E^2+g_{ii}^2\right)-b^2 \vec E^2}\\
f^-_u &=&-\frac{1}{2}(h+1) \frac{u_\star \vec E\cdot \vec h'}{ h \left(b^2 \vec E^2+g_{ii}^2\right)-b^2 \vec E^2}\\
f^+_- &=&-\frac{h_-'}{h_+'}\\
f^-_\tau &=& -\frac{h_-' \left(4 b^2 \vec E^2-g_{ii}^2 (h-1)\right)+2 b^2 g_{ii}'^2 (h+1) h_+'}{2 b^2 g_{ii}'^2 \left(2 b^2 (h-1) h_+'-(h+1) h_-'\right)}\\
f^i_u &=& E_i\frac{ u_\star \left((h+1) h_-'-2 b^2 (h-1) h_+'\right)}{2 \left(b^2(h-1) \vec E^2+g_{ii}^2h\right)}
\end{eqnarray}

{\tiny
\begin{eqnarray}
f^i_\tau &=& \frac{2 u^3 h\left(4 b^4 g_{ii}^2 (h-1) (h_+')^2 - (h_-')^2 \left(4 b^2 |\vec E|^2-g_{ii}^2 (h-1)\right)-4 b^2 g_{ii}^2 (h+1) h_-' h_+'\right)\left(b^2|\epsilon_{ij}E_j|(1-h)\vec E\times \vec h' + g_{ii}^2h h_i'\right)}{b^2 g_{ii}^2 \left( (h+1) h_-' - 2 b^2 (h-1) h_+' \right) \left(h \left(b^2 \left(|\vec E|^2 + 4 \left(\vec E\times\vec h' \right)^2 u^3 (h-1)\right)+g_{ii}^2 \left(4 (\vec h')^2 u^3 h+1\right)\right)-b^2 |\vec E|^2\right)}\,.
\end{eqnarray}
}

After using this transformations the new components of the metric can be written in term of the old one as follows
\begin{eqnarray}
\nonumber \tilde s_{uu} &=&  2 f^-_u ( s_{+-} f^+_u + s_{-y}f^y_u + s_{mz} f^z_u ) + s_{--}  (f^-_u)^2 + 2 f^+_u ( s_{+y} f^y_u + s_{+z} f^z_u + s_{u+} u_\star ) + \\
\nonumber &&  s_{++} (f^+_u)^2 + 2 f^z_u ( s_{yz} f^y_u + s_{uz} u_\star ) + 2 s_{uy} u_\star f^y_u + s_{yy} (f^y_u)^2 + s_{zz} (f^z_u)^2 + s_{uu} u_\star^2 ,\\
&&\\
\nonumber \tilde s_{\tau\tau} &=& 2 f^-_\tau ( s_{-y} f^y_\tau + s_{-z} f^z_\tau + s_{+-} ) + s_{--} (f^-_\tau)^2 + 2 f^y_\tau ( s_{yz} f^z_\tau + s_{+y} ) + \\
&& s_{yy} (f^y_\tau)^2 + 2 s_{+z} f^z_\tau + s_{zz} (f^z_\tau)^2 + s_{++} \, , \\
\nonumber  \tilde s_{--} &=& 2 f^+_- (s_{+y} f^y_-  + s_{+z} f^z_- + s_{+-}) + s_{++} (f^+_-)^2 + 2 f^y_- (s_{yz} f^z_m + s_{-y}) + s_{yy} (f^y_m)^2 +\\
 &&+2 s_{-z} f^z_- + s_{zz} (f^z_-)^2+ s_{--}  \, ,
\end{eqnarray}
and
\begin{eqnarray}
\tilde s_{yy} &=& s_{yy}\,, \\
\tilde s_{zz} &=& s_{zz} \,,\\
\tilde s_{yz} &=& s_{yz} \, .
\end{eqnarray}

\section{Fluctuations}
\label{ap:fluctuations}

This appendix provides the details on the computation of the equation of motion for the fluctuations. The solutions of this equations are necessary to obtain the retarded correlator among two charged currents.

In order to do so, %To study the fluctuations
 we will set $E_z=0$ and we will introduce the fluctuations $a_M(u,x^-,\tau,Y)$ into the Eqs. (\ref{eq:MaxFluct}). With this dependence in coordinates the system can be decompose  in terms of longitudinal and transverse modes respect to the $y$ direction.  We can also fix the gauge $a_u=0$. Thereafter, we  Fourier transform them as follow
\begin{equation}
 f(u,X^-,\tau,Y) \to f(u)e^{-i(\omega (\tau - \beta b^2 X^-) - k Y)}\, ,
 \end{equation}
where $\beta = \frac{2u_\star^2}{2u_0^2-u_\star^2}$ \footnote{We have identified $x^+$ as the boundary field theory time, however the new temporal coordinate and the former time are related at the boundary through the transformations (\ref{eq:coordinatesTrans}) as $\mathrm dx^+ = \mathrm d\tau - \beta b^2 \mathrm dX^-$. Therefore, we evolve in the Fourier transformation along the direction of $x^+ = \tau -  \beta b^2 X^-$.}. After some tedious computation we get a set of three differential equations plus one constraint in the longitudinal sector $(\tau, -, Y)$ and one differential equation in the transverse one ($z-$direction). The constraint read as

\begin{eqnarray}
\label{eq:constraint}
-k \tilde s^{yy} a_y'(u) + \omega\left( \tilde s^{\tau\tau} a_\tau'(u) - \beta b^2 \tilde s^{--} a_-'(u)\right) &=& 0 \, ,
\end{eqnarray}
 the longitudinal equations are
\begin{eqnarray}
\nonumber a_-''(u) +  a_-'(u)\frac{\left(g_5^{-2}\sqrt{-s}\tilde s^{uu}\tilde s^{--}  \right)'}{g_5^{-2}\sqrt{-s}\tilde s^{uu}\tilde s^{--}} - a_-(u) \tilde s_{uu}(\omega^2\tilde s^{\tau\tau}+k^2\tilde s^{yy})  + &&\\
\nonumber -\beta b^2\omega^2\tilde s_{uu}\tilde s^{\tau\tau}  a_\tau(u) + \beta b^2k\omega  \tilde s_{uu}\tilde s^{yy}a_y(u) - \frac{i\mathcal N Q_z'}{g_5^{-2}\sqrt{-s}}\tilde s_{--}\tilde s_{uu}\left( \omega a_y(u) + k a_\tau(u) \right) &=& 0 \, , \\
\label{eq:a-}&&\\
\nonumber a_\tau''(u) +  a_\tau'(u)\frac{\left(g_5^{-2}\sqrt{-s}\tilde s^{uu}\tilde s^{\tau\tau}  \right)'}{g_5^{-2}\sqrt{-s}\tilde s^{uu}\tilde s^{\tau\tau}} - a_\tau(u) \tilde s_{uu}(b^4\gamma^2\omega^2\tilde s^{--}+k^2\tilde s^{yy}) +  && \\
\nonumber- \beta b^2\omega^2  \tilde s_{uu}\tilde s^{--}a_-(u)   -  k\omega \tilde s_{uu}\tilde s^{yy}a_y(u) - \frac{i\mathcal N Q_z'(u)}{g_5^{-2}\sqrt{-s}}\tilde s_{uu}\tilde s_{\tau\tau} \left( k a_-(u) - \beta b^2\omega a_y(u) \right)  &=& 0 \, , \\
\label{eq:ay}&&\\
\nonumber a_y''(u) + a_y'(u)\frac{\left(g_5^{-2}\sqrt{-s}\tilde s^{uu}\tilde s^{yy}  \right)'}{g_5^{-2}\sqrt{-s}\tilde s^{uu}\tilde s^{yy}} - a_y(u)\omega^2\tilde s_{uu}(\beta^2b^4\tilde s^{--} +  \tilde s^{\tau\tau}) + &&\\
\nonumber b^2 \beta  k \omega   \tilde s_{uu}\tilde s^{--}a_-(u) - k \omega   \tilde s_{uu}\tilde s^{\tau\tau}a_\tau(u) + \frac{i \omega \mathcal N  Q_z'(u)}{g_5^{-2}\sqrt{-s}}\tilde s_{uu}\tilde s^{yy}(a_-(u) + \beta b^2 a_\tau(u)) &=& 0 \, ,\\
\label{eq:ay1} &&
\end{eqnarray}
and in the transverse sector we obtain
\begin{equation}
\label{eq:az}
\left( g_5^{-2}\sqrt{-s}\tilde s^{uu}\tilde s^{zz} a_z'(u )\right)' + g_5^{-2}\sqrt{-s}\tilde s^{zz} \left( \omega^2 \left( \beta^2 b^2\tilde s^{--}  + \tilde s^{\tau\tau} \right) + k^2\tilde s^{yy} \right)a_z(u) = 0  \,,
\end{equation}
notice that the Chern-Simons term vanishes in the last sector, because if $E_z=0$ the only non-vanishing component of $Q$ points in the $z$ direction
 \begin{equation}
 Q = -E_y g_{ii}h_-'dz.
 \end{equation}

Now it is possible to eliminate one of the equations in the longitudinal sector using the constraint (\ref{eq:constraint}) and the gauge invariant fields, ending with the same number of equations as propagating degrees of freedom. To do so, let us redefine the fields in term of its gauge invariant electric fields,
\begin{eqnarray}
\label{eq:E-}
\mathcal E_-(u) &=& -\left( a_-(u) + \beta b^2 a_\tau(u) \right)\, ,\\
\label{eq:Ey}
\mathcal E_y(u) &=&  a_y(u) + \frac{k}{\omega} a_\tau(u) \, ,\\
\label{eq:Ez}
\mathcal E_z(u) &=&  a_z
\end{eqnarray}
now using the following combination for the equations
\begin{equation}
- (M^- + \beta b^2 M^\tau ) \quad , \quad  M^y + \frac{q}{\omega} M^\tau
\end{equation}
and solving for $a_y'(u),a_-'(u),a_\tau'(u)$ from Eqs. (\ref{eq:constraint}), (\ref{eq:E-}) and (\ref{eq:Ey}) we obtain the two linearly independent equations for the longitudinal sector

\begin{eqnarray}
\nonumber \mathcal E_-''(u)  + \frac{ \left(b^2 \gamma  \omega  c^\tau_\tau \tilde c^-+c^-_- (k \tilde c^y-\omega  \tilde c^\tau)\right)}{\omega(b^2 \gamma    \tilde c^- -  \tilde c^\tau)+k \tilde c^y}\mathcal E_-'(u) + (b^2 \gamma  d^-_\tau + d^-_-)\mathcal E_-(u)  && \\
\label{eq:e-} \left(b^2 \gamma  d^y_\tau+d^y_-\right)\mathcal E_y(u)&=& 0 \\
\nonumber \mathcal E_y''(u) +  \frac{\left(\omega  c^y_y \left(b^2 \gamma  \tilde c^--\tilde c^\tau\right)+k c^\tau_\tau \tilde c^y\right)}{\omega(b^2 \gamma    \tilde c^- -  \tilde c^\tau)+k \tilde c^y}\mathcal E_y'(u) + \left(\frac{k d^y_\tau}{\omega }+d^y_y\right)    \mathcal E_y(u)  +&&\\
\label{eq:ey} \frac{k \tilde c^- (c^\tau_\tau-c^y_y) }{\omega(b^2 \gamma    \tilde c^- -  \tilde c^\tau)+k \tilde c^y}\mathcal E_-'(u) + \frac{ \left(k \omega  d^\tau_\tau-k (k d^y_\tau+\omega  d^y_y)+\omega ^2 d^\tau_y\right)}{b^2 \gamma  \omega ^2}\mathcal E_-(u) &=& 0
\end{eqnarray}
where the unknown functions $c's$ and $d's$ are the coefficients of the equations of motion and constraint, read it from Eqs. (\ref{eq:constraint}) - (\ref{eq:az}) as follow
\begin{align}
a_i''(u) + c_i^j a_j'(u) +d_i^j a_j(u)=0\,,\\
\tilde c^j a_j'(u) =0\,,
\end{align}
with $i,j$ taking values $-,\tau,y$. Notice that the equation for $\mathcal E_z$ is no written because it remains the same as Eq. (\ref{eq:az})

\section{Sources and Perturbative Solution\label{ape}}

In what follows we will explain the procedure to obtain the perturbative solution discussed in the subsection \ref{sec:dc_cond}. This solution was useful to obtain the DC conductivity and the relaxation time Eqs. (\ref{eq:sigmaDC}), (\ref{eq:tau_final}).

First of all we plug in the ansatz (\ref{eq:field_exp}) into the Eq. (\ref{eq:az_q0}), so  the system can be rewritten as follows
\begin{equation}
\label{eq:perturb}\partial_u\left(\alpha_0(u)\alpha(u)\partial_u\mathcal A^{(i)}_z(u)\right) = \mathcal S_z^{(i-1)}(u)\, ,
\end{equation}
where the index $i$ takes values $i=0,1,2$.  The particular form of the sources up to second order in frequencies are
\begin{align}
\frac{\ell^3u_0^2}{2\mathcal N}\mathcal S_z^{(0)} =  i\partial_u\left((1+u)\alpha(u)\mathcal A_z^{(0)} + (-1+u^2)\alpha(u)\log{(1-u)}\mathcal A_z^{'(0)} \right)&&\\
\nonumber \frac{\ell^3u_0^2}{2\mathcal N} \mathcal S_z^{(1)} =  \partial_u\left((1+u)\alpha(u)\log{(1-u)}\mathcal A_z^{(0)} + \frac{1}{2}(-1+u^2)\alpha(u)\log{(1-u)}^2\mathcal A_z^{'(0)} +\right.&&\\
 \left.i(1+u)\alpha(u)\mathcal A_z^{(1)} + i(-1+u^2)\alpha(u)\log{(1-u)}\mathcal A_z^{'(1)} \right) -\left(\frac{T_{eff}}{T}\right)^2\frac{ p(u)u}{(-1+u^2)} \, ,
\end{align}
with the zero-th order source $\mathcal S_z^{(-1)} = 0$, and  where we have defined
\begin{equation}
p(u) = -2g_5^{-2}\sqrt{-\tilde s}\frac{(-1+u^2)u_0}{b^2\ell \mathcal N u} \tilde s^{zz}\left( \frac{\tilde s^{--}}{\rho_-^2}  + \tilde s^{\tau\tau} \right).
\end{equation}

The general solution for eq. (\ref{eq:perturb}) can be written as
\begin{equation}
\mathcal A_z^{(i)}(u) = C_1^{(i)} + \int \mathrm du \frac{1}{\alpha_0(u)\alpha(u)}\left(C_2^{(i)} + \int \mathrm du \, \mathcal S_z^{(i-1)}(u)\right)\, ,
\end{equation}
the integration constants $C_2^{(i)}$ are fixed demanding regularity for the solution at the horizon and $C_1^{(i)}$ is directly identified as the non-normalizable mode.

If we start solving order by order, we get for the zero-th order solution
\begin{equation}
\mathcal A_z^{(0)} =  C_1^{(0)} =a_z(0)\, ,
\end{equation}
$C_2^{(0)}$ must vanishes in order to have a regular solution at the horizon. At the next order the source $\mathcal S$ doesn't vanishes, in consequence, regularity at the horizon demands that

\begin{equation}
C_2^{(1)} = -i\frac{\alpha(1)\mathcal N }{ \pi \ell^3 T_{eff} u_0^2 }a_z(0)\, .
\end{equation}
Finally the first order solution can be written as
\begin{equation}
\mathcal A_z^{(1)}(u) = i a_z(0)\alpha(1)\left[ \frac{ 1}{\alpha(1)}\log{(1-u)}  - \frac{1}{\alpha(u)} \log{ \left(\frac{1-u}{1+u}\right) } -  \int_0^u\mathrm dx\, \frac{\alpha'(x)}{\alpha(x)^2} \log{\left(\frac{1-x}{1+x}\right) }  \right] \, .
\end{equation}

If we repeat the same analysis at second order, regularity demands that
\begin{equation}
C_2^{(2)} = a_z(0)  \left(  \frac{\sigma_{DC}  }{2  \pi   T_{eff}u_\star}\log (2) -  \frac{1}{4} \frac{\ell^3u_0^2\sigma_{DC}^2 }{\mathcal N u_\star^2} \int_0^1\mathrm dx\, \frac{\alpha'(x)}{\alpha(x)^2} \log{\left(\frac{1-x}{1+x}\right) } \right)\, ,
\end{equation}
and in consequence the solution at second order is
\begin{eqnarray}
\nonumber \mathcal A_z^{'(2)}(u) &=& \frac{1}{u-1}\left(i \mathcal A_z^{(1)}(u) + i(u-1)\log{(1-u)}\mathcal A_z^{'(1)}(u) + a_z(0)\log{(1-u)}  + \right. \\
 && \left. T_{eff}^2 \left(\frac{8 \pi ^2 C_2^{(2)} \ell^3 u_0^2}{\mathcal N \left(1+u\right) \alpha (u)} - \frac{2 \pi ^2 b^2 a_z(0) \ell^4 u_0 p(u) \log \left(1-u^2\right)}{\left(1+u\right) \alpha (u)}\right) \right.   + \\
 && -\left. \frac{2 \pi ^2 b^2 a_z(0) \ell^4 T_{eff}^2 u_0 }{\left(1+u\right) \alpha (u)}\int_u^1\mathrm dx\, p'(u) \log{\left(1-x^2\right) } \right) \, .
\end{eqnarray}

\section{Two charges hyperscaling violating geometries\label{f0}}

We will study the following Einstein Maxwell Dilaton system with the particularity of having two gauge fields in order to assess the interaction of different types of charge and their effect on the scaling of AC conductivities.

The action is a simple generalization of the EMD action
\be
 S=\int d^{d+2}x\sqrt{g}\left[R-{1\over 2}(\partial\phi)^2+V(\phi)-{ Z_1(\phi)\over 4}F_1^2-{ Z_2(\phi)\over 4}F_2^2\right]\, .
\ee
The equations of motions for the background (\ref{eq_hypermet}) read
\begin{eqnarray}
\nonumber\phi'^2 + d\frac{C''}{C} - \frac{DC'}{2C}\left(\frac{B'}{B} + \frac{C'}{C} + \frac{D'}{D}\right) &=& 0 \\
\nonumber\frac{1}{2} \left(\frac{C'}{C}-\frac{D'}{D}\right) \left(\frac{B'}{B}+\frac{(2-d) C'}{C}+\frac{D'}{D}\right)-\left(\frac{Q_2^2}{Z_2}+\frac{Q_1^2}{Z_1}\right)\frac{B}{C^d}-\frac{C''}{C}+\frac{D''}{D} &=&0 \\
\nonumber\left(\frac{Q_2^2}{Z_2}+\frac{Q_1^2}{Z_1}\right)\frac{B}{2C^d}-B V+\frac{d C'}{2 C} \left(\frac{(d-1) C'}{2 C}+\frac{D'}{D}\right)-\frac{1}{2} \phi '^2 &=& 0\,,\\
\label{eq_back1}
\end{eqnarray}
where $Q_1$ and $Q_2$ are the charge densities of $A_1$ and $A_2$ respectively. The gauge fields are given by
\begin{eqnarray}
A'_{1,t} &=& \frac{Q_1}{Z_1}\sqrt{\frac{BD}{C^d}} \sim Q_1 r^{\zeta_1 - z}\\
A'_{2,t} &=& \frac{Q_2}{Z_2}\sqrt{\frac{BD}{C^d}} \sim Q_2 r^{\zeta_2 - z}\,,
\end{eqnarray}
where $\zeta_i =  \frac{2 \theta }{q}+d_\theta - \kappa_i$.\footnote{We have assumed $Z_i\sim r^{\kappa_i}$ and the background functions correspond to the hyperscaling-violating Lifshitz  geometry (\ref{eq_hypermet})}

The fluctuation equations  for the two gauge fields are

\begin{eqnarray}
Z_1^{-1} C^{- \frac{d-2}{2}}  \sqrt{\frac{D}{B}}\left(Z_1 a_1' \sqrt{\frac{D}{B}} C^{\frac{d-2}{2}}\right)' - \frac{Q_1 D}{Z_1C^d}\left(Q_1 a_1 + Q_2 a_2 \right)  + \omega^2 a_1 &=&0\\
Z_2^{-1} C^{- \frac{d-2}{2}}  \sqrt{\frac{D}{B}}\left(Z_2 a_2' \sqrt{\frac{D}{B}} C^{\frac{d-2}{2}}\right)' - \frac{Q_2 D}{Z_2 C^d}\left(Q_1 a_1 + Q_2 a_2 \right)  + \omega^2 a_2 &=& 0.
\end{eqnarray}

At this point we shall try to find a hyperscaling-violating metric as a solution of the background  equations. To do so, we substitute (\ref{eq_hypermet}) into (\ref{eq_back1}) and obtain

\begin{eqnarray}
\label{eq_back_eval1}\alpha ^2+2 \theta  z-2 \frac{2\theta ^2}{d}-2 d (z-1) &=& 0 \\
\label{eq_back_eval2}2 (z-1) (d_\theta+z) - B_0 r^{\frac{2\theta }{d} + 2d_\theta}\left( \frac{Q_1^2}{Z_1}  + \frac{ Q_2^2}{Z_2}\right) &=&0 \\
\nonumber B_0 \left(r^{\frac{2\theta }{d} + 2d_\theta}\left(\frac{Q_1^2}{Z_1}  + \frac{ Q_2^2}{Z_2}\right) - 2r^{-\alpha  \delta +\frac{2 \theta }{d}}\right)+\\
\label{eq_back_eval3}\left(2 d_\theta ( (d+2 z-1)-\frac{\theta}{d}  (d+1)) -  \alpha ^2\right) &=& 0\,,
\end{eqnarray}
These equations can be analyzed in two simplified regimes depending on the ratio of $Z_1$ and $Z_2$ in the IR.

\subsection{ Background charged under a single gauge field ($Z_2 \gg Z_1$)\label{f1}}

This configuration makes the gauge field $A_{2,t}$ subleading in the IR region and non backreacting in the geometry. We will parametrize the couplings $Z_1$ and $Z_2$ in the following way
\begin{eqnarray}
Z_1 &=& r^\kappa\\
\label{eq_defZ21}Z_2 &=&  r^{\kappa + \delta\kappa}\,,
\end{eqnarray}
where $\delta\kappa$ will be required to be positive or negative depending whether the IR is at $\infty$ or zero respectively. The lagrangian parameters can be fixed in term of the parameters in the metric as follow.
\begin{eqnarray}
z &=& \frac{(\gamma -\delta )^2+2 d (\delta  (\gamma -\delta )+1)}{(\gamma -\delta ) (\gamma +\delta  (d-1))} \quad, \quad \theta = \frac{\delta  d^2}{\gamma +\delta  (d-1)} \\
Q_1 &=& \sqrt{\frac{2(z-1)}{d_\theta+z-1}} \quad , \quad \kappa = 2d_\theta + \frac{2\theta}{d}
\end{eqnarray}
$\delta\kappa$ is  not fixed, but under the constraint $\delta\kappa >0$ if the IR is at $r\to\infty$ or $\delta\kappa  < 0$ if the IR is at $r \to 0$. The Gubser criterion for this solution is
\begin{equation}
(z-1)(d_\theta+z-1) > 0  \quad,\quad
(d_\theta+z-1) (d_\theta+z) > 0 \quad,\quad
d_\theta(d z-\theta) > 0\,,
\label{gu1}\end{equation}
and the thermodynamic stability condition combined with the existence of a consistent near extremal black-hole is
 \begin{equation}
z (d_\theta-z)<0\,.
 \label{gu2}\end{equation}

Having the background geometry we can write the equations for the gauge field  fluctuations as
\begin{eqnarray}
\label{eq_a11}a_1'' + \frac{(3 - z - \zeta_1)}{r}a_1' + B_0 \omega^2 r^{2 z-2}a_1 -  \frac{B_0 Q_1^2}{r^2} a_1  &=& - \frac{B_0 Q_1Q_2}{r^2}a_2 \\
a_2'' + \frac{(3 - z - \zeta_2)}{r}a_2' + B_0 \omega^2 r^{2 z-2}a_2  - \frac{B_0 Q_2^2}{r^2}\nonumber r^{-\delta\kappa} a_2 &=&  \frac{B_0 Q_2Q_1}{r^2}r^{-\delta\kappa}a_1\,.\\
\label{eq_a21}
\end{eqnarray}
Now we perform the following change of coordinates
\begin{eqnarray}
r &\to& \left(\frac{x z}{w}\right)^{1/z}\\
a_i &\to& r^{\frac{z-2+\zeta_i}{2}}a_i\, ,
\end{eqnarray}
which allow us to write the equations in the following Bessel-like form
\begin{eqnarray}
\label{} x^2 a_1'' + x a_1' + \left(x^2 - n_1^2\right) a_1  &=&  \frac{Q_1}{z^2}   \left(\frac{\omega}{zx}\right)^{\frac{\delta \kappa }{2z}}Q_2a_2 \\
x^2 a_2'' + x a_2' + \left(x^2 - n_2^2\right) a_2  &=&\frac{Q_2}{z^2}\left( Q_1 \left(\frac{\omega}{zx}\right)^{\frac{\delta \kappa }{2z}}  a_1 + Q_2 \left(\frac{\omega}{zx}\right)^{\frac{\delta \kappa }{z}}  a_2\right)\,,\\
\label{}
\end{eqnarray}
where
\begin{eqnarray}
n_1^2 &=& \frac{(d_\theta-z+2)^2+4 Q_1^2}{4 z^2} \\
n_2^2 &=& \frac{(\delta \kappa +d_\theta-z+2)^2}{4 z^2}\,.
\end{eqnarray}

We are interested in the IR behavior of the AC conductivities. With the equations written in this form and using the fact that $z\delta\kappa>0$, we can do perturbation theory expanding the fields as follows
\begin{equation}
a_i^{(0)}(x) = a_i^{(0)}(x) + \omega^{\frac{\delta\kappa}{2z}}a_i^{(1)}(x)+\ldots
\end{equation}
The zero-th order functions satisfy the homogeneous Bessel equations, with infalling boundary conditions. The solutions are
\begin{equation}
a_i^{(0)}(x) = -\sqrt{\frac{\pi }{2}} (-1)^{-3/4} e^{\frac{i \pi  n_i}{2}}H^{(1)}_{n_i}(x),
\end{equation}
where $H^{(1)}_{n_i}(x)$ are Hankel functions of the first kind. The equations for the first order perturbations are again Bessel equations, however in this case they are inhomogeneous
\begin{eqnarray}
x^2 a_1^{(1)''} + x a_1^{(1)'} + \left(x^2 - n_1^2\right) a_1^{(1)}  &=& F_1(x) \\
x^2 a_2^{(1)''} + x a_2^{(1)'} + \left(x^2 - n_2^2\right) a_2^{(1)}  &=& F_2(x)\,,
\end{eqnarray}
with
\begin{eqnarray}
 F_1(x)  &=& \frac{Q_1Q_2}{z^2}   \left(zx\right)^{-\frac{\delta \kappa }{2z}} a_2^{(0)}\\
 F_2(x) &=&\frac{Q_1Q_2}{z^2}\left(zx\right)^{-\frac{\delta \kappa }{2z}}    a_1^{(0)}.
\end{eqnarray}

The solution for this system can be found using variation of parameters, and can be written as follows,
\begin{eqnarray}
a_i^{(1)} = -H_n^{(1)} \int^\infty_x ds\,\frac{1}{s}H^{(2)}_{n_i}(s)F_i(s)+ H_n^{(2)} \int^\infty_x ds\,\frac{1}{s}H_{n_i}^{(1)}(s)F_i(s).
\end{eqnarray}
At this point we are not interested in solving exactly the integrals, but in checking whether this first order solution are in fact subleading respect to the zero-th order one, and the perturbative analysis is not breaking down. To do so, we will use the fact that the IR is at $x\to\pm\infty$ and  the asymptotic behavior of the Hankel functions is of the type
\begin{equation}
H_n(x) \sim x^{-1/2}\,,
\end{equation}
modulo oscillating functions for the amplitude. Using this scaling, we may check that the dominant part in the asymptotic expansion is of the form
\begin{equation}
a_i^{(1)} \sim x^{-\frac{3}{2}-\frac{\delta\kappa}{2z}},
\end{equation}
which is consistent with the perturbative analysis.

Finally, after knowing the system can be decoupled at leading order in the IR, we can follow the computation of \cite{cgkkm}, to  calculate the reflection coefficients of the analogue Shr\"odinger  problem. Hence, the conductivities of each gauge field can be computed
\begin{eqnarray}
\sigma_1 &\sim& \omega^m \qquad , \qquad m = \left| \frac{2 - 3 z - d_\theta}{z} \right| -1 \\
\sigma_2 &\sim& \omega^n \qquad , \qquad n = \left| \frac{\delta \kappa  +d_\theta-z+2}{z} \right| -1\,.
\end{eqnarray}
As the IR geometry is supported  by the charge $Q_1$, the fluctuations of $a_1$ will  feel this charge. That explains the same result for the conductivity as in the single charged gauge field Eq. (\ref{emd1}).  On the other hand the background is neutral for the second gauge field. Therefore, the conductivity depends on the background parameters $z,\theta$ and the exponent of the gauge coupling $Z_2 = r^{\kappa(\theta,d)+\delta\kappa}$, and the conductivity coincides with the one for a single neutral gauge field.

\begin{figure}[t]
\begin{center}
\includegraphics[height=.4\textwidth]{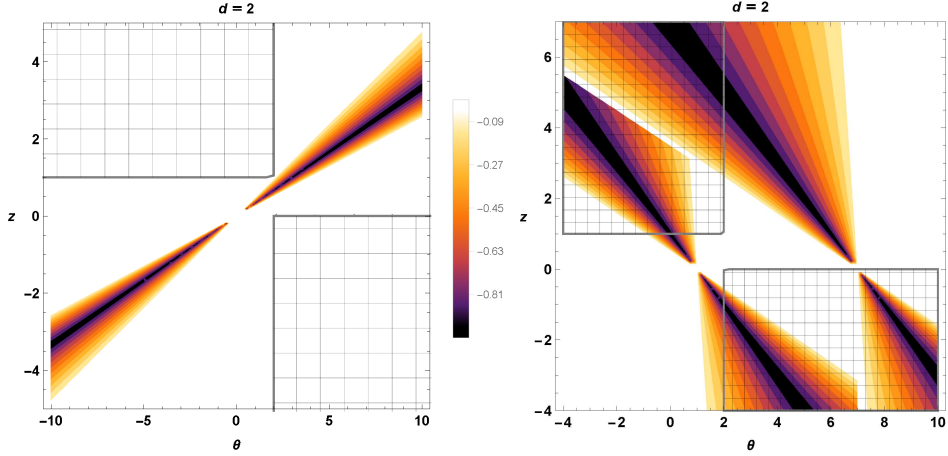}
\includegraphics[height=.4\textwidth]{figures/contoursigma21}
\caption{Contour plots to illustrate the region in the parameter space where the exponents $m$ (left) and $n$ (right) take negative values for the second charge not backreacting the background.   Both plots are made for $d=2$. The negative values for $m$ are outside the permitted region, but there is an allowed region for $n$. We have set $|\delta\kappa|=3$}
\label{fig_2charges1}
\end{center}
\end{figure}

We have plotted in figure \ref{fig_2charges1}, the Gubser bound of (\ref{gu1}), the thermodynamic constraint (\ref{gu2}) and the negative values of $m$ (left plot) and $n$ (right plot) for $d=2$. The value of $\delta\kappa$ has been  fixed to $|\delta\kappa|=3$. We observe that negative values of $m$ are excluded after imposing all the constraints. However the possibility  of having a negative $n$ is not excluded, as we observe in the right-hand side plot.

\subsection{ Both charges equally backreact the IR geometry $Z_2 \sim Z_1$\label{f2}}

In this case we will parametrize the couplings as follows
\begin{eqnarray}
Z_1 &\rightarrow& \frac{1}{2}\left(Z + \bar Z\right) \\
Z_2 &\rightarrow& \frac{1}{2}\left(Z  - \bar Z\right) \,,
\end{eqnarray}

and
\begin{eqnarray}
Z &=& r^\kappa\\
\bar Z &=& r^{\bar{\kappa}}\,,
\end{eqnarray}
the parameters $\kappa$ and $\bar{\kappa}$ will be chosen in such a way that $\bar Z$ is subleading in the IR. In other words, after the substitution of $Z_1$ and $Z_2$ into Eqs. (\ref{eq_back_eval1}) - (\ref{eq_back_eval3}), we obtain the following  set of  algebraic equations

\begin{eqnarray}
\alpha ^2+2 \theta  z-2 \frac{2\theta ^2}{d}-2 d (z-1) &=& 0 \\
\nonumber 2B_0 (Q_1^2+Q_2^2) r^{2 d_\theta-\kappa+\frac{2 \theta }{d}} -2 (z-1) (d_\theta+z) &=&-2 B_0 \left(Q_1^2+Q_2^2\right) r^{2d_\theta - 2 \kappa+\bar\kappa+\frac{2 \theta }{d}} \\
&&\\
 B_0 \left(r^{2 d_\theta-\kappa+\frac{2 \theta }{d}}\left(Q_1^2  +  Q_2^2\right) - r^{-\alpha  \delta +\frac{2 \theta }{d}}\right)+\\
\nonumber\left(2 d_\theta ( (d+2 z-1)-\frac{\theta}{d}  (d+1)) -  \alpha ^2\right) &=& - B_0 \left(Q_1^2-Q_2^2\right) r^{2d_\theta - 2 \kappa+\bar\kappa+\frac{2 \theta }{d}}\,,
\end{eqnarray}
if the following constraint on $\bar{\kappa}$ is satisfied
 \begin{eqnarray}
 c_{\bar\kappa}=\bar\kappa   - \frac{2 \theta }{d} - 2 d_\theta &<& 0 \quad , \quad r\to\infty \\
   \bar\kappa   - \frac{2 \theta }{d} - 2 d_\theta & >& 0 \quad , \quad  r\to 0,
\end{eqnarray}
at leading order the system becomes an algebraic system and the parameters can be fixed for the IR geometry. They read

\begin{eqnarray}
z &=& \frac{(\gamma -\delta )^2+2 d (\delta  (\gamma -\delta )+1)}{(\gamma -\delta ) (\gamma +\delta  (d-1))} \quad, \quad \theta = \frac{\delta  d^2}{\gamma +\delta  (d-1)} \\
Q_T &=& \sqrt{\frac{2(z-1)}{d_\theta+z-1}} \quad , \quad \kappa = 2d_\theta + \frac{2\theta}{d}
\end{eqnarray}
where
$ Q_T^2\equiv 2(Q_1^2 + Q_2^2)$.

The fluctuations in this limit obey the following equations of motion
\begin{eqnarray}
a_1'' + \frac{3 - z - \zeta_1}{r}a_1' + B_0 \omega^2 r^{2 z-2}a_1 -  \frac{B_0 Q_1}{r^2}\left(  Q_1 a_1 - Q_2 a_2\right)  &=&  \\
-r^{\bar\kappa}\left( r^{-\kappa-1}\left(\bar\kappa -\kappa \right)a_1' + 2 \frac{b_0Q_1}{r^2}r^{   - \frac{2 \theta }{d} - 2 d_\theta}( Q_1 a_1 + Q_2 a_2) \right) &&\\
a_2'' + \frac{3 - z - \zeta_1}{r}a_2' + B_0 \omega^2 r^{2 z-2}a_2  -  \frac{B_0 Q_2}{r^2}\left(  Q_1 a_1 - Q_2 a_2\right)  &=&  \\
r^{\bar\kappa}\left( r^{-\kappa-1}\left(\bar\kappa -\kappa \right)a_2' - 2 \frac{b_0Q_2}{r^2}r^{   - \frac{2 \theta }{d} - 2 d_\theta}( Q_1 a_1 + Q_2 a_2) \right) && ,
\end{eqnarray}
The form of the equations suggest the following redefinitions
\begin{eqnarray}
\label{eq:mapaA}A_1 &=& Q_1 a_2 - Q_2 a_1 \\
\nonumber A_2 &=& Q_1 a_1 + Q_2 a_2\,,
\end{eqnarray}
and to combine the equations as follows $Q_1 eq_2 - Q_2 eq_1 $ and $Q_1 eq_1 + Q_2 eq_2$. After doing so, we obtain the following set of equations

\begin{eqnarray}
A_1'' + \frac{3 - z - \zeta_1}{r}A_1' +  B_0 \omega^2 r^{2 z-2}A_1 &=& r^{-2+c_{\bar\kappa}}\mathcal{O}_1 \\
A_2'' + \frac{3 - z - \zeta_1}{r}A_2' + B_0\left(\omega^2 r^{2 z-2} - r^{-2}Q_T^2  \right)A_2   &=&  r^{-2+c_{\bar\kappa}}\mathcal{O}_2\,,
\end{eqnarray}
where
\begin{eqnarray}
\mathcal O_1 &=&\\
\nonumber &&\frac{2}{dQ_T^2}  \left(r \left(2 \theta +2 d^2-d (2 \theta +\bar\kappa)\right) \left((Q_1^2-Q_2^2)A_1'+2 Q_1 Q_2 A_2'\right) - 2B_0 d Q_1Q_2 Q_T^2 A_2\right)\\
\mathcal O_2 &=&\\
\nonumber && \frac{2}{dQ_T^2}\left(2 B_0 d A_2 \left(Q_1^4-Q_2^4\right)-r \left(2 \theta +2 d^2-d (2 \theta +\bar\kappa)\right) \left((Q_1^2-Q_2^2)A_2' - 2 Q_1 Q_2 A_1'\right)\right)
\end{eqnarray}

Notice that the factor $r^{-2+c_{\bar\kappa}}$ in the right-hand side of the previous equations  is even more subleading in the IR than the right-hand side of the analogous equations (Eqs. (\ref{eq_a11}) and (\ref{eq_a21}))  of the previous part.  Hence,  these set of equations can be written in terms of the Bessel equation and the conclusions will be exactly the same, as in the previous case. We skip this analysis and we will directly work with $\mathcal{O}_1$ and $\mathcal{O}_2$ set to zero at leading order.

In consequence, both redefined gauge fields diagonalize the system and satisfy the same equation, however $A_1$ is neutral and the field $A_2$ has charge $Q_T$.
This fact implies that the normalizable modes of $A_1^{(n)},A_2^{(n)}$ will depend linearly only on its own source respectively, hence,  $A_1^{(n)}=\langle 1\rangle A_1^{(0)}$, $A_2^{(n)}=\langle 2\rangle A_2^{(0)}$.\footnote{The quantities $\langle 1\rangle$ and $\langle 2\rangle$ are only functions of the frequency and the parameters of the system. } Now inverting the change of variable (\ref{eq:mapaA}) we can write the normalizable modes of $a_1$ and $a_2$ as follows

\begin{eqnarray}
\langle J_1 \rangle &=& \frac{2}{Q_T^2}\left(Q_1\langle 2\rangle A_2^{(0)}(a_1^{(0)},a_2^{(0)}) - Q_2\langle 1\rangle A_1^{(0)}(a_1^{(0)},a_2^{(0)})\right)\,,\\
\langle J_2 \rangle &=& \frac{2}{Q_T^2}\left(Q_1\langle 1\rangle A_1^{(0)}(a_1^{(0)},a_2^{(0)}) + Q_2\langle 2\rangle A_2^{(0)}(a_1^{(0)},a_2^{(0)})\right)\,,
\end{eqnarray}
which at the same time can be rewritten as
\begin{eqnarray}
\langle J_1 \rangle &=& \frac{2}{Q_T^2} \left(Q_1^2\langle 2\rangle + Q_2^2\langle 1\rangle\right)a_1^{(0)} + \frac{2Q_1Q_2}{Q_T^2} \left(\langle 2\rangle - \langle 1\rangle\right)a_2^{(0)}  \,,\\
\langle J_2 \rangle &=& \frac{2}{Q_T^2} \left(Q_2^2\langle 2\rangle + Q_1^2\langle 1\rangle\right)a_2^{(0)} + \frac{2Q_1Q_2}{Q_T^2} \left(\langle 2\rangle - \langle 1\rangle\right)a_1^{(0)}\,.
\end{eqnarray}

Finally we can read from the previous formula the two point correlator which reads

\begin{equation}
\langle J_i J_j\rangle = \frac{2}{Q_T^2}\left(\begin{array}{cc}
Q_1^2\langle 2 \rangle  + Q_2^2\langle 1 \rangle & Q_1Q_2\left(\langle 2 \rangle  - \langle 1 \rangle \right) \\
Q_1Q_2\left(\langle 2 \rangle  - \langle 1 \rangle\right) & Q_1^2\langle 1 \rangle  + Q_2^2\langle 2 \rangle
\end{array}\right)\,.
\end{equation}

Using the results of section \ref{acs} we can find the conductivities for each $A_1$ and $A_2$ and then substitute the results in the previous formula to extract the desired conductivities associated to $a_1$ and $a_2$.
We obtain

\begin{eqnarray}
\sigma_{11}  &\sim& \frac{2Q_1^2}{Q_T^2}\omega^{m_2} + \frac{2Q_2^2}{Q_T^2}\omega^{m_1} \qquad\quad , \quad m_1 = \left|\frac{(z-d_\theta-2)}{z}\right|-1\\
\sigma_{12} &=&\sigma_{21} \sim \frac{2Q_1Q_2}{Q_T^2}\omega^{m_2} - \frac{2Q_1Q_2}{Q_T^2}\omega^{m_1},m_2 = \left|\frac{(2-3z-d_\theta)}{z}\right|-1 \\
\sigma_{22}  &\sim& \frac{2Q_1^2}{Q_T^2}\omega^{m_1} + \frac{2Q_2^2}{Q_T^2}\omega^{m_2}\,.
\end{eqnarray}

It is important to emphasize that we are studying the IR behavior for the conductivities ($\omega\ll 1$), therefore, if the constraints on $z,\theta$ allow a negative value for $m_i$ this term will dominate in the frequency dependence for sufficiently small $\omega$. Notice also that this computation has been done at zero temperature. On the other hand, at finite temperature we would expect this scaling, to appear as an intermediate scaling  if $T\ll\omega\ll\Lambda$, with $\Lambda$ the interpolation scale for which the IR geometry flow to the conformal UV $AdS$ geometry.

\begin{figure}[t]
\begin{center}
\includegraphics[width=.49\textwidth]{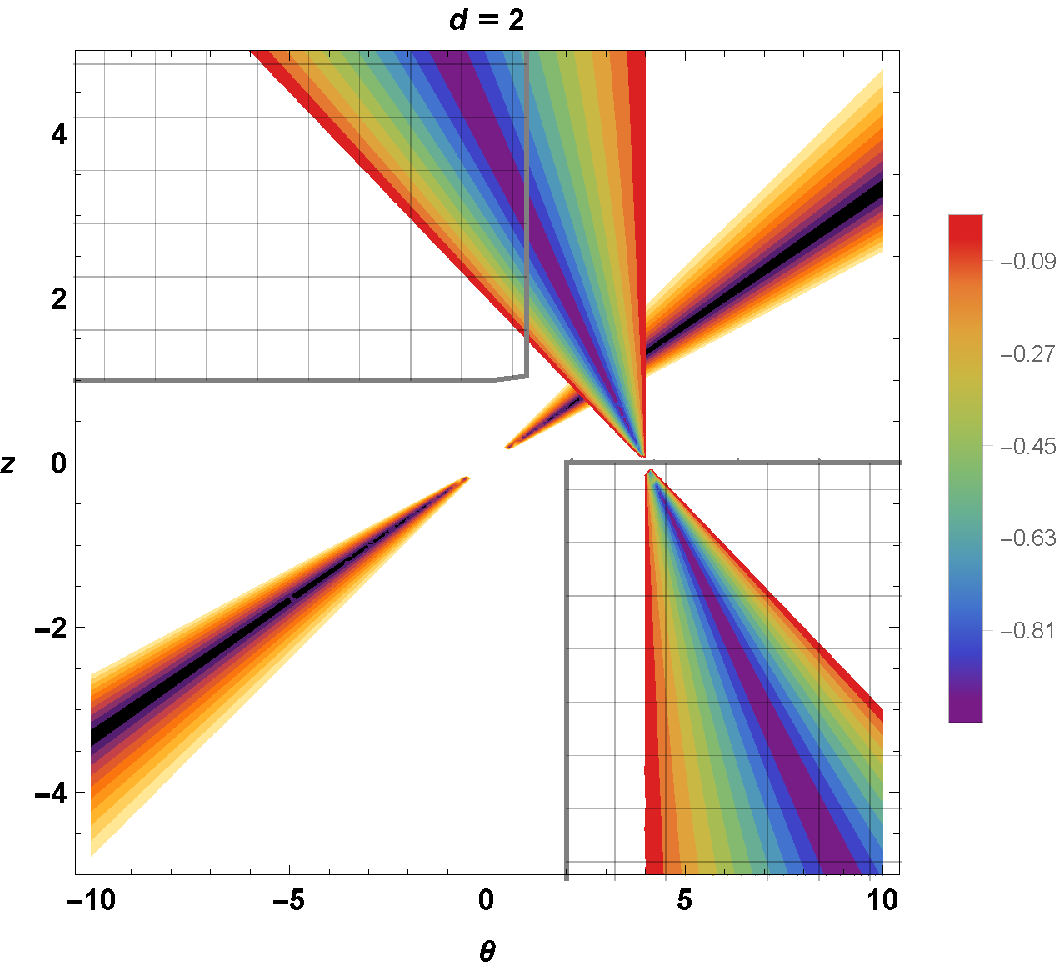}
\includegraphics[width=.49\textwidth]{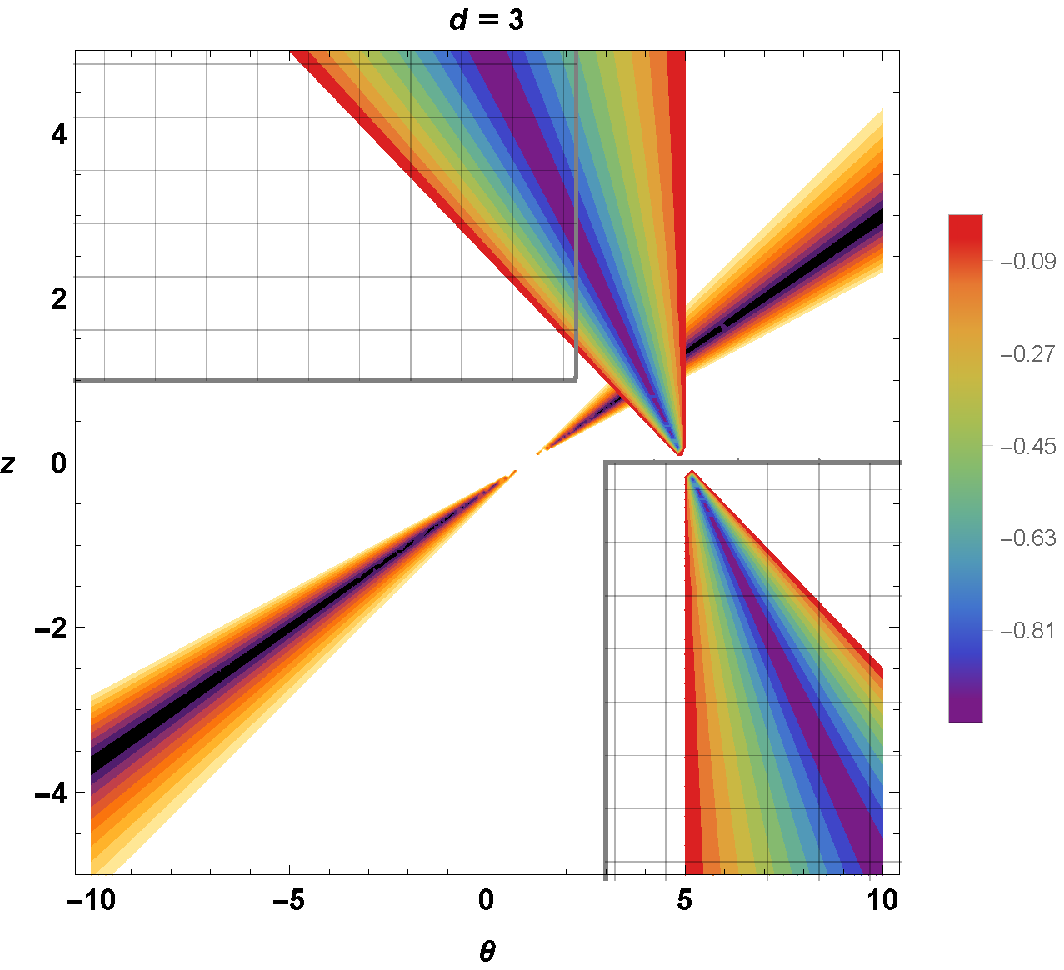}
\caption{Contour plots to illustrate the region in the parameter space where the exponents $m_1$ (rainbow coloured) and $m_2$ (scales of brown) take negative values for the two charges model.   Left: Conductivity for $d=2$. Right: Conductivity for $d=3$. The allowed values for the parameters are bounded by the gray mesh. The negative values for $m_2$ are outside the permitted region, but there is an allowed region for $m_1$.}
\label{fig_2charges}
\end{center}
\end{figure}

In figure \ref{fig_2charges} we show the regions of negative $m_1$ and $m_2$ for $d=2,3$. We observe that for both values of $d$ there is an allowed region of negative $m_1$
whereas that $m_2$ has to be always positive, in consequence we conclude that for small enough frequencies the conductivities will behave as

 \begin{eqnarray}
\sigma_{11} &\sim& \frac{Q_2^2}{Q_T^2}\omega^{m_1} \quad , \quad m_1 = \left|\frac{(z-d_\theta-2)}{z}\right|-1 < 0\\
\sigma_{12} &\sim& -\frac{Q_1Q_2}{Q_T^2}\omega^{m_1} \\
\sigma_{22} &\sim& \frac{Q_1^2}{Q_T^2}\omega^{m_1}\,.
\end{eqnarray}
Notice that $m_1$ and $m_2$ coincide with $n$ and $m$ respectively in the previous subsection, after setting $\delta\kappa=0$ in the formula for $n$. The reason for this similarity originates in the fact that in the previous configuration one of the gauge fields is charged and the charge of the second one is subleading in the IR, whereas in the present case the redefined gauge fields correspond to a configuration in which the background has charge $Q_T$ associated to the first gauge field, but is neutral respect  the second one.

\end{document}